# The Wealth of Nations:

# Complexity Science for an Interdisciplinary Approach in Economics

Klaus Jaffe

# Table of Contents

# *SUMMARY*

Classic economic science is reaching the limits of its explanatory powers. Complexity science uses an increasingly larger set of different methods to analyze physical, biological, cultural, social, and economic factors, providing a broader understanding of the socio-economic dynamics involved in the development of nations worldwide. The use of tools developed in the natural sciences, such as thermodynamics, evolutionary biology, and analysis of complex systems, help us to integrate aspects, formerly reserved to the social sciences, with the natural sciences. This integration reveals details of the synergistic mechanisms that drive the evolution of societies. By doing so, we increase the available alternatives for economic analysis and provide ways to increase the efficiency of decision-making mechanisms in complex social contexts. This interdisciplinary analysis seeks to deepen our understanding of why chronic poverty is still common, and how the emergence of prosperous technological societies can be made possible. This understanding should increase the chances of achieving a sustainable, harmonious and prosperous future for humanity. The analysis evidences that complex fundamental economic problems require multidisciplinary approaches and rigorous application of the scientific method if we want to advance significantly our understanding of them. The analysis reveals viable routes for the generation of wealth and the reduction of poverty, but also reveals huge gaps in our knowledge about the dynamics of our societies and about the means to guide social development towards a better future for all.

## *Caveat*

I must warn the reader about some quirky characteristics of this work, which might influence its interpretation and understanding. The aim of this work is to motivate students to engage in critical thinking and in doing their own research. This book strives to achieved learning by discovering information by oneself rather than through magisterial lessons of specialists. This book is not intended as a classical scholarly or comprehensive study. It is oriented to open windows to new ideas to people and students who want to explore novel knowledge in economics, evolution and complexity sciences. Each topic addressed in a chapter or section of a chapter, is an introduction to an extensive field of knowledge whose understanding and treatment require several volumes specifically devoted to the subject. In many cases, the state of the art in the field is highly developed and is the focus of a particular discipline. In other cases the field awaits development by researchers in the future. Only phenomena that have empirical evidence behind them are treated and statements about facts made in this book are based on scientific publication. The book is written for open-minded readers who want to explore novel heuristic possibilities of classical ideas with no claim to be possessors of the truth. The reader must share the conviction that there is progress in the understanding of ourselves and of the world around us, but that this progress follows circuitous routes that are not always obvious. By the time the book reaches a reader, most of the literature in this fast evolving field will be obsolete. The search for work by other authors to complement or deepen the subjects exposed here counts on the reader's ability to search documents on the Internet with a healthy dose of skepticism, specially the open-peer-reviewed Wikipedia. References in the text were designed so that the reader can search the phrase or key words in smart academic search engines such as Google Scholar. By doing this, the adventurous student will get access to the latest information and may start a fascinating new exploration of knowledge on its own. I encourage hands on study of these subjects. Most data required for repeating the analyses shown here and for designing new ones can be found at the World Bank Database analysis facility **databank.worldbank.org**, its equivalent at the Organization for Economic Co-operation and Development (OECD) **stats.oecd.org**, and at the International Monetary Fund (IMF) **www.imf.org/external/data.htm.** For example, to obtain Figure 3.2 you access databank.worldbank.org; then click successively on; Create Report; World Development Indicators; Country: Select all; Series: Health - Risk Factors - Causes of death by communicable diseases ...; Time: 2012; Map.

# 1. INTRODUCTION

## *What does Science tell us?*

Edward Wilson in his book Consilience argues that today, the biggest division within humanity is not one that exists between the races, between religions, or even, as many believe, between the educated and the illiterate. It is the gulf between scientific and pre-scientific cultures. Without the instruments and accumulated knowledge of the natural science, humans are trapped in a cognitive prison. They invent ingenious speculations and myths about the origin of what they see and about the meaning of their existence. But they are wrong, very wrong, because the world is too remote from ordinary experience for the human mind to imagine by simple intuition and rudimentary logic.

Economic and social thinking is in many cases based upon our intuitions. The attempts to analyze economic phenomena using the scientific method are still very rudimentary. Why is much of the modern world immersed in poverty? Why are some countries rich and others poor? What is the recipe to overcome poverty? Is there a key to success when implementing public policies? These questions were analyzed for the first time in a systematic way, by the Scottish economist Adam Smith (1723-1790) in his seminal book *The*

*Wealth of Nations*, more than 200 years ago. Despite the ancient search for an answer to these questions, we still lack a convincing answer. This search has been performed mainly from the cultural perspective of the richest and most powerful countries. Perhaps for this reason, a consensus that meets the expectations of politicians and economists in the various regions of the world has not yet appeared. Part of the difficulty in getting a satisfactory solution to these questions is their complex nature. By complex, I mean the problem - or the phenomenon - has many components and involves many different processes. The more complex a reality is, the greater the number of its components is. But our mind is unable to track reliably the relationships of more than 5 events, or follow the dynamics of more than 7 objects simultaneously. Therefore, our mind automatically processes the information of complex systems, selecting arbitrarily parts of the whole. For a cause-effect analysis of a reality of many components, there are plenty of possible explanations, each based on an arbitrarily selected subset of variables. Practically any explanatory theory of a perceived phenomenon can be validated with a selected number of data taken from the total available that makes up the complex reality. That is, a small number of evidence may be consistent with an arbitrary explanation of a complex phenomenon.

This state of affairs dominates the relationship of our mind with complex systems, and significantly affects the analysis of important questions such as the origin of the wealth of nations.

Almost every human on earth has his own explanatory theory of the causes for the occurrence of poverty, and explanations of what makes a person or a nation prosperous. Accordingly, a wide variety of contradictory theories exists. This does not help in solving the problem. An example may illustrate this point. You are an expert car mechanic and your neighbor asks for advice for a long night trip on a rough road he needs to make with his car. You recommend that he fix the front lights of the vehicle before the journey. The neighbor ignores you advice and drives off without front lights. The next day he calls you very angry to tell you that he had an accident completely unrelated to the lack of lights. The engine meted because it lacked oil, and you did not tell him that the car needed oil. He blames you, claiming that you are complete ignorant regarding cars and should refrain from giving advice on these matters.

The example of the stranded car illustrates the working of a complex system. In a car, there are many elements that ensure the proper operation of the system. It is clear that a car should not be driven at night without lights. But, it is also evident that it requires a well maintained engine, in addition to tires, brakes, steering systems, gearbox, battery, glass windows and many other components. This example illustrates that in a complex system there are a large number of components requiring attention from the operator. If this is the case of a relatively simple mechanical system - a motor vehicle - we might expect that the complexity will be much greater in the case of modern ever-changing national economies. But, modern economies are often managed with little regard to the intricate complex working of their processes and actors. This seems especially true of economics, as most people feel confidently expert in judging economic phenomena and dismiss the advice of professional economists. This lack of respect for experts does not seem to occur among technical professions such as medical doctors or shoemakers. Interestingly, a similar complaint was made by Plato when he referred to the way politics was handled in Athens some 2400 years ago.

Unfortunately, there is no single, infallible recipe to ensure a harmonious and uniform operation of a complex system. Certainly, continuous attention over many details is needed to maintain some control. But the complexity of a system makes it likely that no simple solutions exist. We analyze the small part of the problem we are seeing at a given moment and which we think understand best, even if it is not the most relevant to explain the problem. An illustrative story is the drunk looking under the lamppost for his lost car keys. A friend tells him that he probably lost the keys elsewhere. The response is that below the lantern there is light and therefore it is much easier to look for the keys there.

This past millennium was extremely significant regarding the evolution of ideas and construction of rationality. This was mainly due to the emergence of science as a fundamental tool for understanding ourselves, our history, and the world around us. Science is defined here, not in its various post-positivist conceptions, but as a method that increases the practical knowledge of our surrounding world that allows us, for example, to build planes and cure diseases successfully. The contributions of science have been primarily in the area of natural sciences, as these areas are less affected by our emotions and are more easily dealt with objectively. Many researchers that apply inductive and empirical analysis to phenomena that characterize human societies have been largely unaffected by the new knowledge, techniques and ideas produced by the natural sciences. During the last two centuries, social sciences and natural sciences have had a rather independent and relatively autonomous development.

The impression of many intellectuals is that in the XXI century we will achieve the synthesis of the social and natural sciences. This synthesis, popularized by Edward O. Wilson with the term "consilience", refers to the principle stating that evidence gathered by scientific disciplines has to converge in order to draw strong conclusions. Without consilience, the evidence is comparatively weak, and a strong scientific consensus is unlikely to emerge. This synthesis necessarily requires interdisciplinary approaches and attitudes that can explore problems at the interface of two or more disciplines, transcending the boundaries of classical academic disciplines.

Wilson says for example that "... the intellectuals, when they approach the study of behavior and culture, have a habit of talking about different types of explanations: anthropological, psychological, biological and others, appropriate to the prospects for each of the disciplines. I have argued that, intrinsically, there is only one kind of explanation. It crosses the scales of space, time and complexity to unite the disparate facts of the disciplines by consilience, the perception of a subtle web of cause and effect ... ". This view of science was born out of the experience of natural sciences. This is the case, for example, of physics, chemistry and biology. Although each studies different aspects of reality and employ different theories and research tools, they produce sets of theories that complement each other. Moreover, biological theories have their roots in chemical knowledge; theories of chemistry can be formulated as extrapolations of physics; and physics may be viewed as a reduction and abstraction of chemistry. At present, a consilient interdisciplinary approach of social sciences cannot be achieved by simply extrapolating our views from the physical and natural sciences. We must follow a path that crosses ecology, anthropology, the neurosciences and complex system sciences, in addition to biology, chemistry and physics. We have to be more assertive in the study of social dynamics, by adapting the scientific method and techniques to the study of societies in a much broader context. To do this, we must understand how science emerged and develops. We need a better understanding of the dynamics of the analytic functions of our mind and of the heuristics of the scientific method. Only with better scientific methods will we advance our knowledge of social phenomena.

Some of the disciplines that have a particular relationship to the study of human societies and their phenomena have been described as pre-scientific by philosophers of science. Many of them base their source of knowledge on long verbal descriptions of phenomena that are poorly understood. This is the stage in the development of knowledge that we might call scholastic. It focuses narratives that classifying and describing reality, but lacks integrative explanatory models that are universally accepted. An analogy can be found in human history, when mythological stories about stars and constellations gave way to astronomy and chemistry eventually replaced astrology and alchemy.

To transform the present day sociology, economy and psychology to more robust heuristic constructs, they need to grow abundant roots that are firmly grounded in natural sciences. A more advanced science will be able to address and understand phenomena of interest to humans and their societies, affecting positively our power to design tools that will benefit science and humanity in general. A broader view encompassing fundamental disciplines, achieving a conceptual continuity with these, will give more explanatory power and consilience to disciplines that address complex problems. It is therefore highly advisable for explanatory theories of psychological, economic and social phenomena, to seek common ground with the physical, chemical and biological sciences. It is also desirable for social science to place more emphasis on empirical approaches using the scientific method, which can address problems that occupy the social sciences today. These interdisciplinary bridges will also fertilize natural sciences, and will allow humanity to confront complex issues and ideas, accelerating the future development of all sciences.

This exercise will lead us, without a doubt, to stumble upon theoretical, moral, conceptual and ideological obstacles and pitfalls. Being true to our scientific insights, we know that we will make many mistakes in this exercise. But there is no progress without action, no action without mistakes, no mistakes without consequence, no learning without errors and no advance without costs.

Let us start action by proposing a simple preliminary pre-scientific procedure for our understanding of the development of science, inspired by the terminology used in the study of the evolution of insect societies. Here is a simplified narrative of multiple levels of scientific development of a discipline or field of knowledge.

- Most human intellectual activity, although very important for

humanity, has nothing to do with science. We may call this activity non-scientific or *unscientific* and I will avoid analyzing it further. Extensive analyses of the structure and evolution of literature, religion, music and the arts, and many other non-scientific academic disciplines, can be found elsewhere. However, not all disciplines classify unambiguously as scientific or non-scientific. For example, parts of mathematics and logic are tools used in science, but are not experimental natural science as such, as they do not use empirical falsifiable methodologies to test their hypothesis.

- A first level of development in a scientific discipline could be defined as the scholastic or *pre-scientific* level. At this level of development of the heuristic exercise, the focus is on efforts in building descriptive models. Rational thinking is used to follow logical steps, using common language as the main tool. Example of this are the works of Pythagoras, Socrates, Plato and various areas of contemporary knowledge, such as sociology, political science, cognitive science, management, among others.

- A second level could be called *para-science*, where the human rational effort focuses on describing the observable nature using language specially designed for this work. This level of scientific development is exemplified by natural philosophers, natural historians, social statistics, taxonomists and classical anthropology. Aristotle is often cited as the founder of pre-experimental empirical science, as he was a superb observer but never got into the habit of using experiment to test his ideas. For example, he maintained that women had fewer teeth than men, a hypothesis that he could easily falsify or verify by just looking into the mouth of the respective subjects, but requires to accept that empirical validation is more important than beauty when working with theories. This type of shortcoming is common also today. Salvador Dali, for example, discovered that ants have 6 legs instead of the 4 he painted in almost all of his paintings, very late in life. I do not know how he finally learned about this easily observable fact in a time many schoolchildren were already taught that insects have 6 legs.

- A third level could be called real or *eu-science*, exemplified by the version of science popularized by Galileo some 1900 years after Aristotle, where theory and experiment are essential parts of the rationalization of reality, and where the experimental results take precedence over human authority, religion, moral concepts and unscientific theories. The disciplines that exemplify this level of development of science are most parts of physics, chemistry and biology. As proposed by the philosopher Karl Popper, these eu-scientific theories should be falsifiable. Falsifiability is the property of a theory that allows it to be demonstrated as false by observation of natural phenomena or by experiment. Theories that cannot in principle be shown to be false are not considered to be eu-science.

- Many heuristic rational efforts seek to copy individual elements of the scientific method (i.e., using precise mathematical language and/or sophisticated language) but do not pay attention to the experimental confirmation of the theory or its falsifiability. Such efforts can be classified as *pseudo-science*.

- The recent emergence of the so-called science of complex systems, forces us to define a new level of scientific development. Computational models that explore artificial intelligence, global weather and climate models, the analysis of turbulent phenomena, cosmology and biological evolution, for example, cannot be described as eu-science. This is because the theories underlying these computational models are complex constructs that cannot be disproved with simple experiments. Their falsifiability is often indirect and partial. This kind of research activity could be called *meta-science*.

The consolidation of the meta-sciences as eu-sciences requires additional efforts from the researchers and practitioners. They need to combine in novel and creative ways, experiments with empirical observations, in order to eventually demonstrate a complex theory as false. Seeking consilience or demonstrating its absence, is a viable complement to establishing the falsifiability of complex theories. Demonstrating a lack of consilience in a complex theory is a way to falsifying the theory. This kind of research requires complex interdisciplinary and collaborative activity. In this sense, a complex phenomenon can be considered as scientifically understood, only if it can be described in a consilient way at different levels of complexity by physics, chemistry, biology, ecology and social sciences related to the phenomenon. In addition, empirical evidence at any level of complexity, which contradicts a part of the explanation of a complex phenomenon, is also proof that the explanation is false or incomplete.

The meta-sciences are still an emerging construct. For example, one of the many dimensions of interdisciplinarity is provided by the different degrees of spatial discrimination with which we study or perceive a phenomenon. It is not the same to study an isolated individual than aggregates or masses of them, regardless of the organism in question (Figure 1.1). Aggregates might show properties that could be independent of the character of the constituent part, such as network structure or viscosity, and other properties which will strongly depend on the behavior of the individual, such as economies. Sciences such as sociology and psychology study different levels of humans. Some of the phenomena sociologists and economist study might be independent of the psychology of the individual in the populations, but many other phenomena are strongly dependent on characteristics of the individual constituents. The historic development of these disciplines has been quite independent until now despite the fact that most of them seek to understand humans. To understand in depth socio-economic phenomena we will need to unite these approaches consiliently.

**Figure 1.1: The Individual and the Masses**

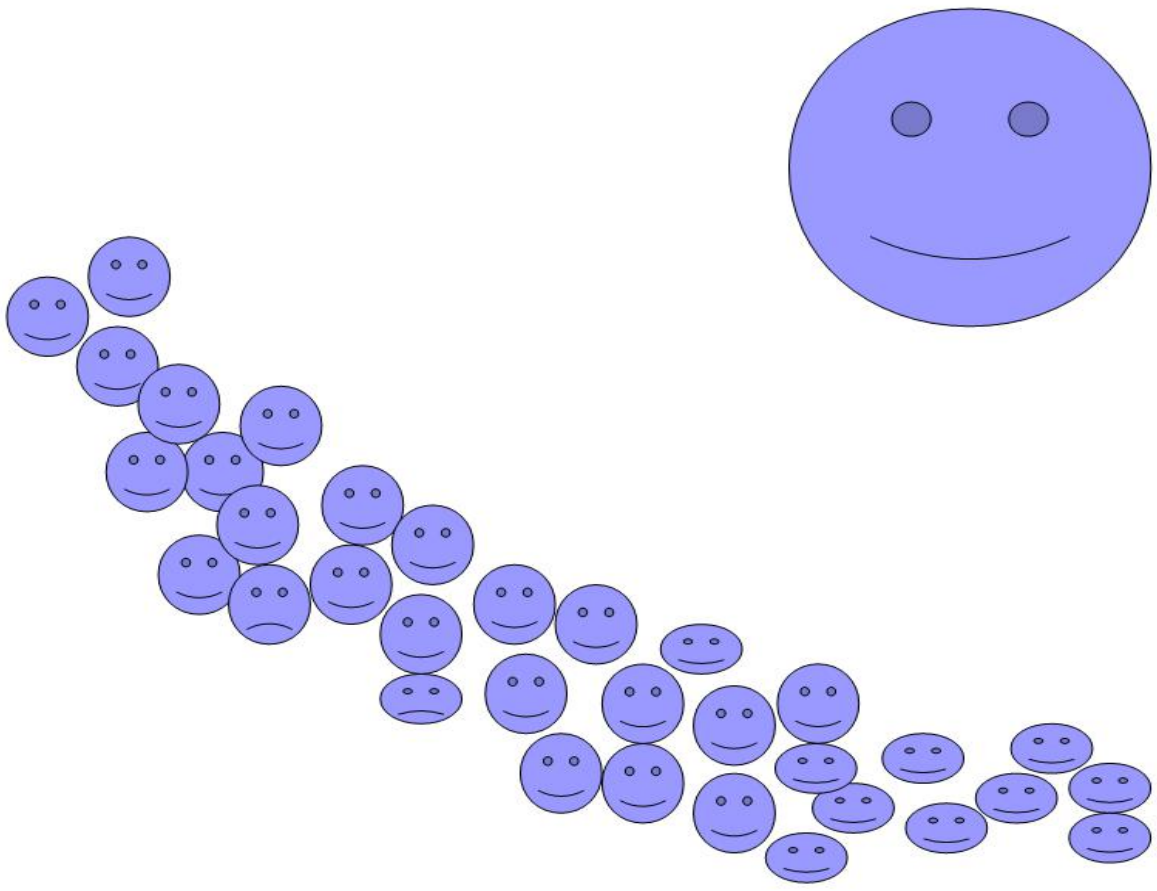

The Forest and the Tree or the Individual and the Mass, Two different levels of perception that require different types of analysis.

The simultaneous analysis of features at various levels of spatial segregation, despite the many facets that may exist at each level, is common practice in various sciences, such as statistical mechanics and thermodynamics, the science of emergence, the sciences studying self-organization, and what has been called complex system science. Complex systems consist of multiple components and assemblies and exhibits nonlinear dynamics, where the interaction of elements at a lower level produces properties or "emergent" phenomena at a higher level of aggregation (see Figure 1.2). These emergent properties require new disciplines or new tools to be understood. Some of these disciplines have emerged among the natural sciences, such as quantum mechanics, thermodynamics and complexity science, but many more are needed to understand real complex systems at the biological, ecological and social level.

**Figure 1.2 An emergent phenomenon we do understand**

The properties of water (H2

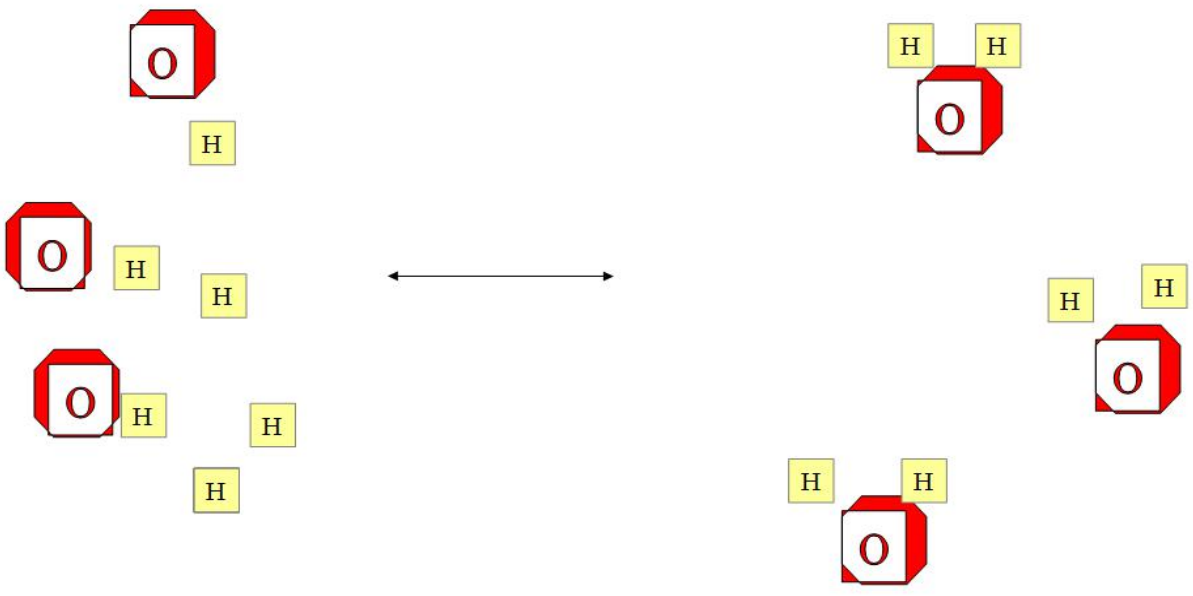

$6\,H + 3\,O = 3\,H_2O$

O) cannot be deduced in classical chemistry by knowing the properties of Hydrogen (H) and Oxygen (O), the constituent parts of water. It is only with the advent of quantum mechanics that we can explain the properties that emerge from this synthesis.

Complex systems have also a complex temporal behavior, which cannot be predicted by simple extrapolation of behavior observed in the past. In thermodynamic terms we say that these systems show irreversible processes. Most historic processes are irreversible and sciences that want to understand historic processes have to consider this. This insight opens a possibility of studying spatial and temporal aspects of complex social phenomena with fully accepted and validated methods borrowed from natural sciences, which might reconcile natural and social sciences. Methods for studying irreversible, complex dynamic systems were developed in parallel by several historical sciences. Each of the historical sciences, however, uses different time windows for their analysis (Figure 1.3). For example, cosmology handles time units in the range of billions of years. Geology is happy considering time spans of hundreds or thousands of millions of years. Evolutionary biology may study events in the range of hundred thousand to hundreds of millions of years. Paleontology and archeology use time ranges around hundreds of millennia, whereas the history of humanity is described in centuries. Economists think in quarters or in years. Psychology quantifies life stories in years and studies events that last less than an hour, but relies on physiology which counts time in the range of milliseconds. Each time window produces different understandings of reality and each of these understandings is achieved using different research tools. However, all these temporal visions overlap and aim to explain phenomena we observe at present, based on the analysis of the past, while dreaming to be able to predict the future. An interdisciplinary approach enriches our understanding of this complex web of temporal dynamic processes. If we take care to maintain consilience in this interdisciplinary analysis, the development of the disciplines involved will be complementary and synergistic. Contradictions that will appear in this interdisciplinary analysis will show the weakness of our reasoning and pinpoint gaps in our knowledge of these complex problems.

**Figure 1.3 Different time windows in the study of our world**

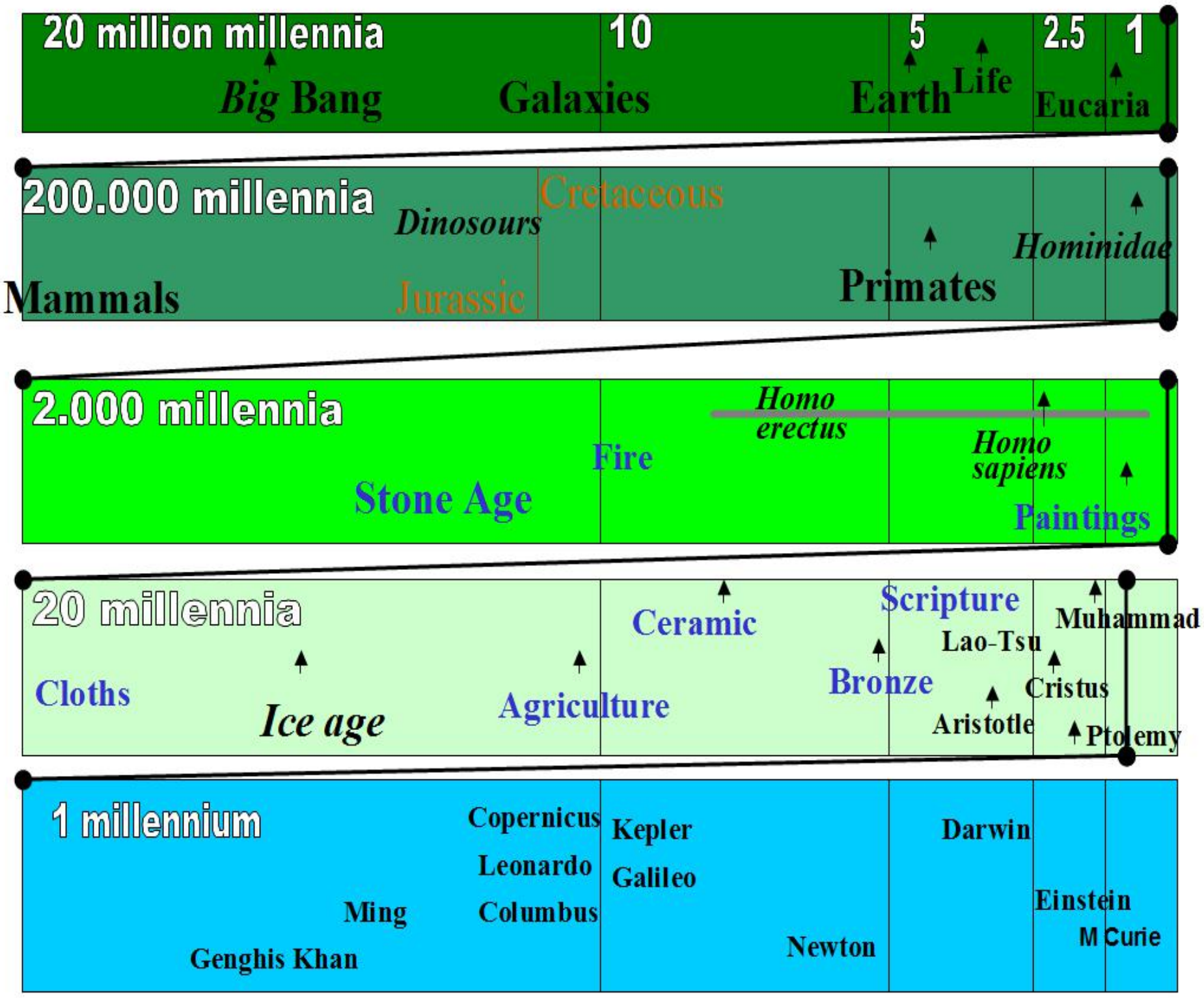

Here we will explore an analytical approach that can be classified as meta-science. In order to maintain scientific rigor in this interdisciplinary analysis of the complex, we will assume three fundamental premises as basic of any modern experimental science:

a. Theories have to be rational and logical, so that any human, and even a computer, can understand and verify them.

b. Our mind was formed by the forces of biological evolution to produce behaviors that allow us to reproduce efficiently, move and orientate in tree-dimensional space, and socialize. Our mind, however, has strong limitations to understand complex phenomena, and needs additional help (i.e. the experiment, mathematics, technology) so that we can capture the details of the physical, chemical, biological and social world around us.

c. Scientific theories must be falsifiable. They have to allow self-denial when their predictions are compared with reality. The answer to the questions formulated by our mind is in the hands of nature. The experiment, empirical observations and manifestations of nature, must prevail over any product of our mind. That is, empiricism must prevail over any dogma or creation of our mind, regardless of their intuitive beauty, antiquity, authority that proposes it, or prevailing intellectual fashion.

Experimenting and trying to show that a meta-theory is false can be very difficult. There is no magic recipe for it and we cannot anticipate the different ways and methodologies that will be used for this purpose in the future. In this book, we will walk a few steps in that direction, exploring a complex problem, from an interdisciplinary perspective, staying true to the fundamentals of the eu-science; and showing how methodologies commonly used in the study of complex systems, such as modeling and computer simulation; help us to advance the understanding of important social and economic phenomena.

With these clarifications, I want to address the central question of this book: Why are some societies rich and others poor? This question was formulated and discussed by Adam Smith in 1776 in his book The Wealth of Nations. Important advances in our understanding of this question have been made since then. The use of tools of natural sciences in this quest, although very rare, has been instrumental in advancing our understanding of social phenomena. Although it may be premature to speak of a science of the social and of economics in the same terms that we can speak of physical, chemical and biological science, the ambition of complexity science is to achieve a new synthesis that incorporates the humanities and social science in the same body of knowledge than natural science. In this search of an interdisciplinary and consilient science that will allow us to understand the dynamics of human societies, we have a number of eminent thinkers and scientists that preceded our effort. The Chinese Shang Yang (400-338 BC), Han Feizi (c280-233 BC) and Mozi (C429 BC) and Europeans Hugo Grotius (1583-1645), Thomas Hobbes (1588-

1679), François Quesnay (1694 - 1774), David Hume (1711-1776), Jean le Rond d'Alembert (1717-1783), Anne Robert Jacques Turgot (1727-1781), Joseph Louis Lagrange (1736-1813), Nicolas de Caritat or the Marquis de Condorcet (1743-1794), Adolphe Quetelet (1796-1874), and Auguste Comte (1798-1857), are just some of those who adapted methodologies of physical sciences to study social phenomena. The reader who is willing to explore the lives and work of these thinkers in any book or encyclopedia will discover how fruitful the exchanges between natural and social sciences can be.

Here just three examples of how ideas from natural sciences fertilize efforts in social sciences. Hugo Grotius took the idea of inertia of a moving object as described by Galileo, as inspiration in his effort to find basic laws governing the interactions in human societies. Modern social theories and specially sociobiology are still looking for general laws that describe social dynamics among humans and other animals. In the case of Quetelet, the flow of ideas was inversed. Quetelet developed statistical methods to understand the socio-political dynamics of his country. These statistical methods were then applied successfully to astronomy. Another example of this flow of ideas is Vilfredo Pareto (1848-1923). He showed how ideas and statistical tools developed to study social phenomena can be successfully applied to natural sciences. Although these last two examples are not exceptional, most examples of cross-fertilization between natural and social sciences involve ideas that started in physics or physiology. The centuries of history passed since Grotius and Galileo served to mature and improve a growing number of scientific tools at our disposal. Especially thermodynamics in conjunction with statistical mechanics, modern biology and our knowledge of the forces that drive the evolution of living beings, are relevant in building new foundations for better bridges between the sciences.

Here, we will take as a case study the phenomenon of the generation and accumulation of wealth and the occurrence of chronic poverty. Evidently, wealth and poverty are closely related and point to opposite directions. Poverty can be defined as a lack of wealth or an inability to produce or accumulate wealth. The analysis of these problems cannot be separated and will lead to improve our understanding of fundamental issues in economics. I hope that this effort may serve as an example of how ideas from different natural sciences fertilize research efforts in social sciences.

## 2. HISTORY

The Hellenic historian Herodotus of Halicarnassus (485-452 BC) already recognized that nations owe their destiny, largely, to the consequences of the actions of individuals and their interaction with environmental events. That is, history and geography are key shapers of a nation, and without a deep knowledge of them, we cannot understand either its current characteristics or the possible paths open to it.

The interpretations of the history of a nation depend largely on subjective factors that guide the historian in his analysis and search of information. They are, therefore, hardly conducive to achieving consensus among analysts of different ideologies. How can this dilemma be solved? How can subjective reasoning help us understand a phenomenon? We do not have a definitive solution to this dilemma, but we cannot ignore the historical and geographical factors either, if we are to understand the phenomenon of the emergence of the wealth of nations and the appearance of poverty in a comprehensive and rational way. We have no choice but to confront the historical and geographical analysis of the wealth of nations with the tools we have at our disposal. We will do so in a very general way and using very large time windows, to minimize the subjective burden of the analysis; and we shall do it very briefly, just indicating possible ways to guide future research that will eventually allow us to gain a better understanding of the problem.

## *Macro History*

For many authors, from the Roman Epicurean, Titus Lucretius Carus (96 to 55 BC.), to the English historian Michael Cook (*A Brief History of the Human Race*), the study of history is an activity that still does not allow a successful application of the scientific method. This perception is often extrapolated to all the so-called social and human sciences. I intend to illuminate possible paths to break with this paradigm. Some disciplines are more conducive to suffer the rigors of the scientific method than others are. Certainly, economics is the area of the social sciences having the greatest potential to merge consiliently with the natural sciences. Then, let us try a historical-economic analysis of the question: How does the wealth of a nation emerge?

The history of the human population and its particular dynamics, regarding the number of inhabitants on the planet, can be synthesized as in Figure 2.1, inspired by the work of J. Bradford DeLonge in 2002 and other authors. The estimates of these authors are summarized in the figure which shows the changes in total size of the human population on Earth during the last million years.

**Figure 2.1: History of the estimated human population of the planet**

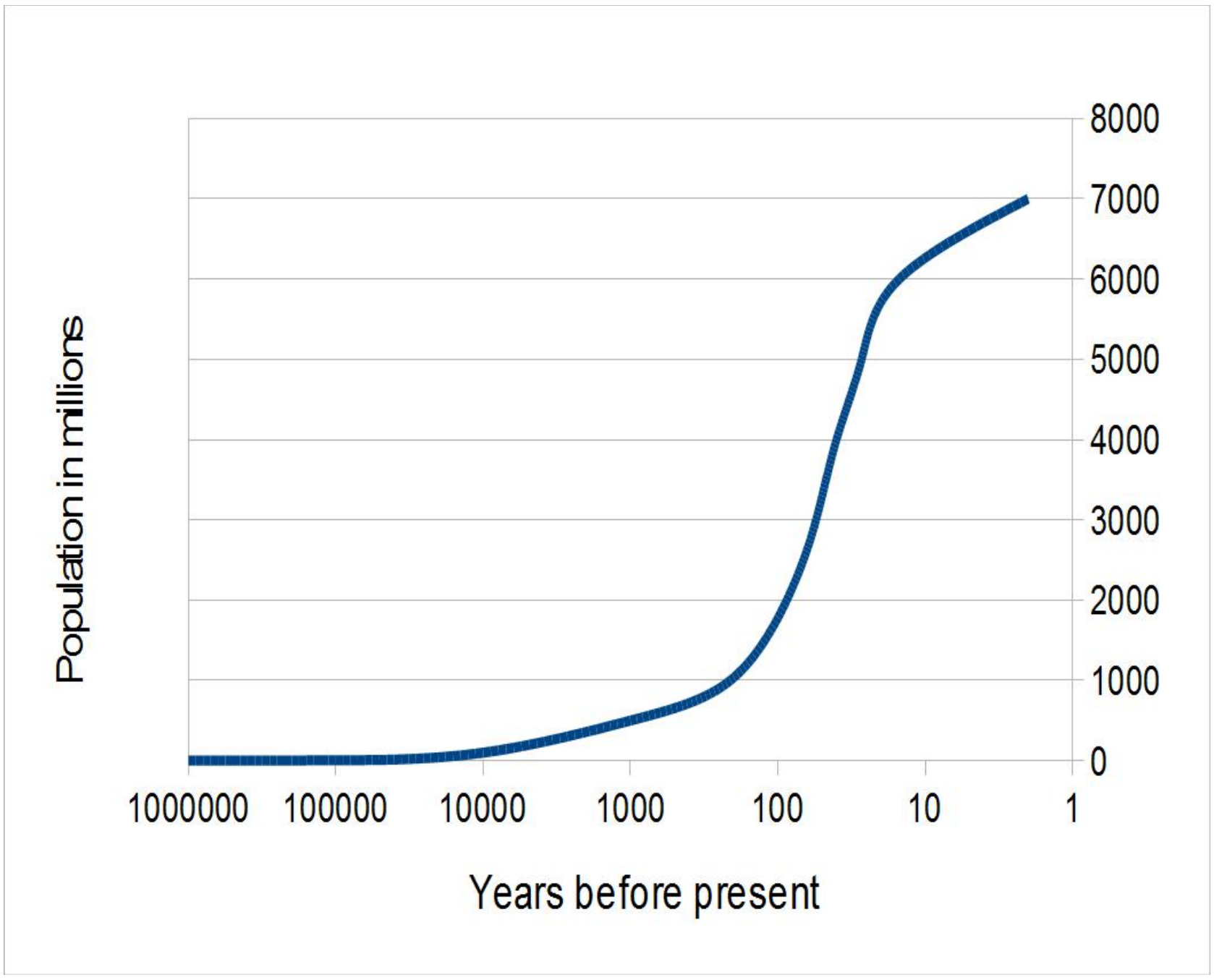

Estimates of the total number of humans on Earth, indicated on the vertical axis in millions of individuals, for different times during the last million years (horizontal axis). Both axes are in logarithmic scale. Data from DeLonge and Maddison Historical Dataset

This graphical representation of the estimated variations in population size of humanity shows clearly discernible stages. We recognize a stable phase that runs through most of human history with no significant change in human population that lasted several hundreds of thousand years. Then a phase with a very slow population growth, that started around 50,000 to 100,000 years ago and culminated some 10 to 5 millennia ago. This phase corresponds to the time when *Homo sapiens* was engaged in hunting, fishing and gathering fruits and tubers. During this time, *H. sapiens* had nomadic habits, was organized in families, which in turn grouped into clans and their social organization was based on the family structure. The end of this period, around 10,000 years ago, was triggered by the agricultural revolution and the domestication of plants and animals by our ancestors. Human aggregates and sedentary settlement in villages and small towns depended on crops and animal stock, which made them more susceptible to epidemic diseases and environmental fluctuations. During this period *H. sapiens,* organized into tribes, clans, and cities, which started building the State. But it was only in the last few centuries that human population exploded.

**Figure 2.2: History of material wealth of humanity**

Estimate of the total wealth produced or managed by humanity, expressed as Real Gross Domestic Product in billions of US $ (GDP on vertical axis) during the last million years. Both axes are in logarithmic scale. Data from DeLonge and Maddison Historical Dataset.

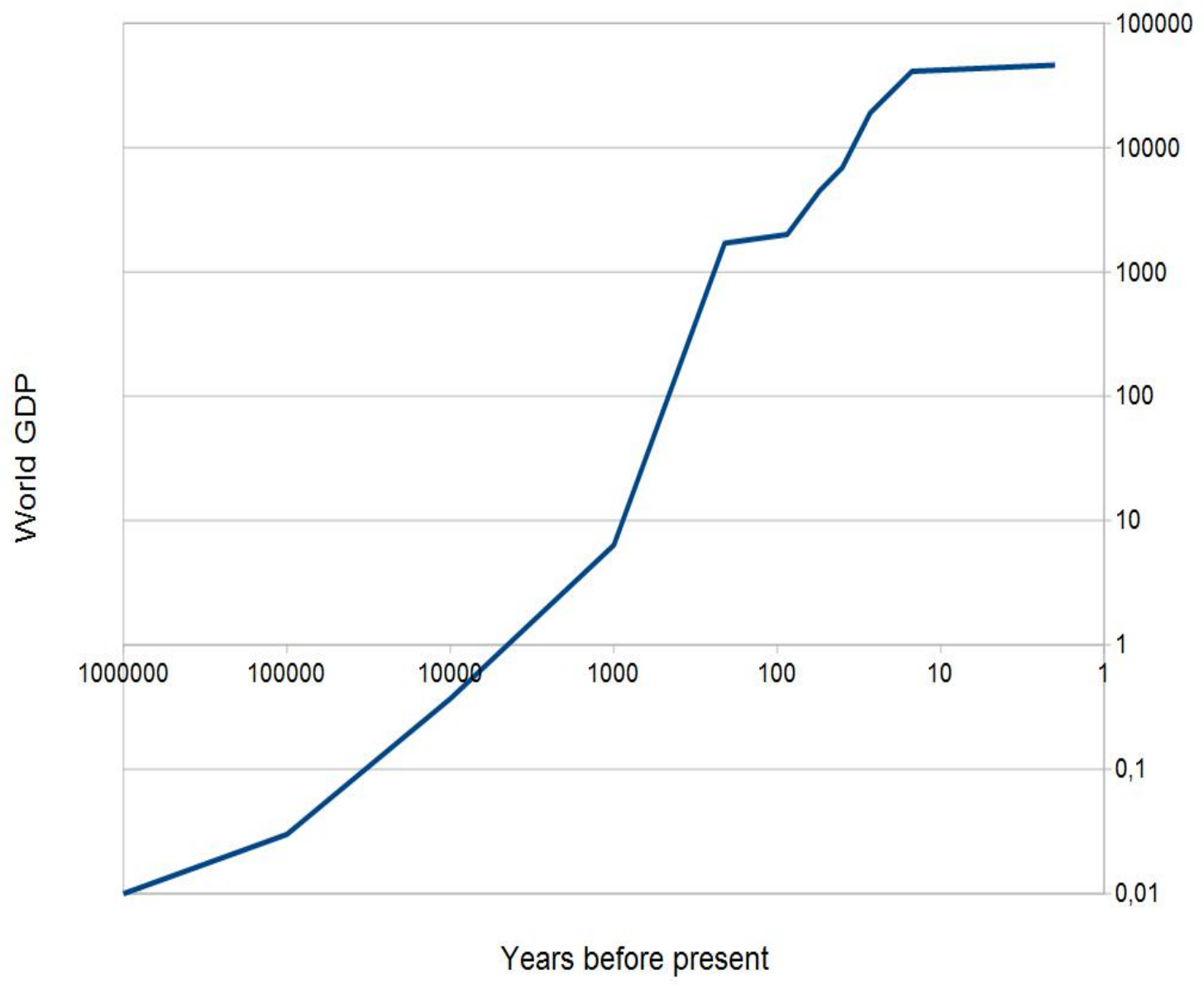

The organizational structures of human societies at all levels are rooted on the behaviors, values and motivations that keep the family working. Behaviors that enable an efficient structuring of the family have ancient biological origins and are shared by many primates and other mammals. This shared motivation allows us to live with pets, as some pets meet human needs. In humans, the psycho-biological forces that allow the operation of the family are extrapolated to support more complex social structures, as described in a very convincing way by the Chinese thinker Confucius (Kong-Fuzi, 531-479 BC). Friedrich Engels (1820-1895) describes the process of the formation of the family, the establishment of private property and the State, as smooth transitions in human history (*Origins of the Family, Private Property, and the State*). The proposals of Confucius, Engels and many others are being improved with modern anthropological and sociobiological information. Unfortunately, we will never know with certainty how this process was developed in detail. The important thing here is to note that each of these social organizations is linked to particular economic structures that require specific behaviors and social values for their optimal performance. These social organizations develop from the family organization, maintaining the primacy of the figure of "head of the family", which eventually morphs into "head of the tribe" or chieftain or leader of clan or nation. This phase in the history of humanity evolves into another phase, about 2000 years ago, with a moderate population growth and important growth in wealth. This phase coincides with the *Pax Romana*, the era of the great empires and the development of global trade. Interestingly, even in the imperial phase, human societies maintain social structures of power analogous to those of the family. More detailed analyses may reveal several additional sub-phases of growth, characterized by different dynamic constants.

In a human society economic growth is not only linked to the increase in population. It depends on the increase of the aggregate capacity of labor and hence also to human productivity. This is presented in Figure 2.2, which shows a dynamic similar to the one discussed above. This figure estimates the total accumulation of wealth managed by humanity. The same phases evidenced in the previous figure are observed in distorted form, which indicates that the increase in the wealth of humanity was accompanied with increases of productivity during the agricultural and industrial revolutions.

**Figure 2.3: Estimated Gross Domestic Product of the Planet**

Estimate of the total wealth produced or managed on average by human inhabitants of the Earth expressed in Real GDP in billions / population in millions (on the vertical axis) during the last million years (horizontal axis). The dominant economies are indicated in each period: gathering, hunting and fishing; agriculture and livestock; industrial production; technology and information. Both axes are in logarithmic scale. Data from DeLonge and Maddison Historical Dataset.

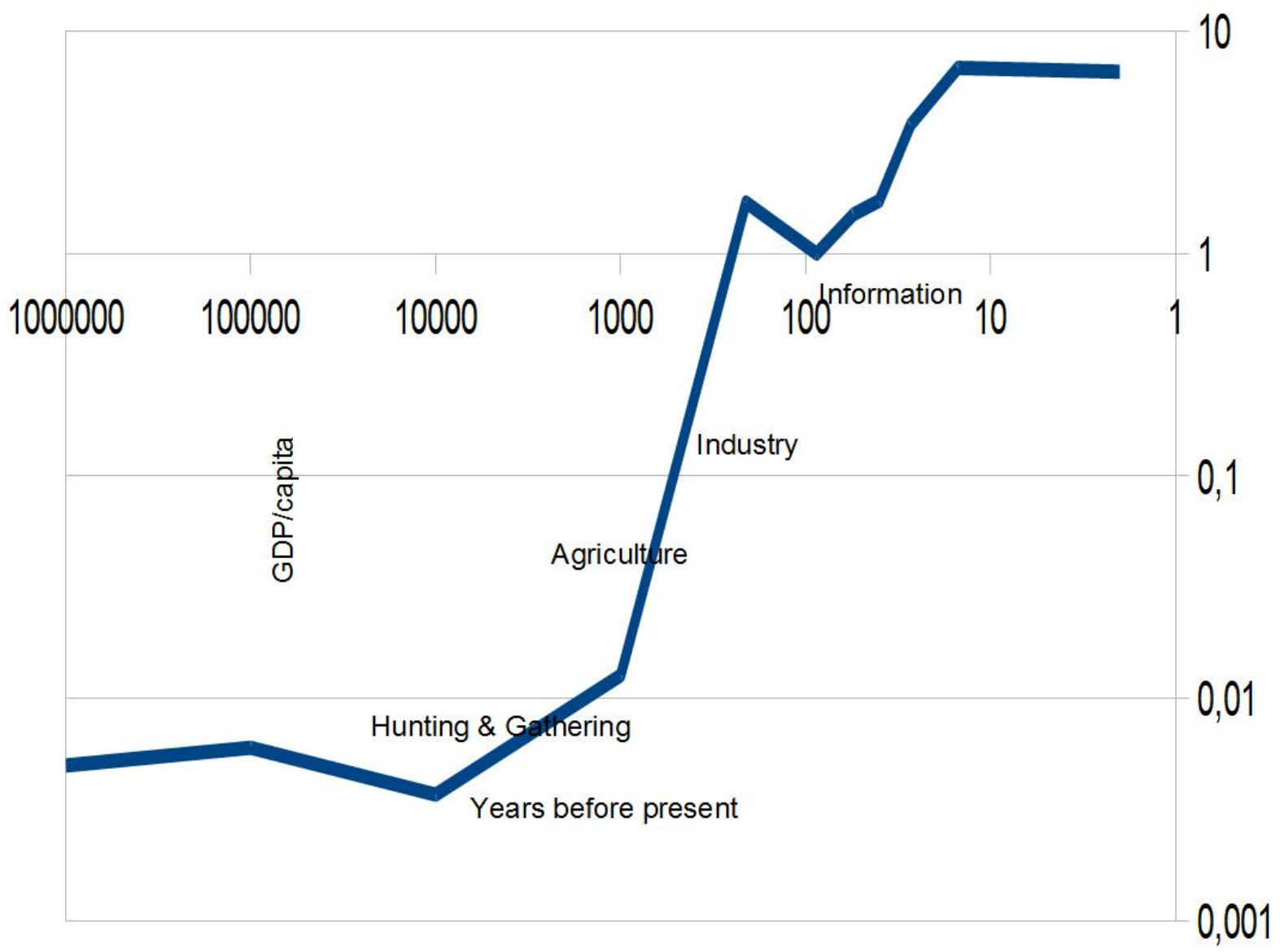

Growth in the human population, and of the total accumulated wealth, does not necessarily imply improvements in the quality of life of *H. sapiens*. If we represent the amount of material goods available to the average human inhabitant on the planet at different times, we can estimate approximately the changes in quality of life. In Figure 2.3, the average quality of life is estimated through a measurement widely used in economics, the gross domestic production per capita, expressed here in the equivalent purchasing power of the American dollar in 1990 (represented on the vertical axis), for periods ranging from a million years before the present until today (represented on the horizontal axis).

The data reveal with dramatic clarity that the phenomenon of accumulation of material goods by the average inhabitant of the planet, and therefore of the wealth of humans, is a very recent phenomenon of only a few centuries of existence. It is with the industrial revolution that sustained accumulation of wealth was made possible, and with it, a dramatic increase in the amount of goods and services accessible to the average person on the planet. Only with the industrial revolution, a significant increase of wealth per capita of the human population became possible in our history. If we accept that the opposite of wealth is poverty, we can analyze the phenomenon of occurrence of poor nations as nations with little or no economic growth that are left behind by others. In this sense, the phenomenon of the relative poverty of nations or of great part of their population arises at the same time as the rapid increase in wealth. This allows us to formulate our initial question more precisely: Why did some nations begin a rapid process of wealth accumulation and others did not?

What these three graphs show us is that the increase of individual well-being or the increase of the wealth per capita of humanity was triggered by the industrial revolution. Moreover, according to this analysis, the only historical event capable of producing a large per capita increase of wealth was the industrial revolution and the concomitant formation of the modern enterprise. The agricultural revolution, although at times could increase the individual well-being of some farmers, did not achieve a sustained increase in world well-being. Agriculture brought with it population growth, which in turn allowed the occurrence of widespread famines when the favorable climatic conditions changed, and the food that could be produced proved insufficient. Symptoms of malnutrition and epidemic diseases in many archaeological remains of the period seem to indicate that the agricultural revolution did not bring about sustainable economic improvement for the entire population.

Under this historical perspective, we could postulate that poor societies, except in very particular exceptions, are those that have not successfully completed - or have not started - their industrial and technological revolution. For some societies, this revolution, together with the emergence of strong public and private institutions, takes longer than for others, thereby producing detectable differences in the accumulated wealth between societies. According to this view of history, it is time and the accumulation of knowledge and technologies, which will enable pre-industrial nations to undertake industrialization and economic growth. However, there may be different routes and rhythms of industrial-economic development of a nation.

We do not have a coherent theory that is supported by empirical historical knowledge, which explains the transition from pre-agricultural societies to agricultural societies, and from these to industrial societies. Classical historic interpretations, like Marxism, proved to be incorrect, as humanity experienced historical sequences of economic systems that Marxism conceived as impossible. The same fate befell the historical theories of Hume, Hegel, Comté, Schmoller, Sombart, Max Weber and Spenger. Even less radical historical-economic interpretations as those popularized by Walt Whitman Rostow in 1960 (*Stages of Economic Growth*), who identified specific stages in the economic development of modern societies, proved to be too simple to interpret historical cases of many nations. This calls for caution in applying new simple models. For example, we could postulate many other forms of economic transition, such as transitions from pre-agricultural societies to industrial societies led by trade and/or mining, but many counter examples can then be used to falsify the theory. The first advance in our understanding of economic history is that there are processes and historical sequences that are irreversible and that the process of evolution of economies and of societies has certain stages that must be developed before others.

One of the interesting observations of historian Arnold Toynbee (*A Study of History*, written from 1934 to 1961) is that advances in human civilization usually occur in the periphery of the geographical area where the last streak of civilizational development thrived. This would explain why countries like Egypt, Iraq, Iran, and Syria, the cradle of Western civilization and the agricultural revolution, did not start the industrial revolution. These civilizations have values favoring the agricultural revolution that are strongly rooted. This seems not to be the case at the periphery of the Western agricultural civilization, as may be argued in the case of Germany, France, the Netherlands and especially England, which started the industrial revolution.

Another view was recently raised by the biologist Jared Diamond in his book *Guns, Germs and Steel*, who with a novel approach that uses concepts and knowledge taken from ecology and biological sciences, analyzes the emergence of contemporary civilizations. He suggests that favorable geo-climatic factors that produce great biodiversity suitable for the domestication of plants and animals, and the lack of diseases, enable societies advantaged with these benefits to produce more sophisticated societies. These societies might then produce more advanced technology, and with it dominate their fellows and nature, producing and accumulating wealth.

## *Industrial Revolution*

We might postulate that the relative poverty of a nation arises as a consequence of differences between industrial developments of societies. The industrialization of some nations and the slower rate of this process in others produce a situation where some nations are rich and others relatively poorer. The future prediction of the relative state of wealth of the nations of the world depends on whether there is a limit on the level of industrialization that can be reached in a given human society:

a- If the degrees of industrialization of a country does not have a top limit, then there will always be poorer countries and richer countries. As poor countries become rich, rich countries may continue to advance in their process of wealth accumulation.

b- If industrialization is limited by factors of the social or economic dynamics of the industrialization process, or by environmental and ecological factors, causing the economic growth rate to decrease as the process of industrialization advances, poor countries may eventually diminish the gap with the richer ones.

The industrialization of a nation also depends on the coherence of the society that supports it. A modern economy cannot be sustained by a tribal society. If we accept that all advanced societies developed a state based on the family, then modern States require historical roots that allow them to pre-adapt to the demands of the organization of a large society, based on rules and institutions that transcend the family and the clan. This means that a modern state cannot be improvised and requires time to develop. Thus, societies that initiate their industrialization starting from an agricultural economy will have it easier than societies of collectors or rent seekers.

The analysis presented here explains much of the economic growth experienced by humanity, and possibly, in a very general way, the difference in the wealth of nations. If this historical-economic analysis grasped some factual truth, we might deepen our understanding of the processes of development and industrialization and its temporal dynamic in order to find answers to the question initially posed. Interfering with the growth processes of the poorest countries, eventually managing to accelerate this dynamic could eventually reduce or eliminate the difference in wealth among nations. However, this analysis neither provides the details to understand the difference in the riches accumulated by specific nations, nor explains the poverty to which much of the world's population is subject to. Therefore, we must look for other factors that help us understand the dynamics of the process of wealth creation.

## *Emergence of the Modern State*

The wealth of a nation also depends largely on the soundness of its institutions, the strength of its businesses and its technological capabilities. Thus it is important to learn about the history of the development of the State and its institutions to understand in turn its economic history. What happened to human societies during the two most important economic revolutions? Did the passage of the economy from gathering and hunting to an agricultural economy, bring about a fundamental social change? Or, is it that this fundamental social change permitted the development of agriculture?

We cannot answer these questions to full satisfaction and with the entire rigor desired. However, we have important elements that outline the shape of a likely answer. We know that societies of collectors, fishermen and hunters are constituted by a very small number of individuals. The ties between individuals are founded on family relationships and the power in a family group emanates from biological instincts, where the oldest ancestor infuses more respect and where individual strength enables the exercise of authority. These societies have little differentiation of tasks. All must participate directly in the economic activities of the family group. The group is mobile and discards members who are not able to follow it. On the other hand, agricultural societies are sedentary and allow mixtures or peaceful coexistence of several family groups in a single region. This social structure that includes a larger number of individuals requires more complex social organizations. These are the clans and kingdoms that emerge as a political solution to these societies. The sedentary lifestyle and the more sophisticated political structure allow to group a larger number of individuals in areas increasingly densely populated, resulting in the communal house (castles, shabonos, fortifications, for example) or the community of houses (towns and cities).

Economic revolutions, both agricultural and industrial, had their effects on the society that saw their birth. Prior to the industrial revolution, possibly because of intense agricultural development, societies developed a national solidarity, as a replacement for solidarity to the clan and a deep experience of centralized government, with a development of a pool of professionals in administration, an educated and literate citizenry, which allowed the implementation of social standards by administrative laws by a government, rather than by force or arbitrariness of the strongest individual.

Another feature that seems to emerge from the historical analysis of several of the time windows shown in Figure 1.2 is that biological evolution in general, and human and social development in particular increasingly refine more the division of labor and the specialization of tasks performed by individuals. More recent or modern societies are much more structured and have more diversified productive sectors than ancient or primitive societies. That is, the evolution of societies, like the cosmological and biological evolution favors the emergence of structures that are increasingly more ordered or negentropic.

The modern state was born when the industrial revolution consolidated, along with the emergence of the private enterprise and the corporation. So far, it has required highly developed agricultural societies as a substrate for its formation. So is it possible to create modern societies from societies that have not completed their agricultural development? This question cannot be answered with certainty. What we do know is that the successful development of agricultural or sedentary societies eventually allows the development of industrial societies. The development of the national solidarity from clan solidarity, the deep experience with central government and other institutions, the preparation of large numbers of managers and educated citizens, the prevalence of the law and social structures that administer it, are all elements that develop in a sophisticated agricultural society. Hibbs and Olsson in 2004 managed to compare the wealth of 112 nations with the age of its industrial revolutions and got an impressive positive correlation: Countries that experienced an early industrial revolution are richer now. This would explain the difference between the wealth of Luxembourg, a country with no natural resources and no access to the sea, but with an old agricultural tradition, and Nigeria, for example, a country rich in natural resources and possessing extensive coastlines and natural harbors but without an important agricultural tradition. There are always exceptions: many of the nations of the Middle East were pioneers in the world in terms of the agricultural revolution starting it 10,500 years ago, and yet are poor today.

The conquest of new territories and the expansion of the frontiers of a nation or civilization also affect the underlying economic structures. The establishment of the European colonies in Siberia, Australia, and America began a new expansion that was not possible to achieve at the heart of the old cultures. That, according to Toynbee allowed the structuring of new economic production systems that still drive the technological advances of today. It is not by chance, then, that the center of the new technological advances of the world is in North America, one of the few rich countries whose population size continues to grow and whose territory still has large unpopulated areas.

What we learn from this brief historical review is that the past certainly affects the future and that a rigorous historical examination of the economy of nations will result in huge benefits for the understanding of our societies and their future prospects. For example, some societies have not mastered their agricultural revolution and are undergoing a transition from an economy based on collecting natural resources to one of expansion of information technology. Clearly, these countries will face different problems and limitations from those initiating the expansion of information technology after exhausting their agricultural and industrial growth. It is time to explore the vision of these phenomena from other disciplines.

# 3. GEOGRAPHY

## *Climatic factors*

Said al Andalusi, an Arab scholar who exercised as a judge in Toledo until his death in 1070, ranked the world's peoples into three categories, based on the geographical area where they live:

1. The inhabitants of temperate latitudes, such as Hindus, Persians, Chaldeans, Greeks, Romans, Egyptians, Jews, and Arabs who managed to develop cultures that promote science. (He did not include the Turks or the Chinese in this list, because despite having developed a very sophisticated artisanship, he did not consider that a mere technological development counted as science).

2. Inhabitants of cold northern areas, people who were blond and stupid because the sun's rays were scarce in those regions.

3. Inhabitants of hot southern areas, people who were black and dumb, due to the excess of solar radiation to which they are subjected.

According to the Tunisian scholar, Ibn Khaldun (1332-1406), this differentiation of human characters identified by Said al Andalusi was not due to race, religion or culture, but was caused solely by the climate to which individuals were subjected. That is, a black person that settles in a country of blond people eventually becomes blond. Nowadays, few people hold such radical ideas about the impact of climate on the personality of the people, but certainly, the climate and geography that constrains it must have an impact on the people and their culture.

The distribution of the human population on the planet is not uniform. We humans prefer to be near the sea, or close to large rivers and lakes, and like to settle in intermediate climatic ranges. Economic activity is higher when population is denser and so the distribution of populations should be related to the distribution of wealth. However, this relationship is not perfect. Several highly populated areas are poor (Bangladesh for example), whereas areas with relatively low population densities are rich (France, Australia or Canada for example). One geo-economic feature stands out. Countries with no access to the sea and located in the tropics are poor (Bolivia, Nepal, Chad, for example). Somehow, geography and climate determined by the geography, affects the economic development of a society.

The graphs in the former chapter suggested that industrial development somehow correlates with population growth. Here we have to remark that this correlation varies in different geographies. This phenomenon had already been observed by Arnold Toynbee (*A Study of History*, written from 1961 to 1934). The geographic latitude is one of the most important determinants of climate. Consequently, in Figure 3.1, the per capita income achieved by the countries is shown (each country is represented by a dot) relative to its latitude or distance from the Equator. That is, on the vertical axis, the country's per capita income is quantified, while on the horizontal axis, the absolute latitude or distance from the country's economic center to the geographical Equator of the planet is shown (0 in the horizontal axis).

**Figure 3.1: Relationship between GDP per capita and geographic latitude**

Examples of relationship between the wealth of a nation given as GDP in PPP estimated by the International Monetary Fund for 2013 (vertical axis) and distance from the Equator in latitude units of the economically most active area of that country (horizontal axis).

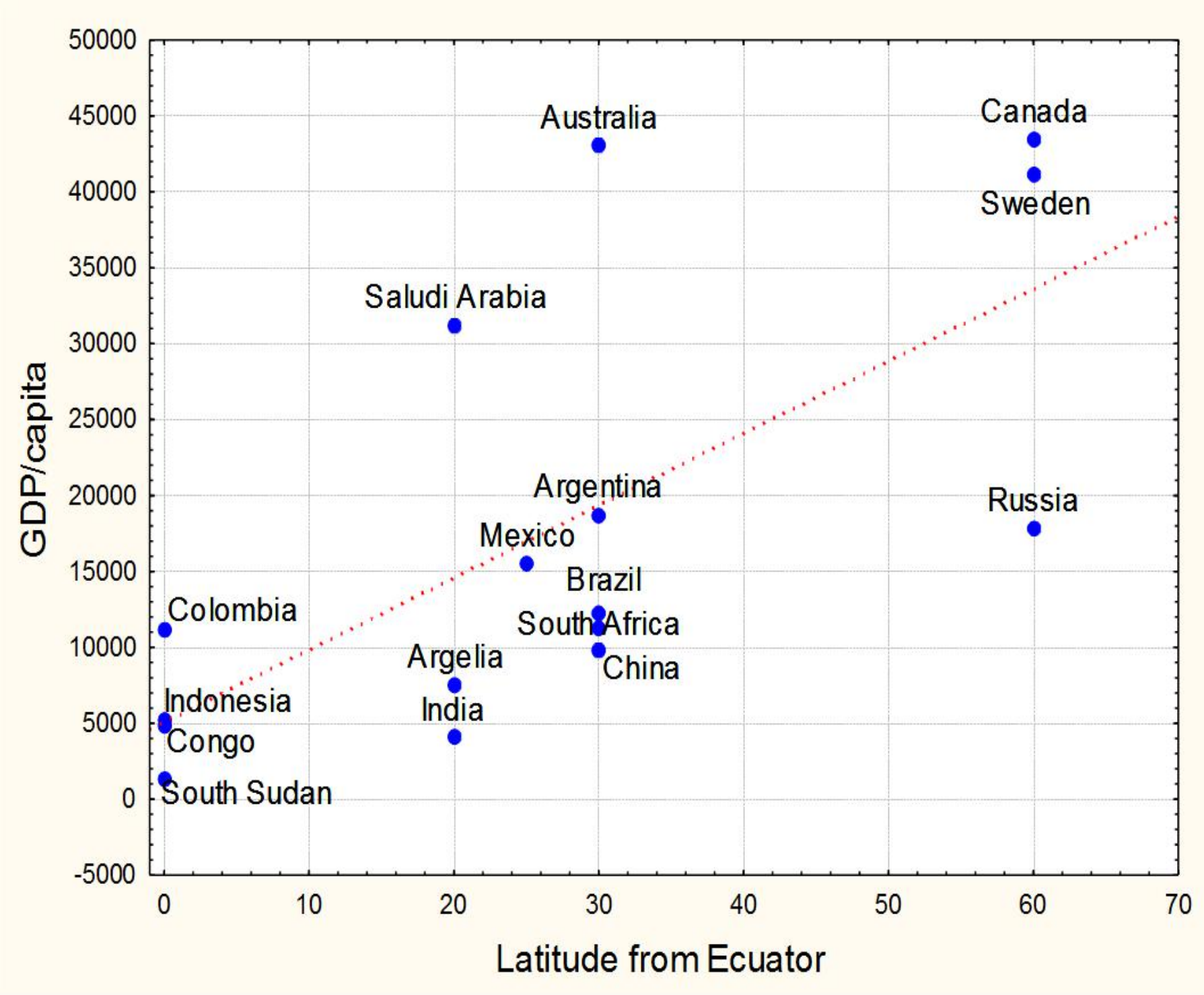

We see a positive correlation between these two variables in the figure: Countries located at away from Equator tend to be richer than countries located close or under the Equator. This result suggests the existence of a strong correlation between climate, determined by the latitude, and wealth of the nation. Expressed in another way, countries with high per capita income are located far from the geographic Equator, and therefore have temperate climates. Countries located in areas were warm climates prevail (near the Equator), tend to be rather poor. Some authors have involved the races inhabiting the different climates as the underlying cause to explain this difference in the ability to accumulate wealth. However, the history of past and present migrations shows us conclusively that the various races can adapt to nearly any climate they migrate to.

Why then climates that are more temperate favor industrial development and wealth accumulation? Several factors have been proposed, I will discuss four of them.

1- The existence of different seasons during the year, especially the presence of a cold winter, motivates savings and the planned use of resources. These attitudes, when shared by large segments of the population, favor economic institutions that promote accelerated economic growth. In contrast, societies that inhabit tropical climates do not require as much saving, and planning of the future use of resources is less urgent. In tropical climates, many possibilities exist for arranging a place to sleep without sheltering from the cold, and access to fruits, edible roots, vegetables and other foods is more or less constant during the year. This makes temperate climates, along with the practice of agriculture, more favorable to the formation of values, such as savings and communal work, essential for the successful operation of an industrial economy. Curiously, a similar effect on ant societies has been reported. It is mostly species living in temperate regions, where colonies need to accumulate food reserves for the winter, that ants have evolved slave maker and slave species. Slave maker species are very rare or absent among tropical ants. In contrast, ants applying a king of agriculture when breeding nutritive fungi need constant warm climates. They strive only in the tropics.

2- Entropy is another factor that may explain differences in economic growth between different climates. In warm climates, all systems have more entropy. Resources degrade more rapidly, fabrics fall apart earlier, weeds invade crops sooner, roads require more maintenance, machinery wear out faster, humans get sick more often, and working in hot environments is more tiring. That is, the temperature of the environment affects the rate of capital replacement. A higher capital requirement to produce a given amount of a utility produces a lower rate of economic development, when compared to that of climates that are more temperate, and entropy is lower.

3- The biogeography or the distribution of natural species on the planet predetermines the natural resources that *H. sapiens* can tame. Jared Diamond postulated that the geographical location and its biological potential affect the ability of accumulation of wealth of a nation. Diamond correlated the areas where humanity has domesticated species of plants and animals with the emergence of human societies and cultures, and obtains an amazing correlation. It is the presence of edible grain producing species - grasses with high capacity of cereal production - which allowed the domestication of these plants and thus enabled the growth of cities and their associated cultures. This occurred in a few places only. Wheat, lentils and barley are domesticated in Anatolia (the Asian part of modern Turkey) and then transferred to Europe. In China *H. sapiens* domesticates and develops crops such as soybeans, rice and silk, allowing and empire to flourish. Rice, the cow, the hen and a countless number of plants and animal species form the basis of economic development in the Hindustan.

4- The presence of parasites, diseases and plagues, such as malaria and the *Anopheles* mosquito that transmits it, prevents the successful formation of industrial cultures. Examples of this trend can be found in vast parts of Africa and parts of Latin America and tropical Asia. This means that the distribution of both, species potentially beneficial to man, and those that harm and parasitize him, affect his ability to establish successful economies and societies.

We can evaluate this last proposal quantitatively. For 2009, the 10 countries with the highest child mortality as reported by the World Health Organization were: Chad (20.9%), Afghanistan (19.9%), Democratic Republic of the Congo (19.9%), Guinea-Bissau (19.3%) Sierra Leone (19.2%), (Mali 19.1%), Somalia (18.0%), Central African Republic (17.1%) Burkina Faso (16.6%) and Burundi (16.6%). With the exception of Afghanistan, a country ravaged by war at the time, all are tropical countries on or near the Equator. A visual example for evidence of the relationship between the likelihood of becoming ill from transmittable disease and geography is given in Figure 3.2.

**Fig 3.2: Geography of health**

World map showing the geographical distribution of the likelihood of deaths caused by infectious diseases during 2012. The lightest colored countries have less than 5.4 % of deaths that can be attributed to communicable diseases and maternal, prenatal and nutrition conditions. The darkest colored ones have over 57.2 %.

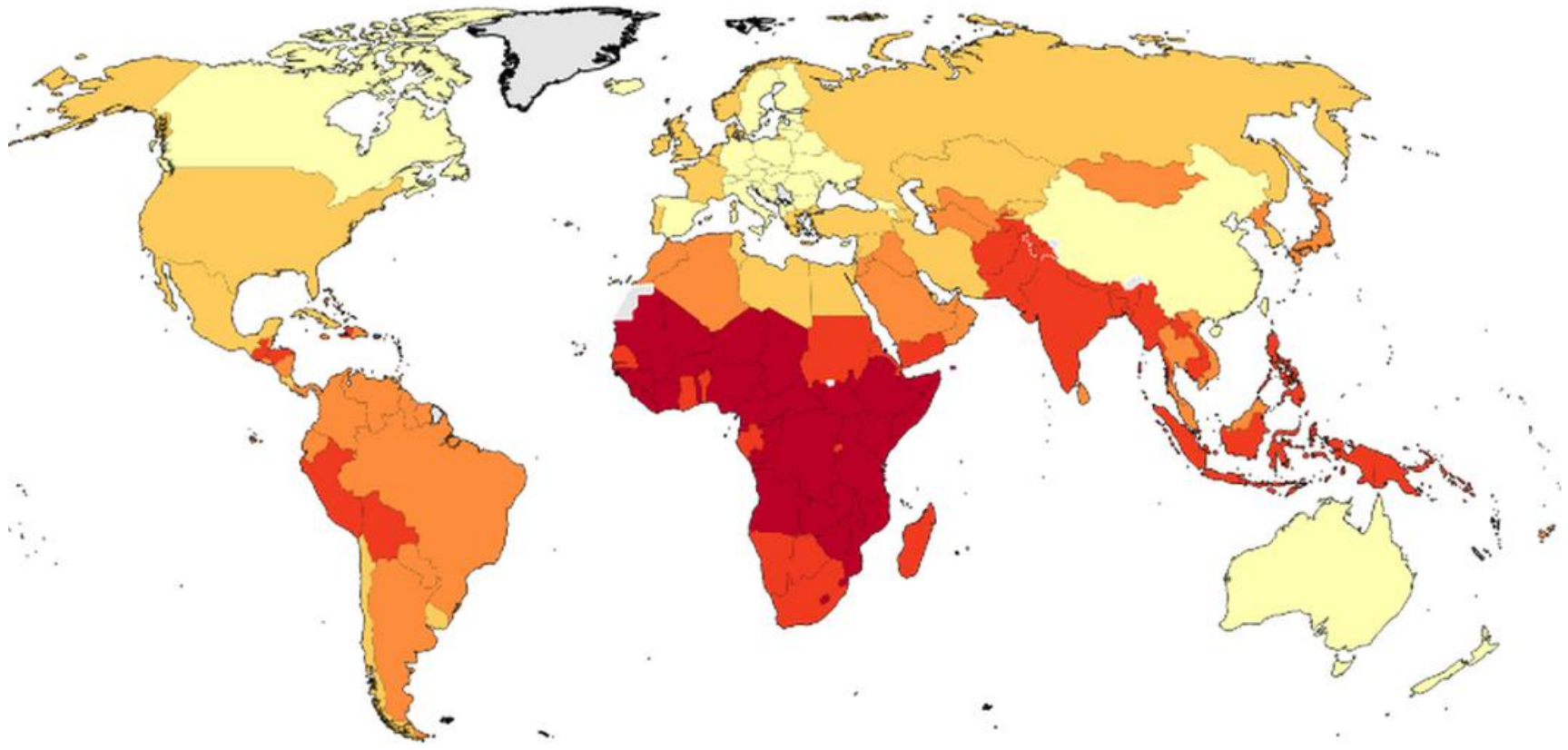

Data and graphs from the World Bank Databank.

Tropical countries like Brazil and the Asian Tigers such as Hong Kong, Malaysia, Singapore, and Taiwan, recognized these limitations of living in the tropics and invested heavily in health, family planning and non-agricultural economic development. Some of them managed to jump the gap from poor countries to reach the well-being comparable to the most developed countries of the world.

Many other factors related to climate, not described here, probably also influence the economic activity of a nation. Some of them might even facilitate economic development in warm areas. It is in warm areas, such as the Serengeti and the Amazonia, where Neolithic societies still survive today. We also know of examples among earlier human societies that consolidated in rather warm climates in Africa. Cultures such as the Maya in Central America and the Cambodian in Indochina flourished in tropical forests.

## *Non-Climatic Factors*

Geographical factors unrelated to climate also determine the economic potential of a nation. Trade promotes economic creativity and industrial production. Countries without direct access to the sea to facilitate its trade, such as Bolivia, Afghanistan, Chad, Zambia, Mali, Mongolia and Laos, are generally poor, while countries with favorable conditions for maritime trade, such as Singapore,

England, Japan and Taiwan, are classic examples of countries managing successful economic growth. Other forms of transport, such as those offered by the rivers Euphrates, Tigris, Nile, Indus, Rhine, Yangtze, Yellow River, etc., were catalysts for the formation of large cultures in the past and are used for transport today in these nations. The deltas of rivers, and canal construction in swamps and other flat areas, have been a key element for the development of river transport networks in Germany, Holland and England, facilitating the initiation of their industrial development.

Nevertheless, we have exceptions: the cases of Luxembourg and Switzerland illustrate that economic affluence can also be achieved by landlocked countries. This is partly explained because geographical barriers that hinder or prevent enemy invasions allow the accumulation of wealth and prevents frequent pillaging by wars, looting and invasions. This seems to be the case of Japan, England, USA and Switzerland.

Flatlands facilitate the construction of large cities and promote the industrialization of agriculture. Fertile lands, generally of alluvial or volcanic origin, are the factor that allowed a healthy agricultural growth that can catapult a subsequent industrial development. Favorable marine currents and large rivers allow for the development of fishing, which can replace or complement agricultural production and create the basis for the successful industrialization of a society.

It is undeniable then that ecology and geography affect human societies and their economic and industrial growth. Despite knowing this truth for centuries, we know very little in particular about this relationship. New tools to visualize, analyze and monitor networks, satellites monitoring continuously human activity on earth, and the emergence and development of interdisciplinary sciences such as ecological-economics, will achieve a deeper understanding of these interesting relationships. For now, it is enough to remember that both the historical processes and the geographical phenomena affect human societies and somehow help determine the wealth that a nation is able to create and accumulate. The multi factorial nature of this dynamic confirms that we face a typical problem of complex systems.

# 4. GENES

The more our knowledge of human nature and of other animals advances, the more we are surprised by the multiple networks of sophisticated physiological mechanisms that drive their behaviors. Physiology anchor our behaviors in biochemical processes regulated in final instance by molecules of deoxyribonucleic acid (DNA) in our genes. This dynamic multilevel process is called epigenesis. The accumulation of information in DNA molecules over time is carried out by evolutionary mechanisms where random mutation, heredity, sex and natural selection are the key players. This dynamics is the object of studies of evolutionary biology.

Biological explanations of social behavior are increasingly numerous. We know genes that regulate the monogamous or polygamous behavior in rodents, or induce them to be sedentary or migratory. We have identified the genetic origin of many human skills that affect social structures. We know genes strategies that use social behaviors of organisms to perpetuate themselves. Therefore, is it not possible that genes determine the potentiality of a society for building a modern state that manages thriving industrial economies?

Evolutionary biology recognizes several selective forces that modulate living organisms. No longer is it interested only about selection acting on the individual, which we know promotes biological evolution. We recognize now that group selection and kin selection can affect the genetic composition of populations. The dynamics of the genes in a population is complex, and so are their effects on the phenotype of individuals and on society and its functioning. It is highly risky to set forth simple causal relationships between a social or group characteristic and genes. Trying to use racial reasoning for explaining genetic social differences, and using such explanations for political purposes, usually ends in tragedies. The massacres of Armenians during the emergence of the Turkish state, of Jews and Gypsies in Nazi Germany, of Turkish Muslims in Bosnia, or of *Tutsis* by *Hutus* in Rwanda attest to this. The existence of these tragedies, however, does not negate that genes produce effects that permeate our societies. Eventually understanding of how they work, will allow a better understanding of the evolution of our societies.

Since the appearance of what we call *Homo sapiens*, probably 250,000 years ago, and the Neolithic revolution about 50,000 ago, there have been some 12,000 and 2,500 generations, respectively, more than enough time for genes to mold any species to its environment. From the Neolithic revolution, where we believe modern man was born, until the dawn of the agricultural revolution, there have been about 2,000 generations. It is hard to believe that in such a long time and with mortality rates as high as that suffered by our hunter-gatherer forbearer, natural selection had no role in shaping our species genetically. Even considering only the 400 generations between the emergence of agriculture and the present, natural selection must have been at work. If experience in the domestication of cattle and dogs is of some value, and we accept as true the reports of genetic improvement made by Brazilian slavers and others a few centuries ago, 400 generations are more than enough to produce significant genetic changes in a human population. If this reasoning is correct, it would imply that *Homo collector* and the *Homo farmer*, differ in more than in their habits of economic production.

I want to clarify that identifying a genetic correlate or a biological basis to a behavior, does not make it immune to the action of cultural or social factors. Knowledge of the genetic basis of a behavior helps us to better understand their limitations and possibilities and eventually handle it at will. For example, the sickness many humans suffer at sea has genetic bases, but it does not prevent sailors and astronauts from overcoming it. Also, the fact that evolution did not provide us with wings has not prevented us from flying further and faster than any animal with wings. Knowledge is not the cause of tragedies. It is the use given to knowledge and mainly human will, which is to beware of. Hence, in future times, after overcoming the traumas ultra-simplified racist concepts have caused, science will expand the study of the genetic basis of our behavior and its impact on societies, which will greatly help us to know ourselves and plan our future with more freedom and skills.

Some illustrative examples may serve to visualize human nature, its limitations and potential, from a perspective of evolutionary biology of social behaviors. The first example illustrates how instincts evolved to optimize sexual reproduction affect the social structure of our industries, business institutions and nations. The second reveals how fundamental elements required for human cooperation, which is needed to construct our societies, are genetically determined. Another example describes the process of how culture dilutes and controls genetically determined behaviors. The last example deals with how behaviors determined by genes form the basis on which conspicuous features of our culture developed.

## *Dynamics of Mate Selection*

The evolutionary advantages of sexual reproduction are not as obvious as it first appears. Computer models and numerical and mathematical analysis of different forms of sexual life show that asexual reproduction (cloning, for example) with a good dose of mutations, produces an optimal balance of innovation and transmission of information that enables the harmonious functioning of the evolutionary dynamics. For example, organisms that use asexual reproduction can produce more fertile progeny. They do not need to produce males, and therefore produce only females. Asexuals do not diluted advantageous genes (they do not mix them with a sexual partner) and therefore transmit to their offspring the genetic information more efficiently. Asexual organisms do not require complicated mechanisms to search for a partner, saving them time, energy a unnecessary risks.

However, a large number of organisms use sex as a mechanism of reproduction. Why does sex exist? What is the adaptive value of sex? Sex enables the modulation of the genetic variability of a diploid population. Computer simulations and the corresponding numerical analysis reveal that sex together with mate selection make sexual reproduction more efficient than asexual reproduction in evolutionary terms. This is especially true for diploid organisms such as humans. When choosing a partner to mix your genes with the aim to produce a better adapted offspring, clearly, the selection of a healthy, strong and successful partner is a better strategy than choosing a weak and sick partner. Unfortunately, not all characters that are important for future challenges are evident at the time of selecting a partner. The amazing insight from the simulations is that the selection of partners with a high degree of genetic similarity to oneself, that is, selection by similitude (homophily or assortation), results in a highly efficient evolutionary dynamics that renders sexual reproduction superior to other strategies including asexual reproduction. The most common way to satisfy homophily is through kin selection, i.e. preferring to cooperate with a genetically related individual (such as family) rather than with strangers, but there are several other forms to achieve homophily such as preferring individuals that resemble once parents, oneself, have similar habits, etc.

When seeking empirical evidence for homophily in mate selection strategies in animals, we find that it is much more common than expected. Even among humans it is extremely conspicuous. The similarities in age, intellectual coefficient, culture and facial expression among freely formed human couples are much higher than expected for a random couple formation. Preferences for homophily are visible in behaviors of more recent evolutionary origin that build upon instincts developed for sexual selection. We not only choose partners with faces like ours, but also pets that resemble us. The dogs we choose to combat loneliness and address parts of our emotional needs tend to look more like us than expected from a random selection. This homophily applies also to our choice of friends and affects our choice in politics. Research has shown that the most successful businesses and the more stable social structures are achieved between individuals who share a large number of characteristics. That is, intuitive behaviors and skills developed to optimize partner selection for reproductive purposes are used by modern humans in sustaining new forms of organization and association. This is an example of how human behaviors are shaped by the evolutionary history of mammals and primates, giving rise to behaviors upon which our present societies are built.

## *Biological Bases of Cooperation*

Cooperation between two unrelated individuals may appear unnatural at first. Why am I going to spend my energies, time, and potentialities to favor another individual who will eventually compete with me and my children, possibly hindering their development and displacing them in evolutionary terms? Is it not more efficient, biologically speaking, to assume selfish attitudes? Is competition not better than cooperation? Is cooperation not a strange invention of humans?

Again, analytical studies, computer simulations and empirical research allow us to untangle elements of this dynamic that illuminate novel perspectives. There are many factors that favor cooperation. One is the hope for future retributions of altruistic acts or expecting mutualism. Cooperation is seen here as a kind of social investment expecting immediate or future returns. Another factor that favors cooperation is to know that there is an eventual punishment for the reluctance to cooperate. This works in structured societies. To accumulate social prestige through the cooperation and to achieve significant economic synergies through cooperation are other ways for evolution to favor the emergence of cooperative behaviors in general.

Cooperation among animals is common. If we take the trouble to count the frequency of occurrence of mutualistic relations as opposed to relations based on exploitation, parasitism or predation, we find that the former are much more common in nature than the latter. This is true even for relationships between entirely different species. For example, among the inter-specific relationships maintained between some butterfly larvae and ants, mutualistic or cooperative interactions prevail over non-cooperative ones.

The impulse to cooperate with others has instinctive roots in humans. Just as we express our joy and sadness on our face when talking on the phone with our friends, or in the darkness of a movie theater, knowing that no one is watching, we often cooperate with others, driven by reflexes, instincts and motivations strongly rooted in our biology.

Humans are not equal in terms of their willingness to cooperate with others. There are the extroverted cooperators that like to start altruistic or mutualistic interactions with their fellow humans; there are the sly ones that accept acts of cooperation from others but do not return them; there are the passive individuals that depending on how friendly their neighbor is or what the ones around him are doing, cooperate or not; and there are the purists that get angry at others that are not willing to cooperate and punish selfish non-cooperators. And there are diverse reactions towards perceived unfairness in interactions. The interesting thing about this variety of personalities is that in many societies, the proportion of each of these is constant, possibly regulated by mechanisms of population genetics that achieves the optimum mix that enables the efficient functioning of our societies. Surprisingly, different societies differ in the balance between the proportions of these different types of altruists. These different balances are related to the values expressed by different societies and are part of the cement forming our cultures.

## *Jus Primae Noctis*

Since the dawn of script and writing in Sumer and the tales of King Gilgamesh collected thousands of years ago, and until the end of the Middle Ages in 1550, we find many stories, laws and evidence of the habit or privilege of the king or feudal lord of being able to demand sexual access to any woman that is getting married, before her nuptials, in the society he dominates. This privilege or right of the alpha male or dominant male for sexual access to females in his troupe is very common among primates and other mammals. It is part of the socio-sexual hierarchies of dominance which structure animal societies. It is very likely to be rooted in a set of genes of ancient origin.

Although this behavior has genetic roots and became part of human primitive societies, it is little practiced in contemporary societies. However, occasionally it reappears: during war it triggers retributive rape; and is practiced by extravagant autocrats and dictators. A modern example is the king of Swaziland and his nuptial customs that draw on these roots. Other behaviors still prevalent in contemporary society, such as the sexual preference of women for men of high social status, may have a common genetic origin with *Jus primae noctis or the Droit du seigneur*, but its social expression has become very different. Often in modern societies the roles are reversed. The male has to acquire a high social status to gain access to the opposite sex, but it is the females that regulate the rules of social interaction that will allow the male to increase his social status.

The custom of *jus primae noctis* exemplifies how societies, through cultural values, can shape, suppress or eliminate behavioral expressions based on genes. The same instincts may produce, by mechanisms of biological and cultural evolution, completely different social structures. The genetic determination of a given behavior does not cause its historical invariance, but it does influence it.

This is just one example of the way genes work to maximize fitness, including reproductive fitness. Instinct evolved to maximize our reproductive fitness work in many ways. The advice of the Bible (Deuteronomy 20:10-20) puts in simple words the working of an instinct, common to most animals when confronted with competitors of the same species: “Put to the sword all the men in it”, and for the women and children, “you may take these as plunder for yourselves” and reproduce. This advice is still followed by many a human culture today. It is a zero sum game that minimizes the reproductive success of competitors and maximizes the fitness of the actor. Organisms with instincts optimizing this kind of zero sum games often express simultaneously genes favoring cooperation, producing a wonderful rainbow of behaviors (see Chapter 11).

## *Shame and Guilt in Society*

Confucius (551-479 BC), Aristocles of Athens better known as Plato (427-347 BC) and Protagoras of Abdera (c 490 - c 420 BC) already recognized the importance of the feelings of shame as fundamental in maintaining social cohesion. Shame is defined here as the instinct or innate need to want to please others and to evoke negative feelings - sometimes very intensely - when one produces a behavior that is disliked by others. This feeling, specially its facial expressions, is detectable in all humans. Alexander von Humboldt (1767-1835) and later in more detail Charles Darwin (1809-1882) described shame as a universal instinct among humans that unconsciously triggers reddening of the face and closing of the eyes. Guilt, a distinct feeling from Shame, seems to be more primitive as dogs and monkeys express it in identifiable form. These two feelings serve to establish and maintain the harmony of social structures. Our modern societies, far from suppressing these instincts, encourage them. All cultures value them, although in different degrees. Guilt and shame cement trust and thus maintain social cohesion. Modern laws and social norms seek to evoke our feelings of civic responsibility and honor, which are nurtured by our instincts of shame. Hence, the efficiency with which admonitions, reprimands and the practice of "bench marking" works in stimulating individuals or companies to apply better practices, by appealing to feelings of shame. Guilt and shame are beautiful examples of how feelings based on instincts mold or present society.

Some of my work published in the Journal of Bioeconomics correlates the importance a culture gives to guilt and shame with macroeconomic variables, reveal a surprising result. Countries that are the mother of languages still spoken today, that have many words and synonyms for guilt and shame, are among the most corrupt. Counties that gave birth to languages with few words for guilt and shame, have better institutions favoring business. That is, countries where acts of corruption are frequent have languages that describe in more detail the phenomenon. Other cultures seem to care less about shame and guilt as they foster individual values more fomenting a different cocktail of values.

These are early times to understand the epigenesis of instincts and their impact on society. Evolutionary biology, experimental economics, human ethology, bioeconomis and economic and social psychology, promise to discover new information relevant to social behaviors that will certainly deepen our understanding of human social behavior in the future. For the time being, the certainty that instincts greatly influence our behavior and that they form the substrate on which culture grows suffices.

## *Genes and Diversity*

Evolution acting on genes trough natural selection works only if genes are diverse. Higher diversity produces faster evolution. Maintaining high genetic diversity is achieved by nature with mutations, with diverse environments and with populations developing independently. Strong uniform selection reduces diversity driving the evolution of organisms to specialize, reducing their potential for future adaptations to novel environments, increasing the likelihood of their eventual extinction. This dynamics has as a result that successful species, such as humans, have genetic traits that are variable. Genetic diversity is achieved by sexual selection and assortative mate selection. This is contrary to popular intuition that genetic predeterminations make life uniform and predictive. Variation, heredity or other forms of transmission of information, and natural selection, drives evolution to produces a huge variety and diversity of forms. Computer simulations show that cultural evolution experiences this dynamics in a form similar to what we see in biological evolution. The same constraints regarding diversity affect both types of evolution. Increased cultural diversity produces societies that are more robust. Increasing diversity that favors evolution in human society is achieved by tolerance, respect to others, promotion of original thinking, and diversity. Totalitarian rule hinders social progress and eventually leads to the extinction of the regime or its society.

The effect of our biological constrains on our behavior in modern economics may be more extensive than appreciated at present. For example, capitalism seeks to maximize the efficient use of capital for the creation of wealth and the achievement of economic growth in an industrialist society. The most efficient use of capital is achieved by highly motivated and skilled individuals, exploiting specialized economic activities in micro-habitats that they know extensively. This knowledge and skills cannot be centralized as they diverge widely and are dispersed heterogeneously among human populations. Experience shows that capitalism prospers best when individual freedoms, including rights to private property, are promoted and enforced. These freedoms are best guaranteed in democracy. The relationship between democracy, individual freedom and economic growth is hard to ignore. Critics of this view hint to to autocratic regimes that auto-classify themselves as communist but achieve fast economic growth, such as China. Prejudice might distort this picture and cause this confusion. For example, contemporary capitalism is more regulated in many ways in a self-proclaimed believer of a free economy such as the USA than in self-proclaimed communist China. China tolerates free economic activity and large unregulated informal sectors of the economy to drive their growth. At the moment, the USA has more legal code regulating economic activity than China.

I know of no empirical evidence that shows socialism to be more efficient in providing wealth and wellbeing to the average citizen than respect for free individual initiatives. In both systems, capital is used to generate wealth, but the outcome in terms of wealth, health and wellbeing of the population is different. A large list of nearly perfect experiments show this, where people with the same culture and long term history, were divided into a relatively more socialist regime, compared to the twin with a system closer to what is referred to as "capitalist", during a certain period of time: Cuba and Puerto Rico; South and North Korea; Austria and Hungary; East and West Germany; China and Taiwan; Tanzania and Kenya (see table in Chapter 8).

Despite of this hard and extensive empirical evidence, humans keep dreaming of socialist egalitarian systems. Simulations with Sociodynamica (see Chapter 11) show that for economies based on harvesting rather than on planting, such as hunter-gatherers, or rent seeking economies exploiting natural resources, the optimal adaptation for individuals is an egalitarian socialist system. If, however, an economy requires synergistic cooperation to produce sophisticated products, such as modern industrial and knowledge economies, heterogeneous unequal societies manage these economies best in the simulations. Human biological adaptations have occurred mostly during the hunter-gathering period and have thus features that are not optimized for modern industrial societies. Ignoring the biological constraints driving human motivation is not the solution. We need to understand better our instincts and biological human nature if we want to build sustainable social systems that promote the well-being of all.

# 5. ECONOMY

Economics began as a science that seeks to understand the management of scarce resources and is not concerned, generally, with resource to which humans had unlimited access such as water and air until a few decades ago. If we assume that wealth means less scarcity, then economics explains the dynamics of the wealth of society. However, economics alone cannot be trusted to lead us successfully in the search for understanding the phenomenon of the emergence of the wealth of nations. Economics is still a young discipline that lacks robust working tools applicable for this job. However, in the light of the rapidly growing complexity of the new financial instruments and the huge size of our economies, we might be surprised by the stability of many currencies and the long-term value of sophisticated financial instruments. Economics has shown some success in explaining these phenomena and might eventually develop increasing its assertiveness in predicting human activity. The immature nature of this science does not justify the great ignorance of most politicians and citizens on key economic issues that regulate a society. Without a well-founded economic vision, it will be difficult to understand the dynamics of our societies. Luckily, this science has recently achieved a major expansion and consolidation, allowing us to better understand fundamental aspects of the subject that afflicts us.

### *Classic Economic Factors*

For neoclassical economics, the wealth of a nation is related to capital and work. This is, in a country where workers are increasingly numerous and carry out their work more efficiently, the generation of wealth should be more pronounced. Moreover, a country with larger capital investment - resources to buy machinery and tools, for example - should generate greater wealth. Therefore, poverty could be explained by a lack of capital and / or work. Labor with adequate capital investment and education can become more productive, producing more wealth per hour and thus accelerating aggregate economic growth (GDP).

This view is a bit simplistic and not entirely consistent with the available data. In Figure 5.1, for example, we assess the productivity of the worker as the GDP produced per hour worked during 1995 to 2013, and relate it with the country's wealth (GDP) for these years. The figure shows data for Austria and Greece. Labor productivity per hour was between 35 and 45 in Austria and between 20 and 30 in Greece. In the year 2008, both GDP and labor productivity started to drop markedly in Greece; and labor productivity recovered somewhat in 2012 despite the fact that GDP continued to drop. In Austria labor productivity and GDP dipped slightly for only one year only. That is, in Austria, changes in labor productivity seemed to be correlated with GDP, whereas in Greece, this correlation disappeared in some periods. Clearly, other factors affect GDP besides labor productivity.

**Figure 5.1: Wealth of a country and labor productivity**

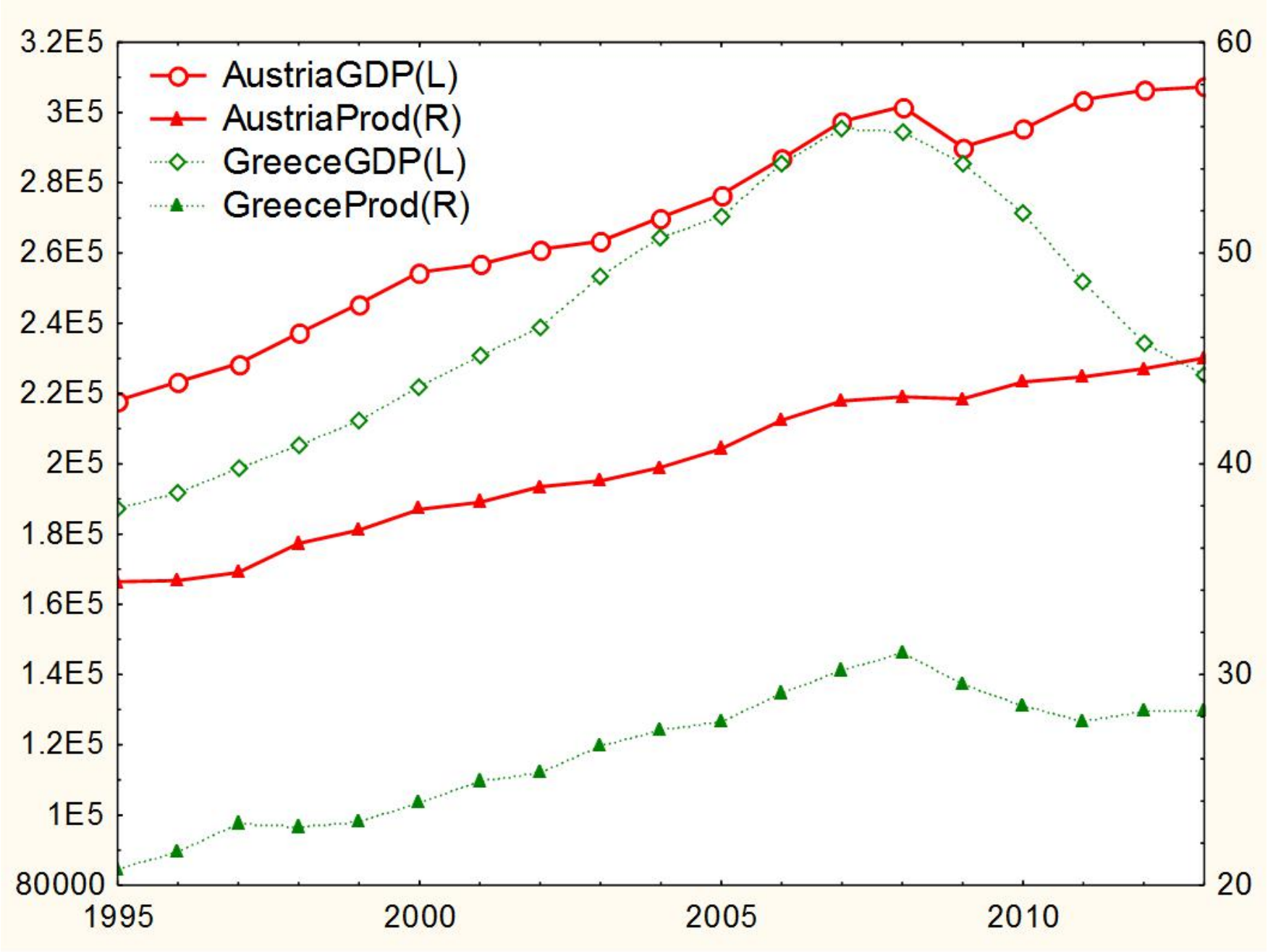

Relationship between the wealth of a nation or GDP (indicated on the left vertical axis) and productivity as GDP per hour worked at US 2005 constant prices (represented on the right of the vertical axis) between 1995 and 2013 (horizontal axis). Data for Austria and Greece from the OECD Statistical database.

We get a similar result when analyzing the capital invested in the economy of a country or if we quantify the contribution of the labor force in other ways. This is the classic way to analyze economies explained as a function of capital and labor in a society or business, found in most economics textbooks. Empirical observations show however that classical economics explains only part of process of the generation of wealth of a country.

The neoclassical theory of economic growth uses equilibrium models to explain economic growth (e.g., Leon Walras 1834-1910) with emphasis on the capital-labor ratio and looking for a production function that ensures sustained and balanced growth. In contrast, Keynesian theories of growth (based on the work of John Maynard Keynes, 1883-1946) emphasize the added value of the economy (the macroeconomic aspects), especially on aspects of capital investment.

Another view often used to explain economic growth is the liberal or neo-liberal theory. This doctrine promotes maximum use of market forces - supply and demand that interacts through prices - and of competition, to coordinate economic activity. The State must reduce its activity in regulating relations between producer and consumer and interfere only in those areas of the economy where market forces cannot act, such as the provision of public goods.

In contrast to liberal theories, Marxism (based on the writings of Karl Marx, 1818-1883) calls for state intervention and centralized planning to protect the workers from market forces, controlling and distributing economic capital gains or surpluses of capital. Marxism bases its premises on recognizing Capitalism as a fabulous force that needs to be tamed by the State.

These theories assume that humans make decisions in a rational and calculated manner, and we only need to handle enough information to make decisions properly. In contrast, empirical evidence shows that in practice, human actions are more driven by emotions than rational thinking. These and other shortcomings of the premises of classical, neo-classical, Keynesian, liberal, Marxists and other economic descriptions, limit their explanatory value to predict specific aspects and often fail in understanding real world economic dynamics. Modern economics, especially behavioral economics, evolutionary economics, bio-economics and complex system science applied to economics, seeks to overcome these limitations.

## *Economic Growth*

The factors that regulate the accumulation of wealth must affect or even determine poverty levels of a society. But is this a direct and linear relationship? The data provided by the United Nations Development Program (UNDP) can enlighten us on this point.

**Figure 5.2: The human development index and the wealth of a nation**

Relationship between the Human Development Index (HDI on vertical axis) with the wealth of the nation as Real Gross Domestic Product per capita (GDP/capita in US $ of 1998 in the horizontal axis). Each data point represents the datum for one country. Data

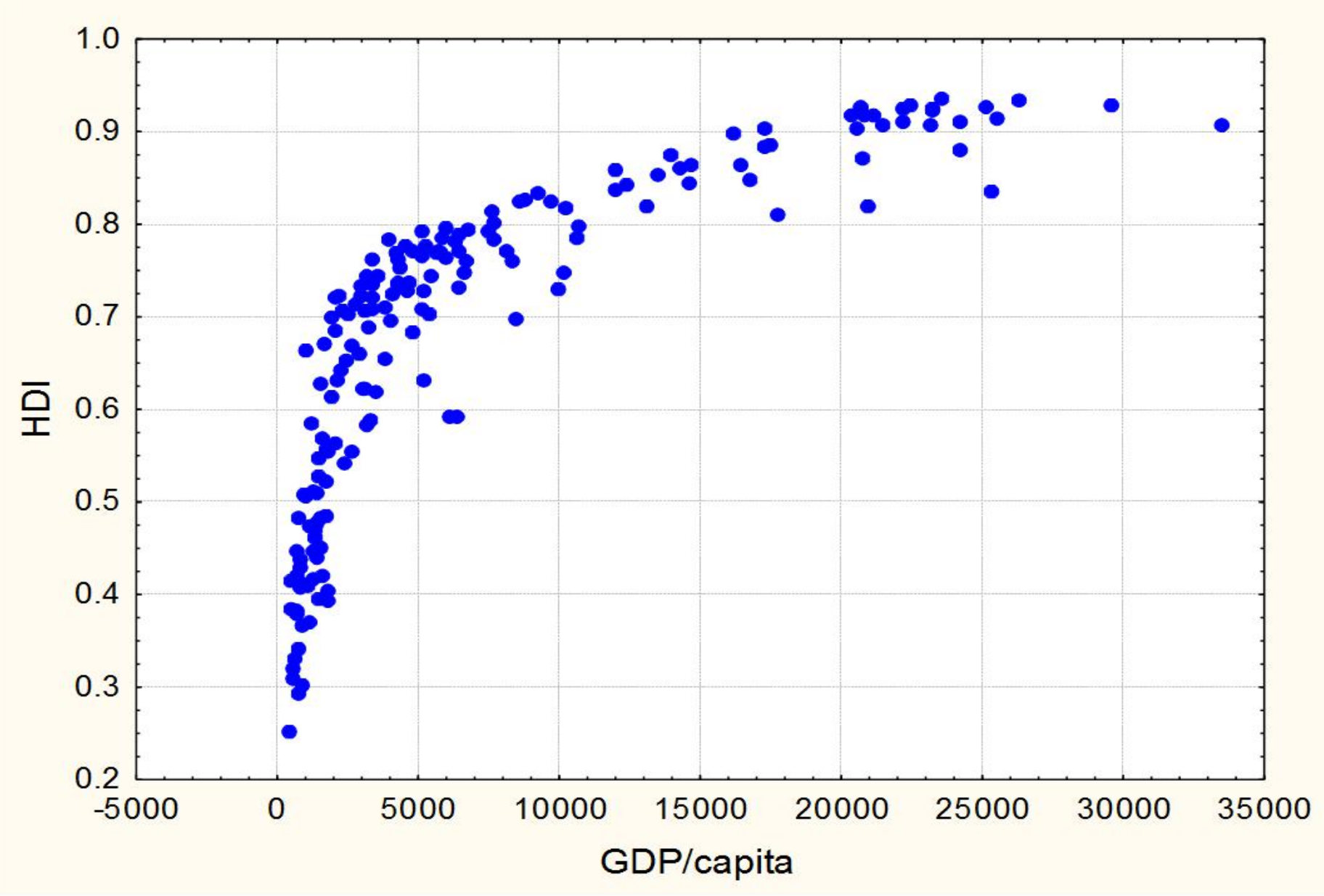

from United Nations Development Program (UNDP ) 2003.

Figure 5.2 shows the score of the Human Development Index (HDI on the vertical axis) calculated by the United Nations for the different countries of the world (data points on the graph) relative to its per capita Gross Domestic Product (GDP per capita on the horizontal axis). The HDI is a composite index that takes into account not only the accumulation of material goods but also access to education, health and recreation, among other factors. The result of this comparison is dramatic. The countries represented can be clearly separated into two groups: those on the left side of the graph with a GDP per capita less than about US $ 5,000 ($ of 1998) per year, and those with higher average incomes (right side of graph). The first are the so-called developing countries, the latter developed countries. The graph suggests that small increases in the average wealth of the population have very important effects on the quality of life of the inhabitants in developing countries as estimated with the HDI; while larger increases in average income in developed countries produce only small improvements of well-being. The graph also suggests that the HDI will no longer be useful to measure increases in the quality of life of rich countries in the future, as these already acquired values close to the maximum possible.

We obtain a mirror image of the relationship just described by plotting the human poverty index (Figure 5.3, HPI on the vertical) calculated by the United Nations for each country against the average income of the country in US $ of 1998 (GDP per capita on the horizontal). The poverty rate is calculated by taking into account not only the amount of income, but also access to drinking water, education and health. It is amazing and very revealing to observe that the levels of minimum poverty seem to converge in the

richer countries to an asymptote with an HPI value significantly greater than zero. Among the riches country the index remains above 10% of the HPI maximum. That is, modern technological society does not eliminate poverty, although it does minimize it.

**Figure 5.3: The human poverty index and the wealth of a nation**

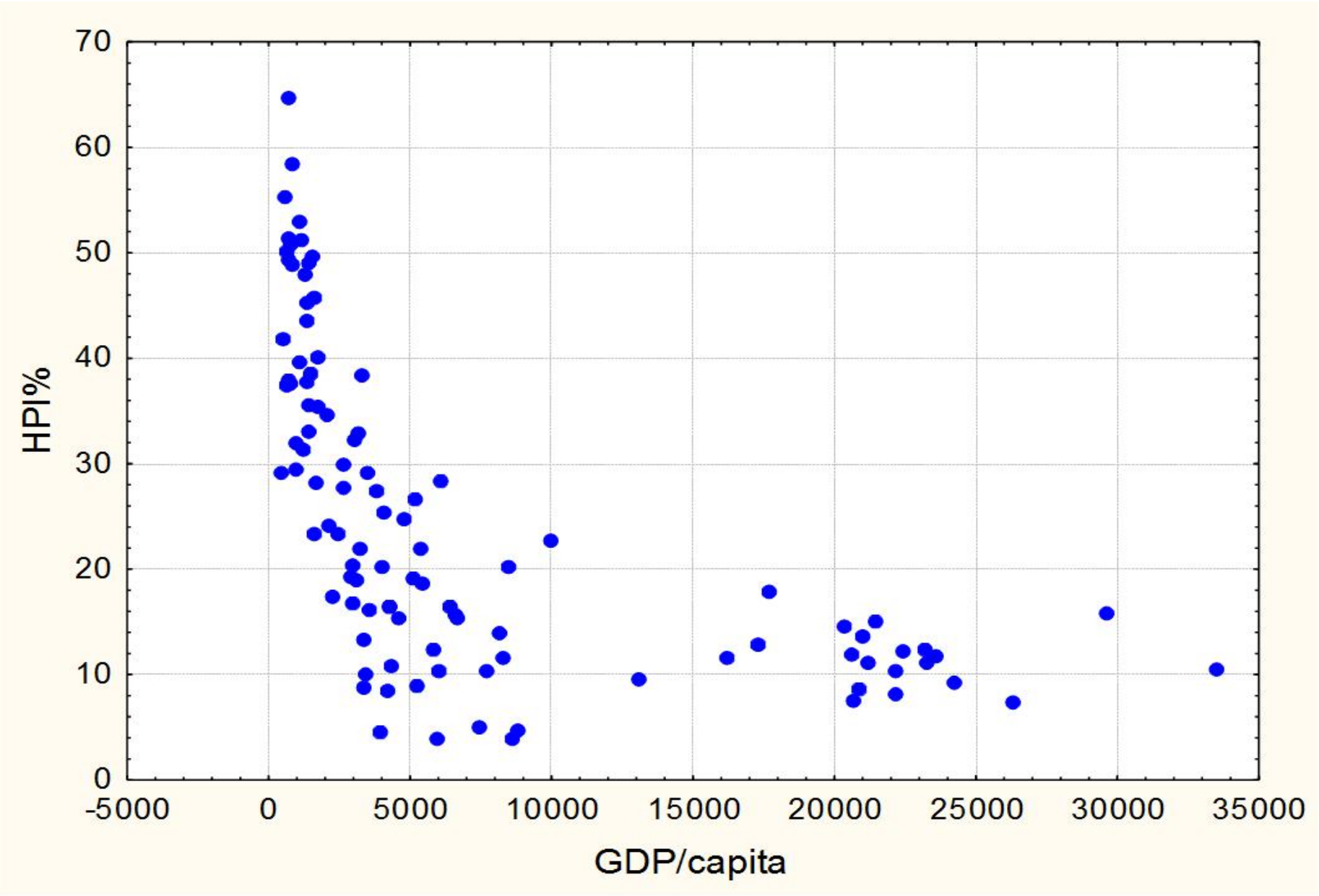

Relationship between the poverty index HPI (vertical axis) and the wealth of the nation as Real GDP (US $ of 1998) per capita (horizontal axis). Each data point represents the datum for a country. Data from Development Program of the United Nations, UNDP 2003.

We can conclude from the analysis of these two data sets that the relationship between the average income of a nation and its poverty levels (or quality of life) is direct, but not linear. At least two phases or distinguishable stages exist that allows grouping countries in two categories with very different economies. In developing countries, economic growth produces the greatest benefits in terms of reducing poverty and increasing the quality of life of citizens. In developed countries, economic growth does not affect the levels of poverty, nor is it so crucial in raising the standard of living of the population, as it is in developing countries. New indexes of human development will be required to guide the evolution of the economies in developed countries.

## *Natural Resources*

We often hear arguments that relate the wealth of a nation with its natural resources. The Gulf nations are rich for their vast oil reserves. Countries with large territories are rich and powerful because of their immense agricultural potential, their vast forests, and their mineral resources. However, among the richest and most prosperous nations on earth, at any historical moment, we find countries that do not have significant material resources, or are small or occupy inhospitable lands. Japan, Taiwan, Iceland, Luxembourg and Switzerland are small countries without large amounts of natural resources but are currently economically powerful. Among the countries with abundant natural resources that are poor nowadays, we can name Venezuela, Nigeria and Congo. Why would a country without natural resources become rich and a country with a good allocation of natural resources remains poor?

A revealing example that contrasts Japan and Switzerland is Venezuela. Venezuela is a country with large deposits of aluminum, iron and oil. It ranks among the countries with the largest oil reserves in the world and is a net exporter of oil for nearly a century, and yet it is poor. Figure 5.4 compares the wealth of Venezuela, reflected by its gross domestic product (GDP), with oil prices in a 50-year period.

**Figure 5.4: History of the price of crude oil and the GDP in Venezuela**

Relationship between the price of oil (Crude oil in US $ per barrel), and Gross Domestic Product per capita (GDP/capita in US $ of 2005) over the past five decades as detailed on the horizontal axis. Arrows indicate two critical periods when oil prices increased and wealth collapsed. Data from the BCV, except for the last 3 years which were estimated using free markets US$ exchange rates.

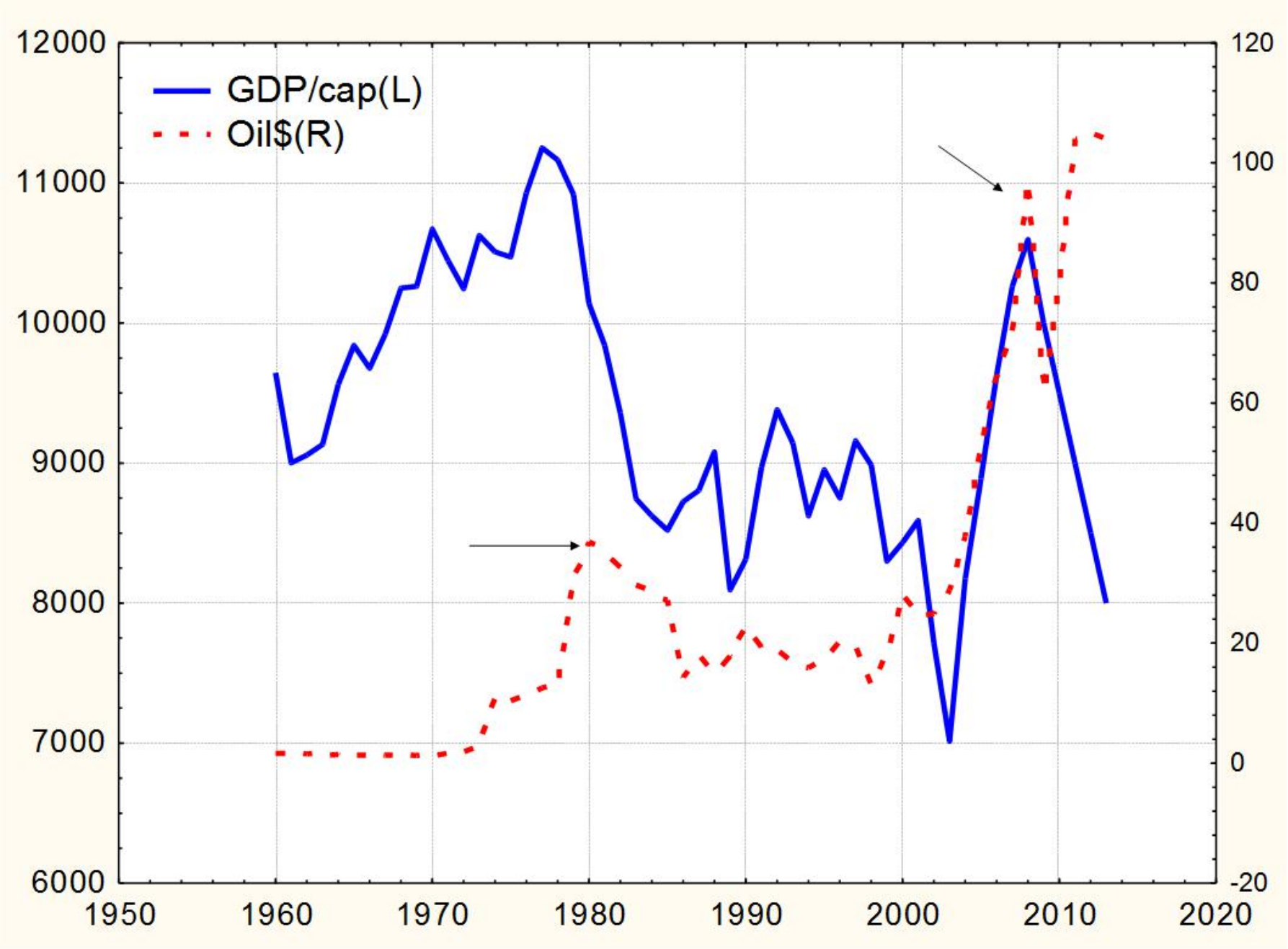

It is surprising to observe that shortly after oil prices increased and therefore the revenues from oil exports jumped, the GDP decreased abruptly. In other words, high oil prices produce less aggregate wealth in this oil exporting country. High oil prices depress the total production of goods and services, whereas in times of low oil prices GDP per capita increases.

We face a case here of intuitively unexpected consequences. The Venezuelan economy has a structure that, somehow, reverses the possible beneficial effects of an increase in the flux of resources into the country, producing a decrease in its aggregate production. Examples of this phenomenon are known worldwide. In Chile, when it was a major exporter of guano and then of copper. In Bolivia, when it was the largest exporter of tin in the world. Libya and many a oil producing countries suffer the same fate. This syndrome is often called the “Dutch Disease”.

Natural resources are exploited by a small sector of the society. The wealth generated by its export affects the exchange rate of the money of country and diverts labor and capital from other productive activities. These cases show that the wealth generated by one sector of the economy can have a negative effect on the overall economy. The effects of these booms in commodity prices are not only economic. They also produce political and social instabilities that have a negative feedback upon the economic activity. Thus, for many scholars, commodity price booms are rather a course to be taken very seriously by implementing countermeasures that might moderate some negative effects on the economy. This apparent contradiction, which shows that the price of the main export product may be inversely correlated with the production of goods and services in the country, has lead to profound reflections. We must examine other aspects related to the wealth and poverty of a nation before we can better understand these relationships and reveal this mysterious contradiction.

# 6. ECONOMIC INEQUALITY

## *Unequal Wealth Distribution*

We often hear that the difference in individual incomes in a human population causes distortions and inefficiencies that produce the impoverishment of large sectors of the population. This thesis was and is widely promoted by many followers of the theories of Marx and Engels. This difference in income can be calculated in several ways. The most widely used index is the GINI coefficient, calculated from the distribution of the wealth of the nation. A coefficient of 0 indicates perfect equality with everybody earning the same; and a coefficient of 100 implies a perfect inequality with one person earning all the wealth. Other ways of representing income inequality is focusing on the proportion of the national wealth assigned to the 10 % of the poorest and 10% of the richest. Figure 6.1 shows an example of wealth distribution among the 10 and 20 % of the poorest and richest inhabitants of a country. It plots the percentage of national wealth consumed by the four different economic strata of the population in increasing proportion according to its wealth. The nations represented are Slovenia, Japan, USA and Nicaragua that have a GINI index measuring economic inequality among its inhabitants of 20, 25, 41 and 60 respectively.

**Figure 6.1: Distribution of wealth in the population of four countries: relative distribution**

Percentage of the wealth of a nation (vertical axis) that is consumed by a given income sector of the country (horizontal axis). For example, all residents that have incomes up to the 20 percentile of the poorest consume in aggregate about 5% of the wealth of the nation in Slovenia and Japan and a fraction of that amount in Nicaragua. Data from UNDP for 1992-1998.

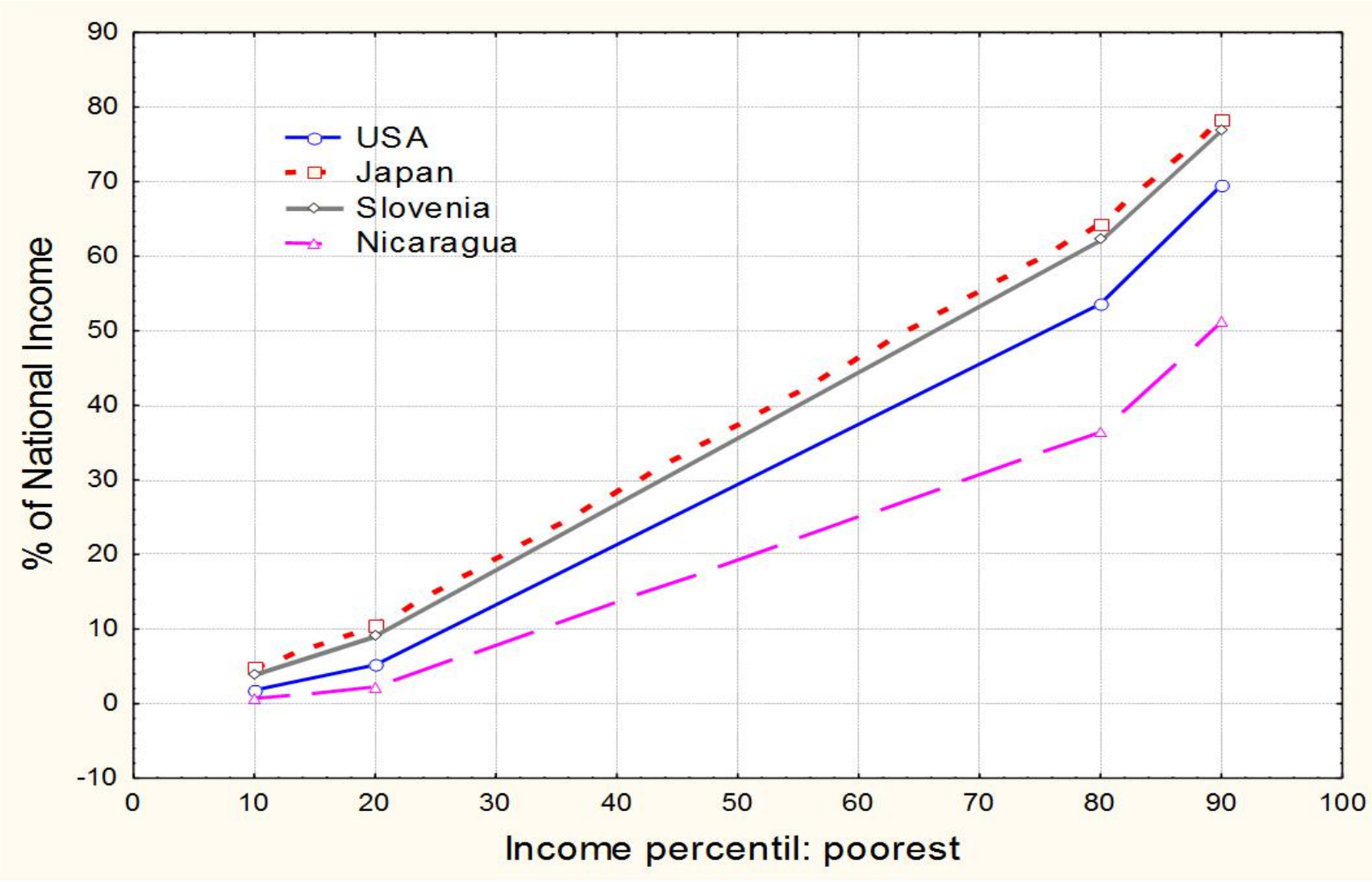

We see from this graph that, for example in Nicaragua, which showed the highest levels of social inequality in the world, 90% of the population consumes about half of the wealth produced in the country and 10% of the population consumes the other half. In the most balanced country in the world, in terms of distribution of wealth, at the time, Slovakia, the line is almost straight and the richest 10% of the population consumes only 18.2% of the wealth of the country. This graph illustrates that in percentage terms, the differences in income between rich and poor in Japan was lower than in the USA.

To view inequality in terms of relative incomes, as the GINI coefficient also does, though useful in many ways, hides other realities. Therefore, in Figure 6.2 we present an income distribution based on absolute wealth earned by the rich and the poor percentiles of the population.

**Figure 6.2: Distribution of wealth in the population of four countries: absolute values**

Income per capita of a country in absolute values (US $ of 1998 on the vertical axis) that is consumed by rising income sectors of the country (horizontal axis). For example, people listed in the richest 10 % percentile located above the 90 percentile of the population (90% in the graph) consume over 60,000 $ per capita in USA and Japan, and a tiny fraction of that amount in Nicaragua. Data of the UNDP for 1992-1998.

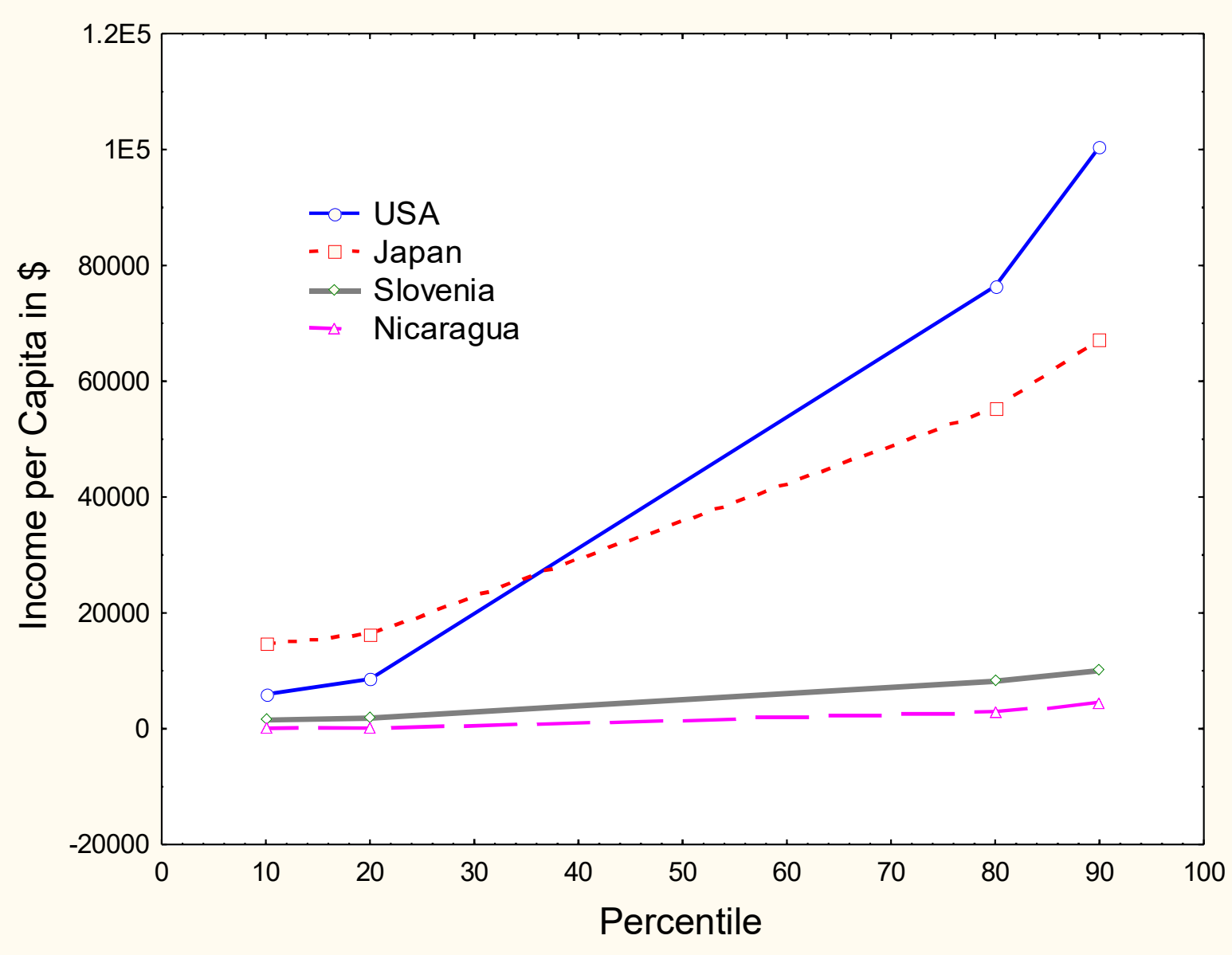

The graph presents data on the income inequality that looks very different from that in Figure 6.1. For example, data points over the 80-percentile value on the horizontal axis indicate the wealth consumed in absolute terms of the 20% richest in the country. In this view of wealth distribution, the rich in the USA have much higher incomes than the rich in Nicaragua have. Rich people in Nicaragua have a rather modest income if compared to residents in the USA. The ratio between the incomes of the 10% richest to the 10% poorest in Nicaragua was 70 times in that year, whereas that in the USA was 20 times. Thus, the widely used GINI coefficient and others reveal different things when comparing inequality distributions between rich and poor countries.

No matter how we measure the difference in the distribution of incomes, the distribution of wealth among citizens of a country varies greatly. According to a World Bank report of 2003 (*Inequality in Latin America and the Caribbean: Breaking with History*), Latin America is one of the regions with the highest rates of inequity in terms of wealth distribution. The reasons for this disparity, according to the report of the Bank, lies in the joint action of four factors: unequal access to education, very large differences in the income of people with high and low education, the high number of children with whom the poorest have to distribute their income, inefficient and misdirected public spending.

Figure 6.3, presents the data of the prevailing inequity in different countries with different average income (GDPc) calculated using three different indexes: The relationship between the wealth that 10% of the wealthiest individuals in the population consumes and the wealth that 10% of the poorest individuals consumes (RI/PO 10%); the relationship between the wealth consumed by the richest 20% of individuals in the population and the wealth consumed by the poorest 20% of individuals (RI/PO 20%); and the GINI index, which measures inequality taking into account all the distribution curve, where 0 indicates total equality, while a value of 100 represents total inequality in the distribution of wealth.

**Figure 6.3: Levels of inequality and wealth of a nation**

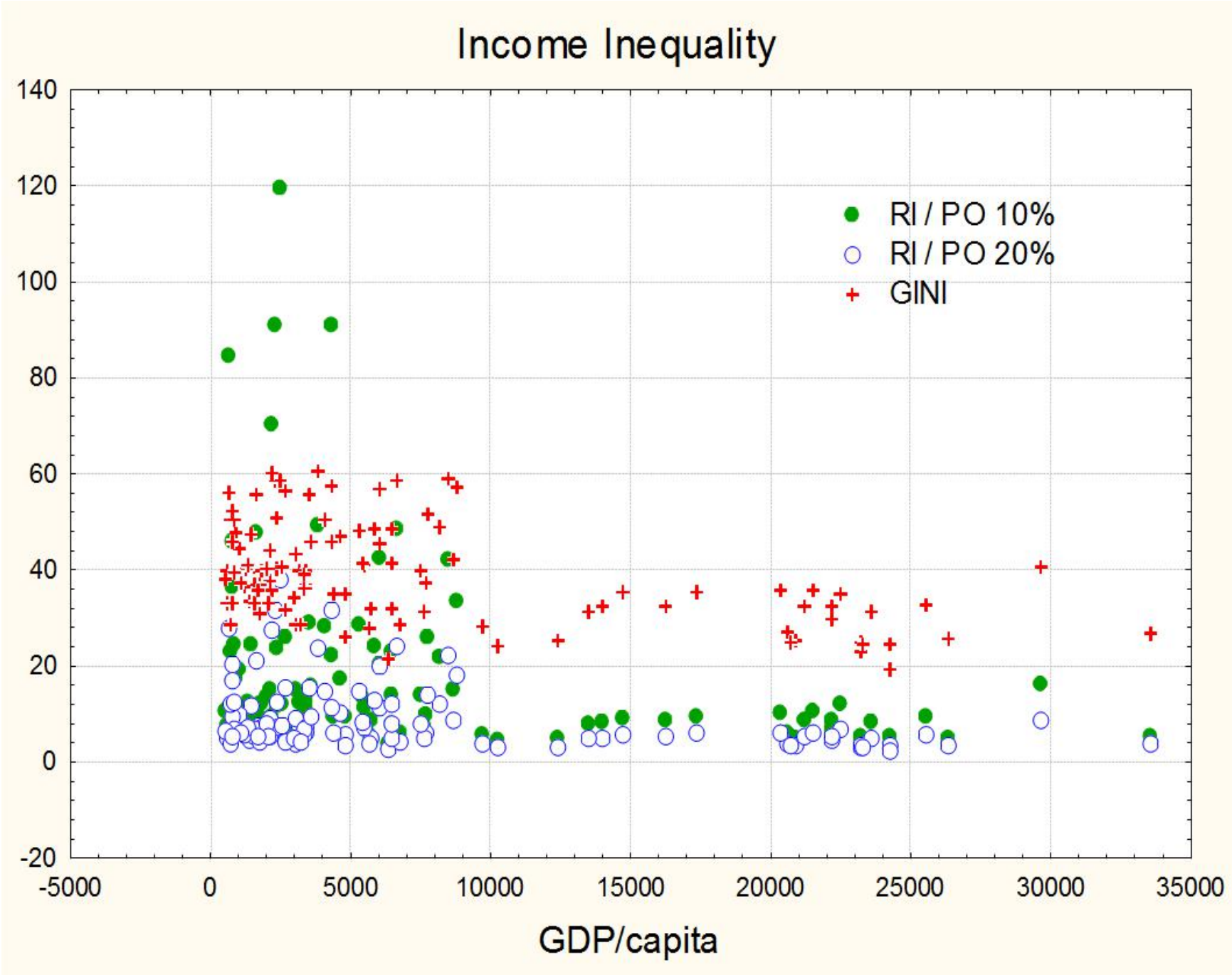

Relationship of the values of three indicators or indexes of economic inequality (vertical axis) with the wealth of nations measured by Real GDP per capita (horizontal axis). Each data point represents the index of a country. Data from UNDP for 2003.

From observing the graph, it stands out that regardless of the index used, the countries that have a higher per capita income (data points to the right of the graph), have lower values of inequality. Each of the three inequality indexes converge in an asymptote significantly greater than zero when per capita income increases. The variability of the values of indexes in low income countries is much greater than the variability of the values of the indexes in high income countries. The graph suggests that there will always exist a residual level of inequality. There are poor countries with inequality indexes similar to that of rich countries and others with much higher values, but rarely with lower ones.

This surprising result suggests that societies, regardless of how much wealth they accumulate, will always keep a minimum and constant degree of individual variation in the wealth of their individuals in percentage terms. That is, the curve that characterizes the distribution of wealth in a population seems constant in developed nations, regardless of their income level. This phenomenon is more easily visualized in the following Figure 6.4, where the percentage of the wealth of a nation that the richest 20% of individuals and poorest 20% of individuals benefit from is presented for different countries sorted by their average income.

**Figure 6.4: Distribution of rich and poor in the countries**

Relationship of the percentage of income of the nation (vertical axis) for the poorest 20% and the richest 20% of each country. Wealth of nations was measured by Real GDP per capita (horizontal axis). Each country is represented by two data points. Data from UNDP for 2003.

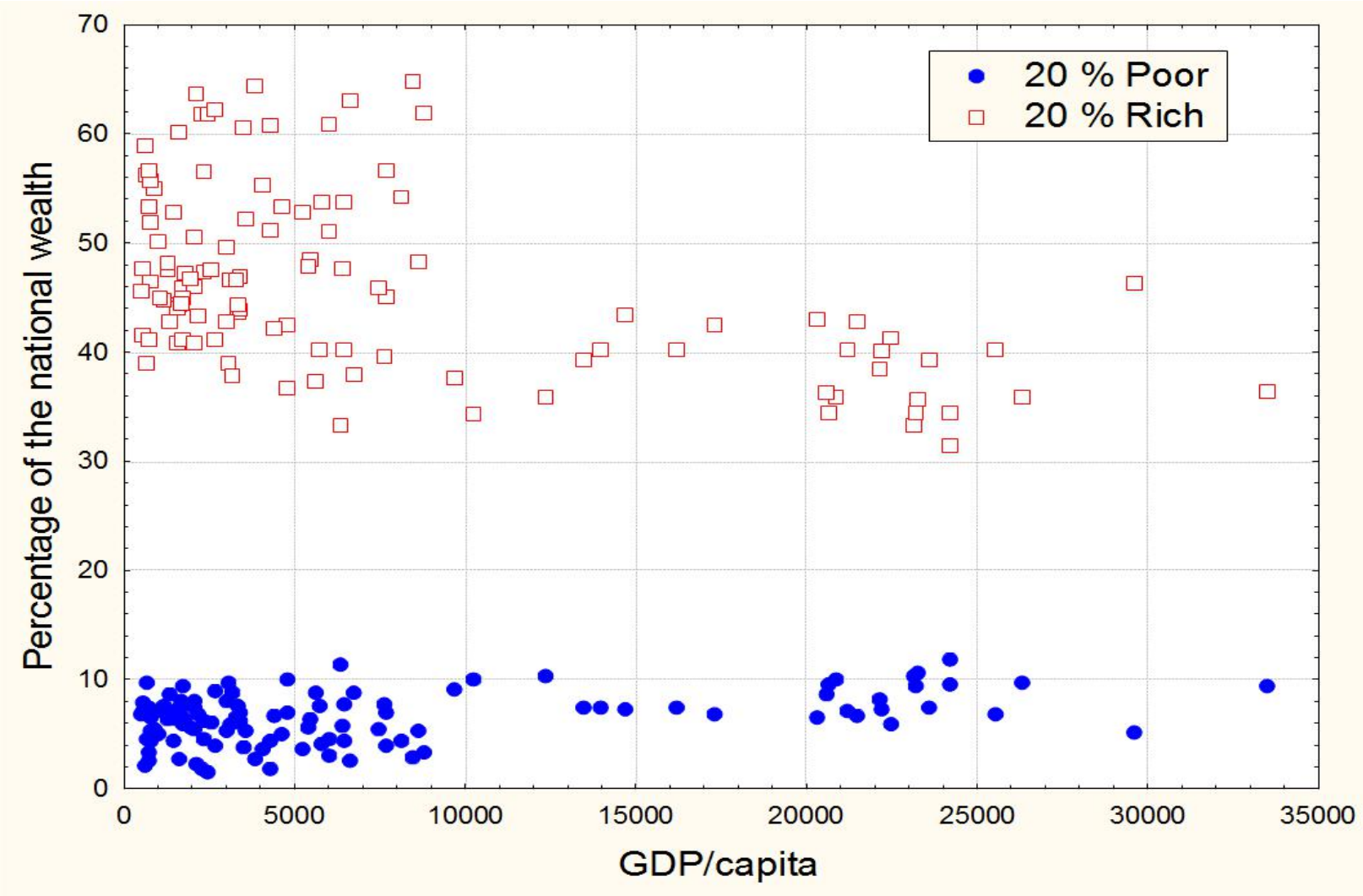

Figure 6.4 shows that the proportion of the 20% poor is approximately a mirror image of the 20% rich, but not perfectly so. This indicates that poverty and wealth do have different dynamics in a economy. The figure also shows that there is an approximately constant separation between the richness of these extreme sub-populations in terms of the sharing of the wealth of the nation, which is maintained independently of the gross domestic product that the nation has. This phenomenon was already identified and partially analyzed in the XIX century by Vilfredo Pareto. We will analyze his contribution in more detail when we discuss econophysics.

The temporal dynamics of different economic strata of the society are different. The 1% richest fraction of the population and the 1% poorest one behave differently from the average citizen (see Thomas Piketty's Capital in the Twenty Fist Century). During the last decade, the 1% richest citizen increased their wealth considerably in most developed countries, whereas the 20% poorer one became poorer. These distortions in the wealth distribution of a country are not reflected in the statistics of the average or median citizen.

## *Statistical Bases for Economic Inequality*

Perhaps a concrete example best illustrates the various ways of understanding the differences in income or variation in the access to wealth that the citizens in a nation have. If citizen A has an income of $ 10 and citizen B has an income of $ 1000, the absolute difference between the two is $ 900 and the relative difference is 99%. If we increase the income of both in $ 10,000 we will have that A now earns $ 10,010 and B earns $ 11,000. The absolute difference is still $ 900 but the relative difference in income of A and B dropped to 9%.

A better approach is to look at several indicators at the same time. For example, we might monitor the wealth and the health of a country and observe its evolution in time. In Figure 6.5, the health of a nation is estimated by the average longevity of its inhabitants and its wealth as GDP/capita. When plotting these distributions for two different points in time separated by a century, a striking image emerges.

**Figure 6.5: Global distribution of wealth and health**

Plots show countries ordered according to their life expectancy in years (vertical axis) and income per person (GDP/capita, PPP $ inflation adjusted on the horizontal axis) for the years 1912 and 2012. Both plots have exactly the same scales on both axes. Data points in dark blue are for African countries, the yellow ones are from the Americas, the light green ones from North Africa and the Middle East, the orange ones from Europe and Russia, and the red ones from south east Asia. The size of the data point is proportional to the population size of the country. Free material from www.gapminder.org.

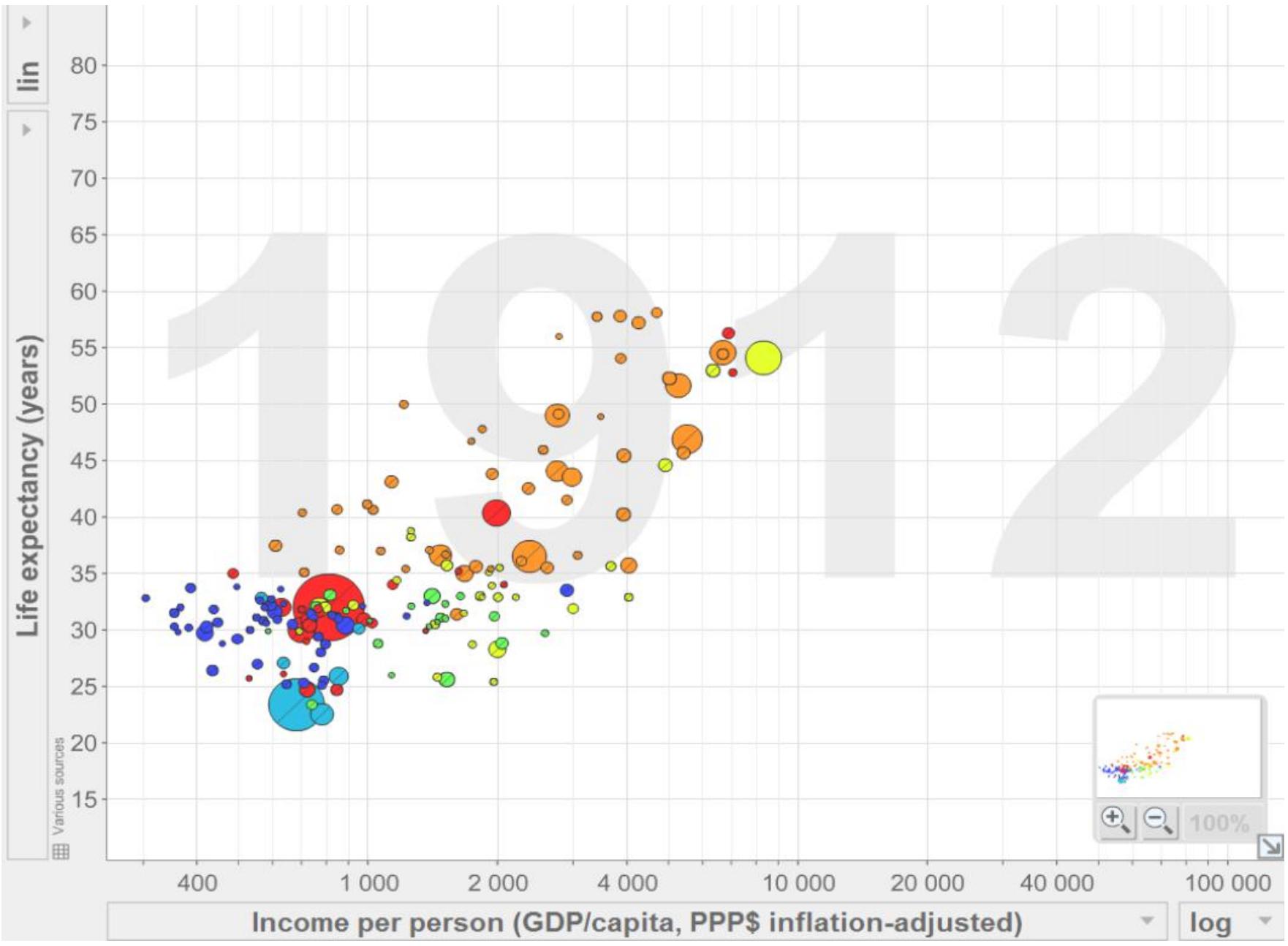

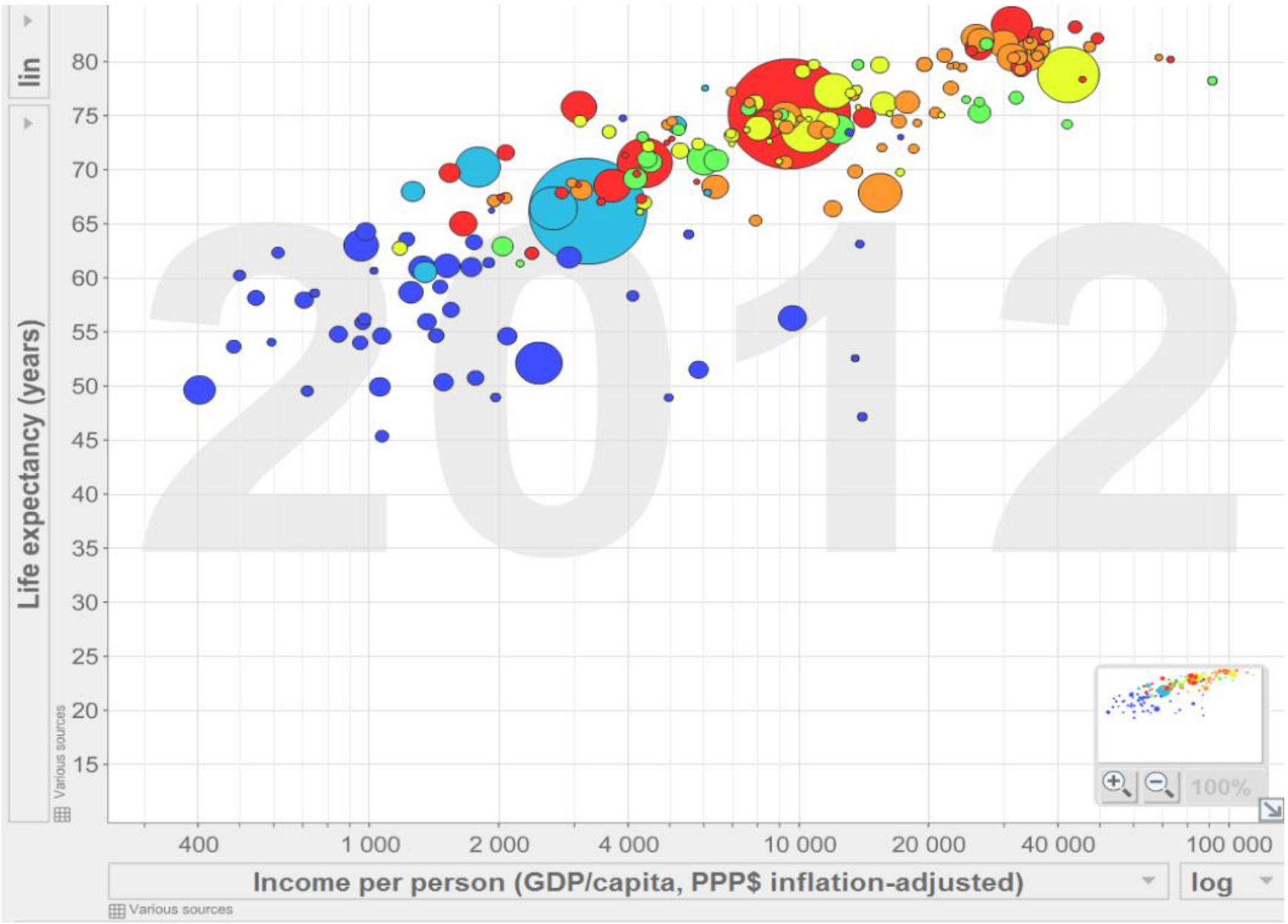

Clearly, during the XX century, a huge change took place. All countries increased their health significantly as life expectancy increased everywhere during these 100 years. No country has now life expectancies below 45 years, whereas in 1912 the majority of countries had life expectancies below this value. Most countries also increased their wealth. That is, the wealth distribution curve shifted to the right, toward higher individual incomes, thanks to economic development. Some countries, however, had a much smaller development than richer ones, making the spread between the poor and the rich countries to increase. This means that although most countries improved their wealth, inequality between the countries increased. Some countries closed or reduced the wealth gap with rich ones during this century. Especially during the last two decade, important reductions in inequality between countries have occurred, but this reduction seems to have reduced its dynamic after 2010, as shown in Figure 6.6.

**Figure 6.6: Closing the Gap**

Difference in annual between the USA and a selected group of countries. The group of countries plotted corresponds to the Least or lowest income countries, Low income countries and Low-Middle income countries as reported by the World Bank.

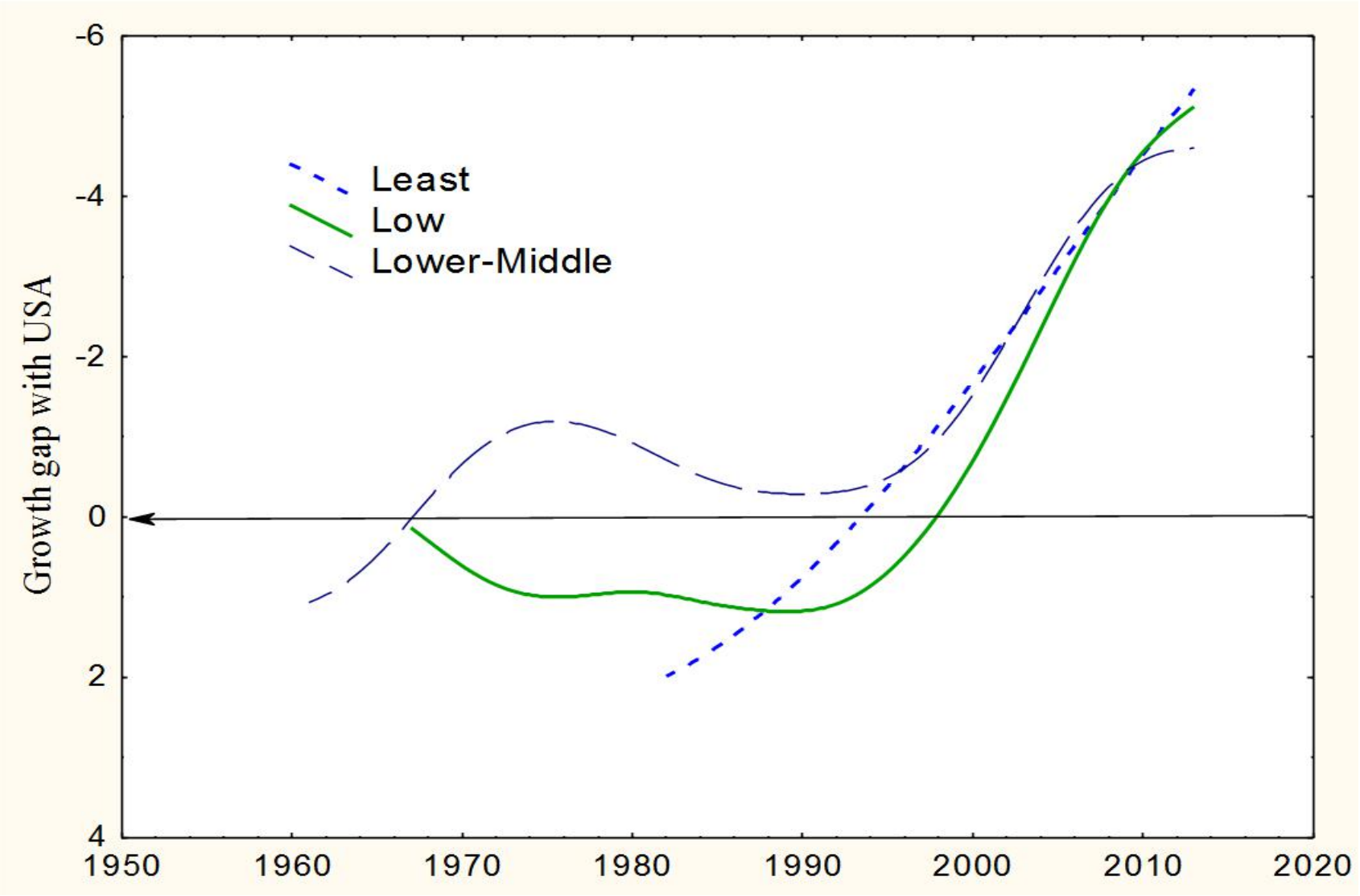

If in addition, we consider the population of each country, this trend is even stronger. The poor in China and India increased their economic income faster than the average citizens of rich countries did. This trend hides the fact that some countries, especially those in Africa, are poor and became poorer between 1980 and 2000 (see X Sala-i-Martin). The countries with the least economic performance were run by autocratic governments that disguised as nationalist their anti-market populist ideologies. They did not manage to produce economic progress nor did they reduce the gap between rich and poor. According to world statistics, the average inhabitant of poor countries, strongly influenced by the population size of China and India, is experiencing faster economic growth than that of the average inhabitant of rich countries, closing the gap between rich and poor globally.

As we see from these examples, the application of statistical methods helps us to understand and quantify the economic phenomena that humanity suffers, but it can also confuse our understanding. A careful and redundant analysis allows us to identify what the changes are and which countries most significantly affect the global economy. In the example just mentioned, it is clear that the economic policies of China and India will be crucial to establish the average levels of poverty in the world, just due to the size of their populations.

Many econometric studies use tools of analysis that are more sophisticated, which can detect more subtle phenomena and minute changes in economic dynamics. In recent years, many economists use tools developed by physicists for the study of complex phenomena; this analytical approach is called econophysics. Other methods are being developed starting from other disciplines that are converging in Complexity Sciences. None of them gives us the whole picture yet. Therefore, when managing a given economic reality, it is still necessary to include a dose of subjective knowledge, which makes economic praxis controversial.

## *Economic Growth and Social Justice*

The data from the United Nations says that the distribution of wealth in percentage terms in several developing countries is more unequal than that for developed countries. This may be due to two causes:

1- Economic growth or development shifts the whole wealth distribution curve of a nation. That is, with a balanced economic growth, the rich and the poor increase their wealth similarly in absolute value. This produces a compression of the curve in percentage terms, reducing the gap between rich and poor. This reduction is even more conspicuous if measured with the Human Development Index (HDI). Expressed in econometric terms, the balanced economic growth of a country produces higher growth rates in poor sectors relative to the rich, which ultimately leads to a narrowing of the gap between rich and poor.

2- The reduction of economic inequalities in a nation promotes enough social harmony to support a sustained economic development. Underdeveloped countries have failed to establish the relevant institutions or implement appropriate social organization to ensure a minimum of equality among its citizens, which prevents the process of economic development from accelerating.

The relative valuation of these two causes differentiates liberal economists from the socialists. The truth is that both phenomena affect the process of accumulation of wealth. It is interesting to note that most advanced contemporary human societies uphold important economic differences among their inhabitants, although to a lesser degree than in developing countries. This seems to suggest that the distribution of wealth among the individuals that constitute a society converges to a characteristic distribution for modern post-industrial societies. This characteristic distribution may vary in different economies, with new dominant technologies or with new environmental or social constraints. In biological terms diversity drives evolution.

Economic efficiency does not have to be hostile to the ambitions of the poor or affect social justice. Faced with the choice between economic growth and the reduction of economic inequalities as a tool to reduce absolute poverty, it is economic growth what will eventually achieve reductions in poverty in a sustained way. Policies aiming to reduce socioeconomic differences without taking into account the needs of the economy have irreducibly led to the collapse of the driving forces of the economy, impoverishing the rich and the poor in the process, and paralyzing or reversing the economic growth of the nation (see next chapter).

Exaggerated differences in economic incomes in a nation may produce social tensions that hinder economic growth. That is, high levels of economic inequity produce social inequity, which in turn produces social resentments, which can lead to political reactions against economic policies favoring growth. This quest for social justice promotes policies that in theory reduce economic differences in the population. In practice, these measures, programs and activities hinder economic development, damaging the prospect of the poor for economic advancement. This phenomenon has been observed on several opportunities in recent history in various parts of the world, and largely explains the emergence of social "revolutions" (communists, socialists, popular, nationalists) that hold back and often reverse the technological progress and economic growth of a country. By reversing economic growth, these policies increase poverty levels and social inequality increases, attaining opposite results to the initial objectives proposed.

To illustrate this interaction between growth, inequality and poverty, let us imagine a cart with rusty wheels that is pulled by a horse by an elastic cord. The elastic cord can pull but not push the cart. The force with which the horse pulls the cart causes the cart to move and stretches the elastic cord exerting greater force on the cart but with a time delay with respect to the force with which the horse pulls. If we oil the wheels, the cord will stretch less. If the horse does not pull, the elastic cord will contract but the cart will stop moving. If we make the analogy with growth, inequality and poverty, economic growth leads to inequality ("stretching the elastic cord”), which can be reduced (“oiling the wheels”) by implementing policies to reduce excessive inequalities in the distribution of wealth; but if we want to pull large sections of the population out of poverty (“moving the cart”), we will only succeed by boosting economic growth (“encouraging the horse to pull stronger”).

A naive and lively mind would say that if we use a rigid bar instead of an elastic cord, the equation is different. Economy, however, is a complex system in which the micro-economic actions and the macro-economy affect each other only with time delay. Reducing the elasticity of the cord to increase its efficiency as transmitter of energy for change is equivalent to maximizing the transparency of public and social decisions and maximizing the efficiency of market mechanisms. This is achieved by liberalizing the economy. Any attempt to control information, centralize decision-making and arbitrarily regulate economic actions that citizens may want to attempt, in practice, will make the cord more elastic. It is the lesson economic history teaches and is resumed in Friedrich Hayek's Economic Calculus, which will be discussed later.

From the empirical data presented here, we can conclude the following:

a- The economic behavior in developing countries and developed countries is different. In the medium term, economic growth does not reduce income inequality in developed countries, while it does in developing societies.

b- Economic inequality among citizens of developed countries is close to the minimum achievable in a post-industrial society.

c- Developing societies show a great variability in inequality of income of their inhabitants. There are poor countries with little income inequality and there are some with a very large economic inequality.

d-A decline in absolute and relative poverty can be effectively obtained with sustained economic growth.

e- Very large differences in the income of the inhabitants of a nation produce political instabilities that impede sustained economic growth.

Now then, economic inequality has been and is an element that affects our sense of justice and its elimination has been the main motivation of socialist and Marxist policies and other ideologies worldwide (an excellent analysis is provided by Karl Popper in *The Open Society*). Economic egalitarianism combined with individual freedom would be extremely desirable. Examples closest to this combination can be found among European democracies, especially in Scandinavian countries. The history of humanity says that pursuing of one of these aims without the other is only possible in a dream, and if we insist on it, it becomes a nightmare. History, thermodynamics and biological evolution teaches us that a certain level of equality is needed to maintain the cohesion of a society. But, it also teaches that freedoms promoting diversity is very important, that attempts to impose equality at all costs threatens freedom, and if freedom is lost, there will be even less equality among the un-free.

A fundamental consequence of inequality is that markets tend to satisfy only consumers who can pay for goods and services. Therefore, a fundamental role of the State is to ensure basic services for those excluded from the market. One mechanism that can help the excluded sectors access to fundamental goods and services, in addition to the optimization of market mechanisms, is the implementation of re-distributive policies. There seems to be consensus among contemporary economists that there is no better way for the developing world than to implement policies that aggressively promote economic growth and simultaneously implement re-distributive policies. The re-distributive policies that have shown greater efficiency to achieve lower levels of poverty are direct subsidies to the most vulnerable sectors of the population. These work best in areas of the economy that have high probabilities of stimulating self-sustained growth, such as education, health, transport and communications infrastructure.

Economic growth also has a moral and psychological component that is important to consider. The value of an increased well-being rests not only on material improvements that the individual takes advantage off, but also includes social, political and moral aspects deemed important to people. Increasing prosperity, history suggests, makes people more tolerant, more willing to settle differences peacefully, and more inclined to favor democracy. Economic stagnation, on the other hand, is associated with intolerance, social tensions, and authoritarian dictatorships. It is only with sustained economic growth that each and every one of the citizens of a country can aspire to a better life for themselves and their children. Economic improvements are far from negligible when aiming at the moral and psychological well-being of people. We should strive to what economists would call a Pareto superior solution to a problem of satisfaction of present needs: A solution that optimally benefits all and each of the members of the community.

# 7. THE STATE

In most countries, and for much of recent history, the State is and was the most important political and economic actor. The State has many forms, and its detailed analysis escapes the possibilities of this work. However, its impact on the problem that concerns us is undeniable and we have to deal with it. We will present a selection of modern approaches that relate the State and its functioning with the creation of national wealth and the emergence of poverty

## *The Size of the State*

The state, and specifically the size of the state in relation to the rest of the economy, affects the growth in wealth of a country. Figure 7.1 presents four different examples of the relationship between the size of the state and economic growth. In these four examples, GDP per capita is represented (in constant US $ of 2005) and State involvement as a percentage of total government expenditure in relation to the whole economy measured as GDP, from 1960-2013. The countries sampled are Thailand, Ireland, Chile and Kenya.

**Figure 7.1: History of State size and wealth production in four countries**

Relationship between GDP/capita (dark line) and percentage of GDP captured by the state (light line) during five decades (horizontal axis). Data from the World Bank.

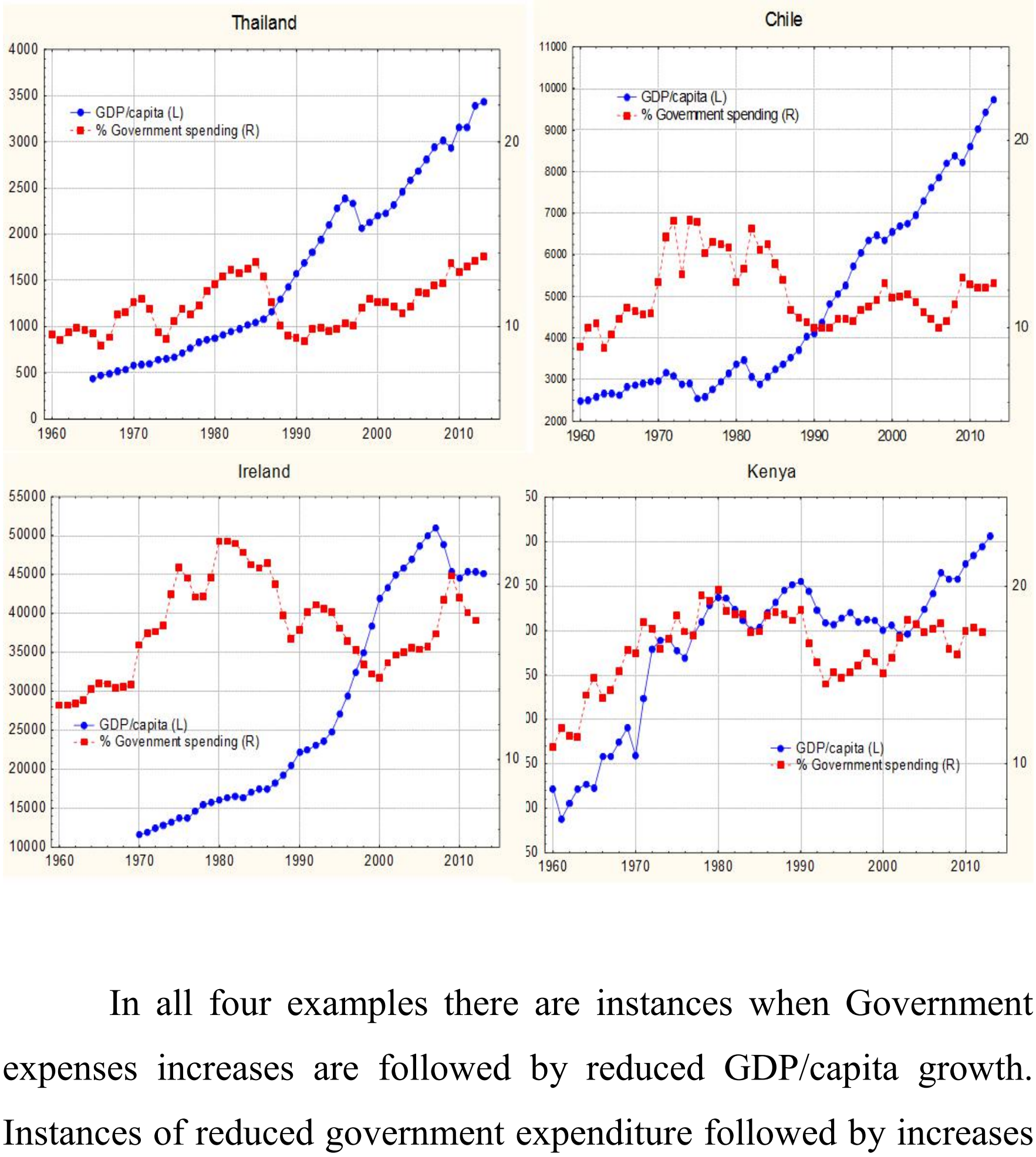

In all four examples there are instances when Government expenses increases are followed by reduced GDP/capita growth. Instances of reduced government expenditure followed by increases in GDP/capita growth are conspicuous in all these examples. The correlation between GDP/capita and the size of the State is analogous to that of the population of a prey and its predator or host and parasite in the ecological literature. When predator populations grow too much, they reduce the population of prey; when prey populations grow fast, they trigger the growth of predator populations; and when predator populations fall, prey populations increase. It is amazing to see how the dynamics reflected in the economic history of Thailand, for example, where the state contracted its spending in 1986 triggering economic expansion, is indistinguishable from any prey/predator population dynamics.

We can find examples of conservative economic policies that achieve during certain time periods continued growth of the State, but at levels that did not stifle the growth of the economy. That was the case in all four countries during decade of 1960 and parts of 1970. These examples suggest that there are states, or activities of these, that are harmful to the national economy, while others are less harmful or even positive. This leads us to conclude that not only the size of the State is important in determining the economic growth of a country, but also its quality or the quality of its activities.

Figure 7.1 showed that the State can exert a positive influence on the economy, that it can be neutral, that it sometimes inhibits economic growth, and that it can act as a parasite sucking out the vitality of the economy forcing it to contract or even to collapse. In the words of Adam Smith, there are two types of States: the proper one, where the government gets its sustenance from society through taxes; and the improper one, in which the most important resources are property of the State and it distributes them to the society. From concrete experiences already described by Adam Smith in 1776 and since then studied in multiple opportunities in several continents, governments in improper States create dependency, monopolies, cronyism, political privileges and corruption, slowing down the economic development of the country.

The role of government in modern society is to regulate the interactions of the various sectors of society and to ensure compliance with the laws. A modern government is responsible for regulating and harmonizing the relationship between producer and consumer and cannot become an actor of the economic production without affecting its essential functions. Seen from the metaphor of government as a referee in a sports game:

- *Every serious sports game has a referee.*
- *A sports game without a referee is a sloppy game, but it is a game.*
- *A sports game where the referee also scores points stops being a sports game.*

But not only do states refrain from exceeding its regulatory responsibilities. Unconstrained, the state tries to interfere in the economy, and its actors search profits for the governing individual, party or clan. This might lead to a parasitic relationship, where small elites monopolize the government. In this case, governments are true parasites, as suggested by the dynamic relationship shown in Figure 7.1, which shows a dynamic equivalent to the parasite-host, or predator-prey dynamics described in ecology textbooks.

### *Bureaucracy and Business*

It is essential to clarify our understanding of how the activity of the State and its legal system influences the formation and maintenance of the most important modern social institution: the company or the modern enterprise. It is the company, which coordinates the work of individuals, allows the synergistic action of these and generates new wealth. Peruvian economist Hernando De Soto masterfully highlights the relationship between state regulations, establishment and operation of businesses and the generation of wealth in a country. He measured how the degree to which the legal system of a nation hinders enterprise affects economic growth of a country. Thus he demonstrated a clear link between government regulations and the possibilities of creation and accumulation of wealth by their citizens. Originally, De Soto and his colleagues estimated in several developing countries the number of steps or errands required to formally legalize a small business with just one employee. Then, they calculated the relationship between the wealth of the nation estimated by its gross domestic product (GDP) and the bureaucratic pain its citizens endure as estimated by the number of bureaucratic steps required. The result of the analysis of this relationship cannot be more eloquent. In countries with high GDP/capita, legal and formal errands counted by De Soto and colleagues would take a few hours to successfully complete. In poor countries, these same tasks took De Soto collaborators months or even years to complete.

Recent data from 130 countries analyzed in a similar way, conducted in 2003 by the World Bank, confirmed this phenomenon. If we plot for each country (each data point on the graph) the days required registering a business (horizontal axis) against the per capita income of the nation (vertical axis), we see an exponentially decreasing relationship, shown in Figure 7.2.

**Figure 7.2: Levels of bureaucracy and wealth of a nation**

Relationship between the days required to legalize a business (horizontal axis) and the wealth of the inhabitants of the nation as Nominal GDP expressed in US $ per capita (vertical axis). Data from the World Bank.

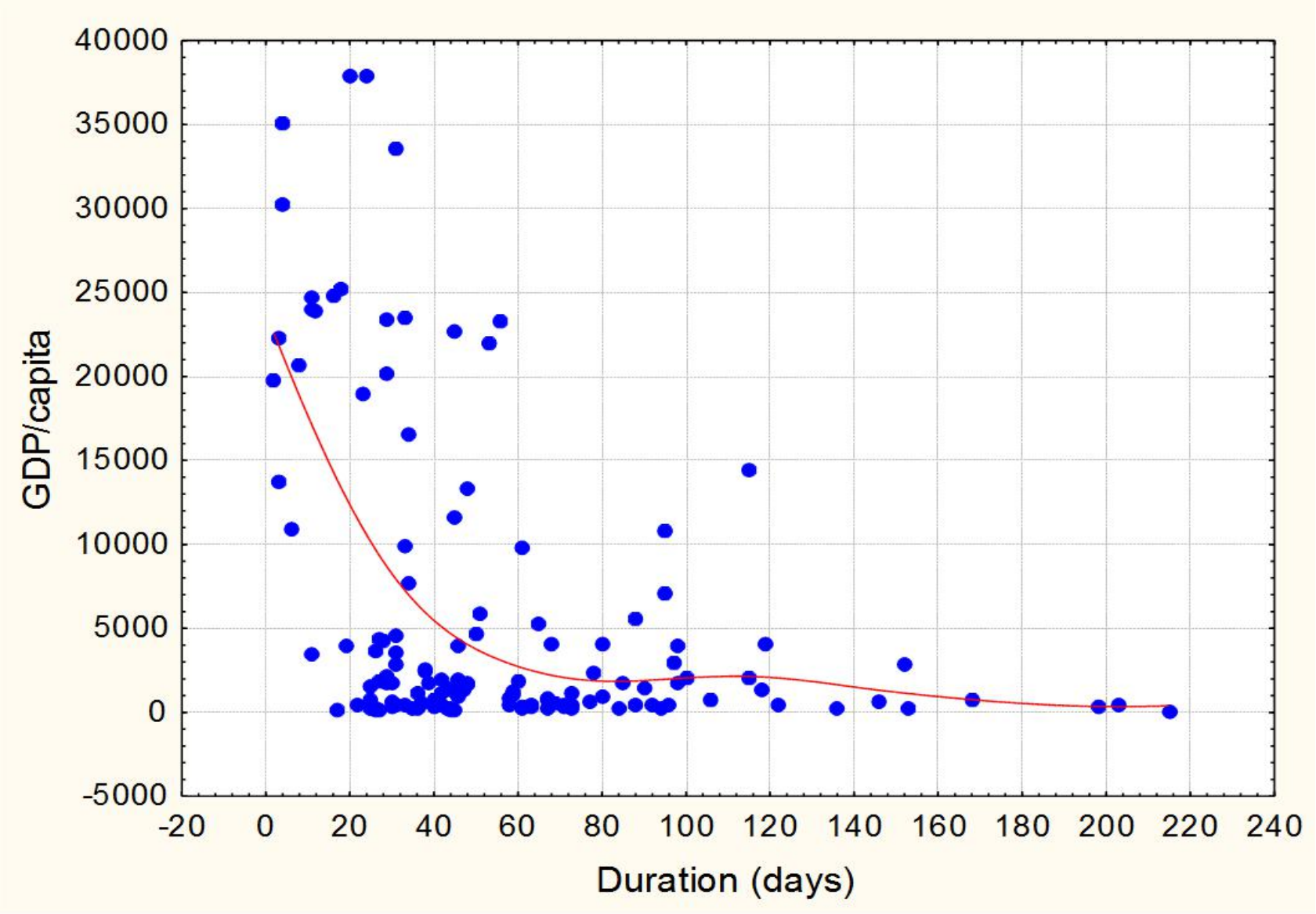

This and several other studies regarding to the size of the State and its effect on the economy of a country, suggest that the State often behaves indeed as a parasite of society, weakening its strengths and potentialities and reducing economic growth and the production of wealth. The more the State interference in the activities of a society, and the more complex and arbitrary its regulations, the greater the damage it does to the economic activity. State interference hinders its citizens from generating wealth, condemning the country to endure higher levels of poverty. In these cases, the State deserves the description of Leviathan.

The State as rentier and dominator of the society was described by Adam Smith in the XVIII century when comparing Spain to England. The gold and riches of America that flooded

Spain created a State that stifled the entrepreneurial spirit of Castile, allowing much poorer kingdoms, which supported the free exercise of economic activities of their subjects, to become rich powers, eventually surpassing Spain in power and wealth, as was the case of the Netherlands and England.

A critical phase in establishing a business is its beginning. This phase is particularly critical in innovative companies and companies that open novel markets. The small innovative businesses are the ones to suffer the most from excess bureaucracy in critical phases of their growth, as the bureaucracy attacks the weakest point of the creative process of the economic production. The harmful effects of bureaucracy that hinder innovation and business creation can be extreme. Often in government circles, bureaucracy is perceived as a necessary and unavoidable evil in a promoter State. The empirical experience of many modern governments that have managed to activate economic growth in their countries teaches us otherwise.

## *Institutions and Private Property*

A desirable characteristic of a modern government is to foment and tolerate strong public and private institutions. The French Emperor Napoleon Bonaparte used to say that only institutions determine the fate of nations. Efficient institutions in the management of legal, social and political aspects are essential to the functioning of a modern State. The efficiency of this operation is correlated with the efficiency of the functioning of society and therefore of the economy. The relationship between the solidity of the institutions and the average wealth achieved by the countries is a suggestive evidence of the importance of this factor in determining economic growth. Institutional strength can be measured in several ways. Several studies detected strong direct correlation between the index of institutionality and the wealth of a nation. Rich countries have strong institutions whereas poor countries do not. These indices are built measuring things like the stability of constitutions, the strength of the legal system, the independence of the legal system, the enforcement of law, etc.

Some studies focused on the strength of the law and its enforcement regarding the protection of private property. They demonstrated the existence of a strong correlation between the per capita income of a country and the strength of the private property. The strength of the private property can be measured, for example, by the level of risk of expropriation of the property. The direct relationship between these variables suggests that at higher levels of protection of the private property, higher levels of wealth are achieved. Or, that low protection of private property relates to poor economies. In both cases, a conspicuous correlation between the strength of institutions and economic growth achieved by the country can be evidenced statistically. This suggests the functioning of basic economic principles. A long-term guarantee of private property that is easily exerted promotes investments that in turn encourage economic growth and the generation of wealth. This finding vindicates one of the main functions of the State: the protection of the weak against the arbitrary exercise of power by the strong. A proper State implements instruments that protect individual freedoms and access to opportunities, which include the private property.

The importance of public institutions to protect private property and facilitate its operation can be understood by comprehending the dynamics of the "tragedy of the common goods". Goods that belong to all do not receive the care they need to remain productive indefinitely. For example, let us consider a communal fishing preserve, or public grassland used as pasture for domestic animals. Each user is interested in extracting the maximum amount of resources possible. If he does not, the others will. This leads to overexploitation of the resources, as no individual in his right mind is going to invest capital and resources to increase the yield of this common good. Any investment made by an individual will be immediately absorbed and utilized by the other users of the common good.

There are two solutions to this problem, the regulation and control of the use of the common resource by a regulatory agency

(the State), or convert the common resource in private property. The latter involves partitioning the common good among several users, or regulating its use so that individual responsibilities can be tracked and rewarded. Then everyone may invest in his land with the assurance that no one will steal his investment. This promotes and encourages investment and ensures the sustainability of the resources in the long-term. The first option implies that the State makes the investment and bills third parties for the resources required to carry it out. It is easy to imagine that third party intervention, by having a more distant relationship between the origin of the funds to invest, the investor and the end user, will be the most inefficient solution from the economic point of view. Therefore, whenever possible, the management or exploitation of resources in the long term should be assigned to private entities. This solution will minimize the dissipation of efforts and resources, and be more efficient in economic terms, favoring greater rates of wealth accumulation.

Governments have fundamental responsibilities in guaranteeing private property and regulating its management. For example, the unregulated implementation of private property may create monopolies, which produce an inefficient distribution of resources. Private property, for its proper application, requires a rational regulatory system, a transparent judicial system that can resolve conflicts in an efficient way, a fair access to resources for investment, and easy access to relevant technical and legal information. Providing these conditions is the role of private and public institutions like the central bank, the judiciary system, the education system, the media, law enforcement and others.

## *Government Policies*

Governments, especially if they are too large relative to the community they serve, if they centralize too many resources, and if their actions are opaque, are inefficient. They provide public services deficiently and at high cost. So what is an optimal amount of government necessary for the proper functioning of a nation? Not all government programs increase poverty. Many examples of successful government programs are known. Eradication of infectious diseases, development of sophisticated science and technology, prevention of natural catastrophes, and many other examples are proudly exhibited by governments to justify their existence.

Some government programs have been successful in reducing the levels of poverty in their countries in relatively short periods (decades). However, there is no agreement on the policies that enable to achieve this goal. According to the CEPAL - Social Panorama of Latin America, 2001-2002 Edition: “The exposed elements reiterate the need for economic and social policies that strengthen the possibilities of expanding the productive base, but at the same time involve the progressive redistribution of income, which allows that economic growth improve quickly the standard of living of the population with fewer resources". That is, they recognize that economic growth is central to reducing poverty levels but also value policies of income redistribution.

Prioritizing or emphasizing redistributive policies over policies that favor growth slows down and even reverses economic growth, causing higher levels of poverty. That is the lesson we learn by analyzing the history of the experiments with communist economies and with ultra-nationalist policies, as detailed in the next chapter.

## *Corruption*

We often hear about, both from ordinary citizens and from the professional politician, that the cause of the high levels of poverty is corruption. Now then, what is corruption? The dictionary defines it as "the practice of using funds and functions of public organizations for the benefit of oneself or of a few". Corruption is difficult to measure objectively and acts that could be called corrupt in one country do not necessarily pass as such in another. A classic argument runs as follows: if corruption makes money flow, from the economic point of view, it is equivalent to any other form of economic activity and any resources subtracted from the economy by corruption gets back to the economy when the corrupt spends those resources, stimulating economic activity in general. The counterpart to this argument assumes that corruption leads the use of scarce economic resources to less efficient economic activities, reducing the growth potential of a country, and thus amplifying poverty levels unnecessarily.

Here, I would like to show with an example, how the relationship between corruption and economic growth of a country can be revealed. Despite the difficulties in defining corruption clearly, a nongovernmental organization based in Berlin, called Transparency International, conducts an annual survey worldwide to measure it. With the participation of entrepreneurs that have activity in the country, a questionnaire estimates on a scale from 1 to 100 the levels of corruption and/or of transparency in each country. On this scale, 1 indicates high levels of corruption while 100 indicates low levels of corruption and high levels of transparency in businesses with the State. Using this indicator of corruption, a correlation between high levels of corruption and low economic development can be demonstrated as shown in Figure 7.3.

**Figure 7.3: Corruption and economic wealth**

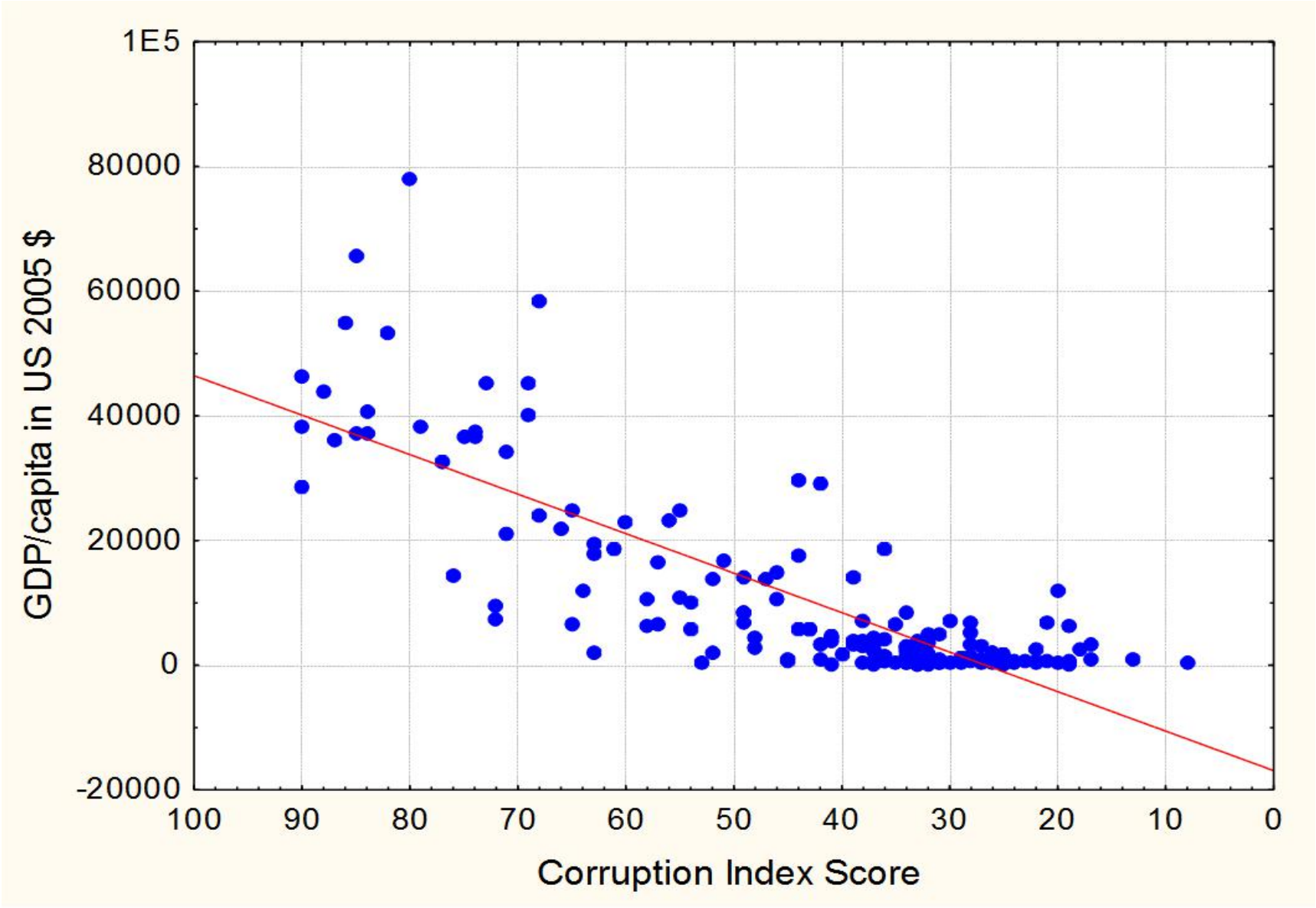

Relationship between economic wealth (vertical axis) and transparency of economic activities as measured by Transparency International (horizontal axis). Data from the World Bank and from Transparency International.

Many other studies have shown dependence between corruption and foreign investment, and the dependence of foreign investment and economic growth. Extrapolating these findings suggest a relationship between corruption and economic growth. Foreign capital inflows help economic growth, especially so in developing countries. Countries open to capital that receive foreign investment in large quantities are more likely to have higher transparency scores. Countries that are closed to foreign investment and that treat foreign and national capital with punitive regulations and bureaucratic and financial constraints are more likely to have low transparency indices. A study published by The Economist showed that the countries most open to capital achieved positive and high growth indexes in the nineties, while developing countries with little foreign capital flows had an economic decline in that period.

Transparency or the level of corruption in a country is not only correlated with the levels of external resources entering the country as investment capital. Transparency affects many other levers of the economic engine. A most important effect of corruption is to slow down and make more costly capital investment. That is, corruption increases the bureaucratic steps initially described by De Soto, reducing the events of successful investment, slowing down private sector initiative and thus decreasing the economic activity of the country. Like bureaucracy, corruption affects small businesses disproportionately.

Another aspect that harms an economy is the dissipation of wealth caused by corruption. These economic losses can be very important. It is possible that part of the misappropriated resources from investment projects and social projects, because of corruption, be reinvested in the economy. Nevertheless, it is more likely that most of these resources are incorporated into economies of richer countries, or be reinvested with a much lower efficiency.

The most efficient formula to fight corruption is the simplification of bureaucratic processes, transparency of decision-making systems and the implementation of clear and simple laws and regulations. Implementation of these recommendations, however, has not been easy, as they often affect interest of the ruling elite.

Countries that are rich in natural resources often have poor populations because the temptation of governments to meddle in the exploitation of these resources, forgetting prudence and exceeding State intervention in economic activities. Oil and other natural resources and the companies that exploit them have often been willing to deal with any person and government that assured them a concession. This has benefited corrupt and repressive governments and has fostered armed conflicts. In Africa, civil wars have devastated resource-rich countries such as Congo, Angola and Sudan. In the Middle East, democracy has failed to materialize. Controlling this curse of the so-called *devil's excrement*, could do much to alleviate the poverty and misery in the world. This can only be done with transparency and responsibility.

Private initiatives such as *The Open Society* from the banker George Soros, or public ones such the *Extractive Industries Transparency Initiative* from the British government, try to alleviate this problem by getting each company or government to publish what it pays, including bribes and other aids to governments or companies and their representatives. The implementation of these initiatives has borne fruit for some time in Azerbaijan, Nigeria, Sao Tome and Principe, Kyrgyz, Ghana and Trinidad and Tobago, Peru and East Timor. But in the end, it is the capacity of the common citizen to control and demand responsibility from their governments that will control corruption. This requires institutions and democratic and transparent systems that allow citizens to be informed and exercise their rights and claims. This is associated with the maturity of a society and is a measure of progress of civilization.

The founders of the republic in the United States of America have a lesson to teach. They were well aware of the dangers an uncontrolled state poses. They suffered abuse, arbitrary rule and humiliation from their colonial rulers and wanted to prevent such things from happening again. They implemented a delicate system of controls and balances so that different government institutions could control each other. This is reflected in the Constitution of the USA which aims to guarantee individual liberties and constrain the power of government through a system of separation of powers. These systems of balances seemed to have worked well for the inhabitants of North America. The Constitution of the USA is among the constitutions in the world that have suffered the least changes. At the same time it has helped to create a world power that manages to provide a large range of opportunities to its citizens, making them among the richest of the world.

## Rational Thinking

There is a long list of further ways to measure the effect of various aspects of the action of a state on the economy. All of them show strong correlation between an index or an estimate and an economic variable related to wealth or growth. Each index seems to explain some economic phenomena better than others do. It is analogous to the example of the complexity of a car. Many different aspects must converge for the smooth working of an economy. But some indices might capture better or more relevant information than others. The preferred tool for economists to tackle these complexities is the use of statistical multiple regression analysis. These analyses produce sharp numbers that load economists with confidence that they have grappled a complex problem. In most cases, however, that is a fatal illusion. Multiple regression analysis requires characteristics of the data used that few datasets comply with. The overconfidence these analysis generate have been the cause of many an economic crises.

Non-parametric statistical tools avoid in most cases, the need for sophisticated analytical tools that should be used only by experienced professionals. Here, I choose a simple cluster analysis to throw some light on the problem we are discussing. In Figure 7.4, we compare a selection of widely used indices as to their degree of correlation with the wealth of a nation measured by its GDP/capita.

**Figure 7.4: Relation between social, political and economic indices**

Dendrogram showing the results of a cluster analysis of a variety of widely used indices. The smaller the linkage distance the closer the indices are related. For more details, see Jaffe et al in Bibliography.

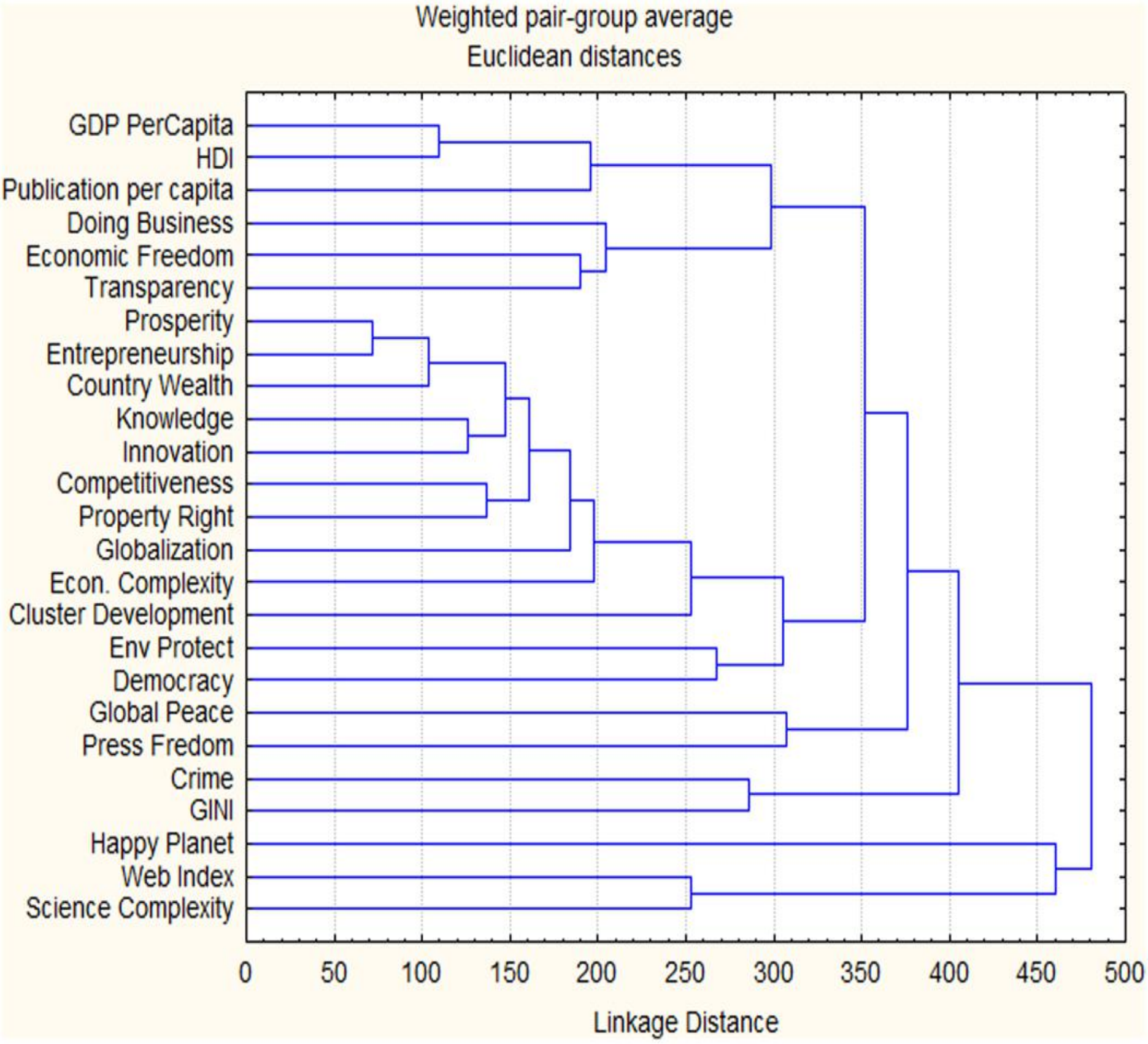

The dendrogram has the index most widely used for economic wealth, GDP per capita, at its root on the top. Very closely correlated to GDP/capita is the Human Development Index (HDI) produced by UNDP. It reveals about the same trends when used to compare differences among countries as GDP/capita. Another cluster of closely related indices is conformed by the Ease of Doing Business index of the World Bank, the Corruption Perception Index by Transparency International, and Economic Freedom estimated by the Heritage Foundation. The indices less correlated with GDP/capita are indices related to Crime such as the Crime Rate Indicator of the World Bank, the GINI indicator produced by UNDP, Happiness as calculated by the New Economic Foundation NEF also referred to as Happy Planet, and other sophisticated indices such as the Web Index produced by the World Wide Web Foundation and the Science Complexity index related to economic growth published by the on-line science journal Plos One.

The index closest to GDP/capita and HDI was an index little used by economist and not tracked extensively by the World Bank: The level of scientific activity in a country as measured by its scientific publications. This index is calculated using the number of scientific publication of a country, as recorded by SCOPUS, divided by the population of the country, to produce the index "Publications per capita". This index is an estimate of the amount of scientific activity in a country. This surprising result shows that other factors besides natural resources, capital and the state affect economic activity and thus economic growth of a country. Wealth is related to production of goods and services, which in turn needs science and technology to prosper. But rich countries have more resources to invest in research, and therefore produce more scientific papers. Our understanding of these processes involving various aspects of a complex network of relations is still incipient. Resolving this network of complex relationships requires expanding our explorations to other realms of human activity. An advance of the phenomena to be discovered in the next chapters is the importance of rational thinking. Counties which value rational scientific thinking seem to be less prone to use emotions for decision making, allowing economies to develop long term plans. In contrast, more emotional policy decision making affects the scientific environment and force economies to adapt to short term horizon, hindering the growth of strong and sustained economic development. This explains the fact that in Figure 7.4, scientific development, measured as scientific publication per capita, is the factor strongest correlated to economic growth.

# 8. CULTURE

For many politicians, scholars and contemporary thinkers, social phenomena that characterize humanity are the result of cultural factors. Culture is understood as the human values, creations, emotions, beliefs, transformation capacities, spiritual concerns, sense of belonging, historical consciousness, ethics, aesthetics, collective images, symbolic frames and any other manifestation of human behavior that has an impact on fellow humans. Culture has been related to the phenomenon of wealth creation, another product of human behavior. However, general definitions do not help to dissect a problem or to extract information from observable fact or from the available data. Therefore, we will analyze different cultural factors separately, knowing that they act jointly. The more we can understand the relationship of specific cultural factors with the problem being analyzed, the more we will be able to perfect our definitions, and the better we will understand the relationship network that relates culture phenomena to the creation of wealth of the nations.

## *Religion*

Religion was one of the key factors in modulating the development of mankind by a significant period of history. Religion,

in ancient times, often constituted the only model of the world that society used. It caused many wars during its expansion and served and continues to serve as a template for the institutionalization of the State in several countries of the world.

Several intellectuals and economists, as for example D.S. Landes, from Harvard University, USA (see *The Wealth and Poverty of Nations: Why some are so rich and some so poor*), argue that the differences between religions explain the difference in the accumulation of wealth achieved by the nations of the world. Landes suggests that especially the Catholic religion does not favor the generation and accumulation of wealth and postulates that Protestant creeds are more conducive to making nations rich. The proponents of this explanation, draw on Max Weber's work of 1905 (*The Protestant Ethic*), and argue that Catholicism foments values related to unconditional obedience, poverty and suffering, and rejects as sin profit and accumulation of wealth. Protestant Christians, in turn, value work and accumulation of wealth, and disdain inactivity and leisure. It is this work ethic and the propensity to save, according to these proponents, which enabled economic growth in the Protestant Anglo-Saxon world. This view explains, for example, the relatively lower economic development of the Latin world by the dominance of Catholic values in these countries.

This argument, although drawing attention to interesting phenomena, is rather superficial. Religions are part of culture and should be included in any cultural analysis, as will be done in the next section. But considering only religion as modulator of the economic behavior of a nation is too simplistic. A more elaborate analysis of religion as a cultural adaptive instrument, using a complex systems approach, is carried out by D.S. Wilson in his book *Darwin's Cathedral* published in 2002. Wilson presents abundant empirical evidence to conclude that the evolutionary value of all the religions he studied consists in maintaining the cohesiveness of societies, enabling the harmonic development of its social and economic activities, and controlling social parasitism.

The argument popularized by Landes was developed more carefully long before by many prominent thinkers, such as the international politician, military man and thinker Francisco de Miranda, who especially in his visit to France, Germany, Switzerland and Italy in 1788, compared Catholic to Protestant provinces. He describes that protestant provinces, by being usually free from the yoke of a feudal lord, are freer and more prosperous. He argues that it is not religion itself, but the dogmatism and fanaticism of a society which discourages economic development. An example falsifying the religious determinism theory of Landes and supporting Miranda's vision can be found in modern Germany. Bavaria is the State of the German federation that is mainly Catholic. Bavaria is among the richest states in Germany. Citizens of most other states of the German federation profess the Protestant religion and yet are poorer than Bavaria. The German example is particularly relevant as the cultural differences between the states of the federation are minimal, while the differences in the religion their inhabitants profess are conspicuous. That means that the effect of religion on economic development can be measured with little interference from secondary factors in Germany and the result is the opposite of the one postulated by Landes.

## *National Culture*

A recurring theme in discussions about the occurrence of poverty and the reasons that explain the difference in the wealth of nations is traditions and culture. Now then, what aspect of culture may be linked to the reasons that determine the economic success of a nation? Are the values of a society reflected in the characters that the society idealizes? Can culture contribute, in terms of the accumulation of wealth is concerned, to the success or failure of a nation?

Culture encompasses many aspects of human social life that are unrelated to the economy or to the creation of material wealth. Culture is often associated intuitively with factors that build individual happiness, and for many people it is more important than economics in determining the happiness of a society. However, happiness, measured through surveys using questions that estimate the self-assessment of the feeling of happiness are very unreliable and the results volatile. Often, people of the same culture are happier if their indexes of human development or their levels of wealth are higher. But many instances of happiness uncorrelated to wealth but correlated with culture have been published. This shows that the relationship between wealth and culture is rather complex. We know for example that culture affects a company's ability to create wealth. We also know mechanisms by which the economic welfare of a society affects its culture and know this relationship involves many aspects in a network of relationships, many of them still to be discovered.

Different cultures in different eras have different values, but some values are universal. Sectors of a society might resemble more a given sector in another society than a neighboring sector in their same country. Values are reflected in the personalities a culture admires. I do not know of any thorough review of the idols of each country. However, some idols are very good in defining values of the people idealizing them. These idols are important positive role models for some, while in other cultures, even in the same country, they are considered as negative models to be avoided.

For example, the Argentine revolutionary Ernesto "Che" Guevara is considered an important figure in much of the world today. He is admired by many followers for his conspicuous role in the communist Cuban Revolution led by Fidel Castro and by his efforts to export Cuban Marxism to Latin America by violent means. His violent death while attempting to change power in Bolivia is considered an ultimate personal sacrifice for his ideals. Modern admirers of the "Che" usually despise iconic characters such as David Rockefeller or Bill Gates, some of the richest men of their time, for considering them to be capitalists, motivated by profit, who accumulate wealth at the expense of the suffering of others.

Others consider David Rockefeller or Bill Gates characters with unique skills to produce wealth and of exceptional kindness. They appreciate their efforts to improve the wellbeing of millions around the world. They applaud the contributions of these idols in improving the incomes and allow a decent job to a huge number of workers worldwide. Some of these people, who appreciate Gates or Rockefeller, despise "Che" Guevara. They emphasize that "Che" was an egocentric idealist with no ability to understand the Latin American reality or basic economic principles, who failed in all his undertakings, leading to the deaths of hundreds of naive people who followed him in his dreams.

This difference in the valuation of characters reflects different underlying values. These values are influenced by the culture that embraces them. For example, a culture idealizing Che

Guevara values pain, sacrifice, suffering and idealism; the culture admiring Gates may value more success, wealth, work and pragmatism.

Values can be investigated with quantitative tools. A good example is provided by the “World Values Survey” directed by Ronald Inglehart from the University of Michigan. Their research tries to capture culture through the values associated with life in a society. The results of this investigation show that quantitative measures separate different cultures sharply. The groupings of cultures thus achieved are very similar to what the intuition of a knowledgeable person would do, and to what historians, sociologists and politicians have indicated on several occasions. Scandinavians resemble each other; countries speaking languages derived from Latin share many values; North Europeans are different from the ones living in the south of Europe; Arab-speaking countries share many values; and English-speaking cultures resemble each other in many ways.

The probability of finding rich or poor countries in different cultural groupings is not uniform. The culture of the Nordic countries of Europe, for example, is more associated with economic wealth in our days than cultures common in the Mediterranean basin, contrary to what prevailed during the times of Said al Andalusi (see Chapter 3). However, it is interesting to note that there are examples of countries that have managed to accumulate significant wealth in modern times in almost all cultural groups. Clearly, the relationship between culture and economic success as a nation is far from clear and will require major research efforts to be elucidated in the future.

One of the cultural characteristics that have been correlated with the ability of economic progress and entrepreneurship is trust. In his book *Trust*, Francis Fukuyama argues very convincingly that social capital establishes ties among individuals based on trust, and this is a factor that accelerates or promotes the dynamics of accumulation and creation of wealth in an industrialized society. Mistrust and the inability of being able to count on the support of others in the creation and development of a company constitutes a very high cost, which often prevents the industrial and economic development of a nation. Interpersonal trust may create networks, based on family ties as seems to be the case of the Chinese and Latino cultures; or it may develop a solid mesh of laws and social regulations as seems to be the case of many Anglo-Saxon cultures. These differences in the trust networks provide large differences in the modes of wealth creation and the economic structure of society. Societies with high levels of corruption, preoccupy themselves with naming fine differences in feelings of shame and guilt (see Chapter 4), are more likely to have rudimentary trust networks, are the least likely to have economic growth, and are prone to suffer poverty and misery.

As trust is a cultural element that can be acquired, nurtured or developed, formal and non-formal education could weaken or invigorate it. This suggests that education has a preponderant role in modulating the growth as will be explained in the next chapter.

## *Law and Rules*

One of the most important cultural aspects is the way society organizes itself and values its institutions. Several of these values and systems are reflected in the laws and codes of conduct of society. The relationship between rules, laws and economics is very old. The first known laws, such as the Code of Hammurabi and the Ten Commandments of Moses, regulate among other things, private property and trade, cornerstones of any economy. Every law somehow regulates economic exchange and aspects affecting the forms of production and generation of wealth. Little is known quantitatively about this relationship.

In Chapter 7 we saw that the State may acquire excessive powers that stifle the economy and that Constitutions and the Law should place controls and balances so as to maintain a harmonious relationship between the powerful and the week. Unfortunately, the borders between regulation favoring harmony, restrictive authoritarian rule, and straight exploitation of the citizen by a parasitic government, are fuzzy.

Two systems of law that differ in their origin and underlying logic are currently widely spread around the world. These legislations differ in terms of the philosophy on which they are based, and have had a major divergent impact on the economic growth of modern nations. They are known as the "Common Law" and the "Constitutional Law". Although a study of the legal history of these systems escapes the capabilities of this work, a very simple representation of them, bordering on caricature, may convey an idea of the importance of the legal framework in the development of cultures and civilizations.

As a legacy from the Roman Empire, many countries adopted constitutions that promote comprehensive legal systems. These systems, designed by legislators comfortably gathered in the capitals of the countries, and mostly unaware of the problems suffered by the average citizen, instruct and attempt to regulate the activities, relationships, obligations and rights of citizens, including those living in the remotest corners of the country. That is, the law is born at the center of power and is aimed at citizens at the base of the pyramid of power.

In contrast, as a result of the Nordic, Barbaric, Viking, and Anglo-Saxon tradition, the Common Law assumes that problems should be resolved when they arise in specific setting, adapting the decision to the local circumstance. The exercise of power through the law is carried out locally by juries, judges or local councils, which consider local tradition to resolve differences and conflicts. The accumulation of these decisions is used to guide future decisions and in this way a framework for legal reference is being established. In contrast to the Constitutional system, this system builds the legal framework from the bottom up and from the periphery to the center.

This brief description is not fair with the complexity and richness of existing legal forms. Countries that have developed its legal system from the Common Law have constitutions and laws devised by professional legislators. Likewise, many countries with constitutions based on Roman law have relaxed their governance systems allowing incremental degrees of freedom to the provinces, districts and municipalities. However, it seems that the effect of diverging conceptualizations of the law: centralized vs. aggregate has had a lasting impact on the political-economic performance of a country. Or, is it more the product of an ideological difference that characterizes these countries? In any case, this difference markedly affects the potential of a country to produce wealth steadily over time.

A multitude of legal systems is common today. They might be grouped into two diverging groups. We may caricature the extremes or poles of some of these visions of legal thought as follows:

*- The law limits power (Anglo-Saxon tradition) vs. The law is an instrument of power (Autocracies)*

*- The common law favors individual responsibility vs. Central government seeks to implement social responsibility*

*- The Individual vs. The King*

*- Free market vs. Government regulation*

*- Decentralization vs. Central government*

*- Law as product of trial and error vs. Law reflecting unchangeable moral principle*

*- Emergence of spontaneous structures vs. Central planning*

*- Regulation vs. Prohibition*

One of the effects that the legal system has on a country is that of determining the degrees of freedom of its economic system. Countries vary significantly in terms of the freedom granted to the entrepreneur to perform economic activities. These differences are followed annually by the Heritage Foundation. Some of the correlations between economic and political and freedom and economic efficiency were presented in the previous chapter. It is not difficult to detect the relationship between the Scandinavian and Anglo-Saxon societies that originated from Scandinavian or Anglo-Saxon settlers, and the levels of economic freedom assigned to these country. Canada, USA, Australia, New Zealand and the United Kingdom are all countries where the Common Law prevails and which are classified among the economies with greatest economic freedom in the world.

Now then, a system with large economic freedoms does not necessarily have to encourage the creation of wealth of a country. Or, will it be that economic freedom is the broth of sustenance of a healthy modern economy? Data from the World Bank, comparing Economic Freedom measured in 1995 with the economic growth achieved by 142 countries for the period 1995 to 2002, reveals that the countries grouped in the first quintile of countries with the highest degree of economic freedom, achieved an average annual growth in this period of 4.9%. The values for the next quintiles of decreasing economic liberties were 3.8, 3.4 and 3.1 %. Counties in the quintile with least Economic Liberties achieved only 2.5 % of annual growth.

That is, the group of countries that were grouped by having the highest economic freedom between 1997 and 2004 (first quintile) experienced the highest economic growth in the period 1995-2002. In contrast, the group of countries with less economic freedom, grouped in quintile 5, showed the least growth of all. In other words, economic freedom in a country, which we know is correlated with its political freedom, its democracy and its legislative autonomy, is also correlated with economic growth. The greater the economic freedom is, the greater the potential for growth in a modern and diversified economy.

This relationship was tested experimentally worldwide with the communist revolution of the XX Century. This experiment allows us to compare the economic performance of centralized economies where the state plans the economy following the vision of Karl Marx, with countries where the state grants economic freedoms. The former are the “Improper States”, according to Adam Smith, and the later are liberal "Capitalist" economies where the State plays a more regulatory role and is less involved in the production of goods and services. We can enumerate pairs of countries with equal geography and history that differ radically in the economic welfare their citizens enjoy, thanks to the different economic systems practiced. These are the cases of North and South Korea; the former East and West Germany; Haiti and the Dominican Republic; Puerto Rico and Cuba; Hungary and Austria; Taiwan and mainland China; Palestine and Israel and many more.

**Table 8.1.** A small selection of examples is presented in the Table where increases in GDP/capita are compared between the years of 1950 and 1990. Here GDP per capita is estimated using 1990 International Geary-Khamis dollars from the Angus Maddison

historical dataset.

| | 1950 | 1990 | % Difference |
|---|---|---|---|
| | | | |
| **Hungary** | 2480 | 6459 | 160 |
| **Austria** | 3706 | 16895 | 356 |
| | | | |
| **North Korea** | 854 | 2841 | 233 |
| **South Korea** | 854 | 8704 | 919 |
| | | | |
| **Tanzania** | 424 | 549 | 30 |
| **Kenya** | 651 | 1117 | 72 |
| | | | |
| **Cuba** | 2046 | 2957 | 45 |
| **Puerto Rico** | 2144 | 10539 | 392 |
| | | | |
| **China** | 448 | 1971 | 318 |
| **Taiwan** | 916 | 9938 | 985 |

Examples of sequential experiments, where the same country shows a rapid growth after the liberalization of the economy, can be studied in the case of England, the USA, Japan, Germany, Chile, Singapore, South Korea, China, India, etc. In all known cases of countries with states that regulate the economy rather than interfering in it, produce benefits that are orders of magnitude higher, both in economic terms and of the welfare of the population, than those of economic systems with centralized economies. The lesson we learn from these examples is that strong and independent institutions, the prevalence of the law over the whim of the powerful, honest and efficient governments, economic contracts that are impartial and apolitical, economic freedoms, and adequate protection of private property, are magical stimuli for modern economic development and technological revolutions.

Statistics, of course, can be used to support the other side of this controversy. If the relation in GDP/capita between the USA and the USSR is calculated for 1950 and for 1990 using Maddison Historical Dataset, we obtain in both cases a relation of 3.36. That is, a less developed USSR developed as fast as a three times more developed USA in these 40 years. Less developed countries are supposed to develop faster than developed ones. Well, the USSR did not archived that under communism, but Russia, a successor of the USSR, achieved faster growth than the USA much later under a somewhat more open economy.

Another effect legal frameworks have on economies is their permanence or temporal stability. Jose Luis Cordeiro for example found that countries that have enforced a larger number of constitutions, or have constitutions with a greater number of articles, are among those who achieved less economic growth. This finding is consistent with what we will describe as long term sustainable economic growth in Chapter 10.

The main role of government regarding economics in a society that seeks economic growth is not to produce goods and services or be an economic actor. A government has the role of regulator, of arbiter between the weak and the strong, of custodian over the dynamics between consumer and producer, protector of the common goods, and facilitator of the market dynamics. If we see the economy as a game of sports between producers and consumers, the role of the state is to be the referee. In a sports game that is considered as such, the referee only calls the faults but never kicks the ball.

# 9. EDUCATION

## *Basic Education*

Education is possibly the most influential element on values, and in general, on the culture of the inhabitants of a country. The relationship between the level of education of a country and its ability to generate wealth is very striking. Educational levels can be measured in several ways. Perhaps the most reliable indicator is the percentage of the population that is enrolled in a formal education program. The data from the United Nations in this regard are revealing. As an example, Figures 9.1 shows the number of male students finishing secondary education in 1980 and in 2009 in rich and poor counties.

**Figure 9.1: Advances in educational enrollment**

Percentage of the population with secondary education (males) in 1990 and in 2009 (vertical axis) in countries with different GDP/capita as assessed in 2010. Data from the World Bank.

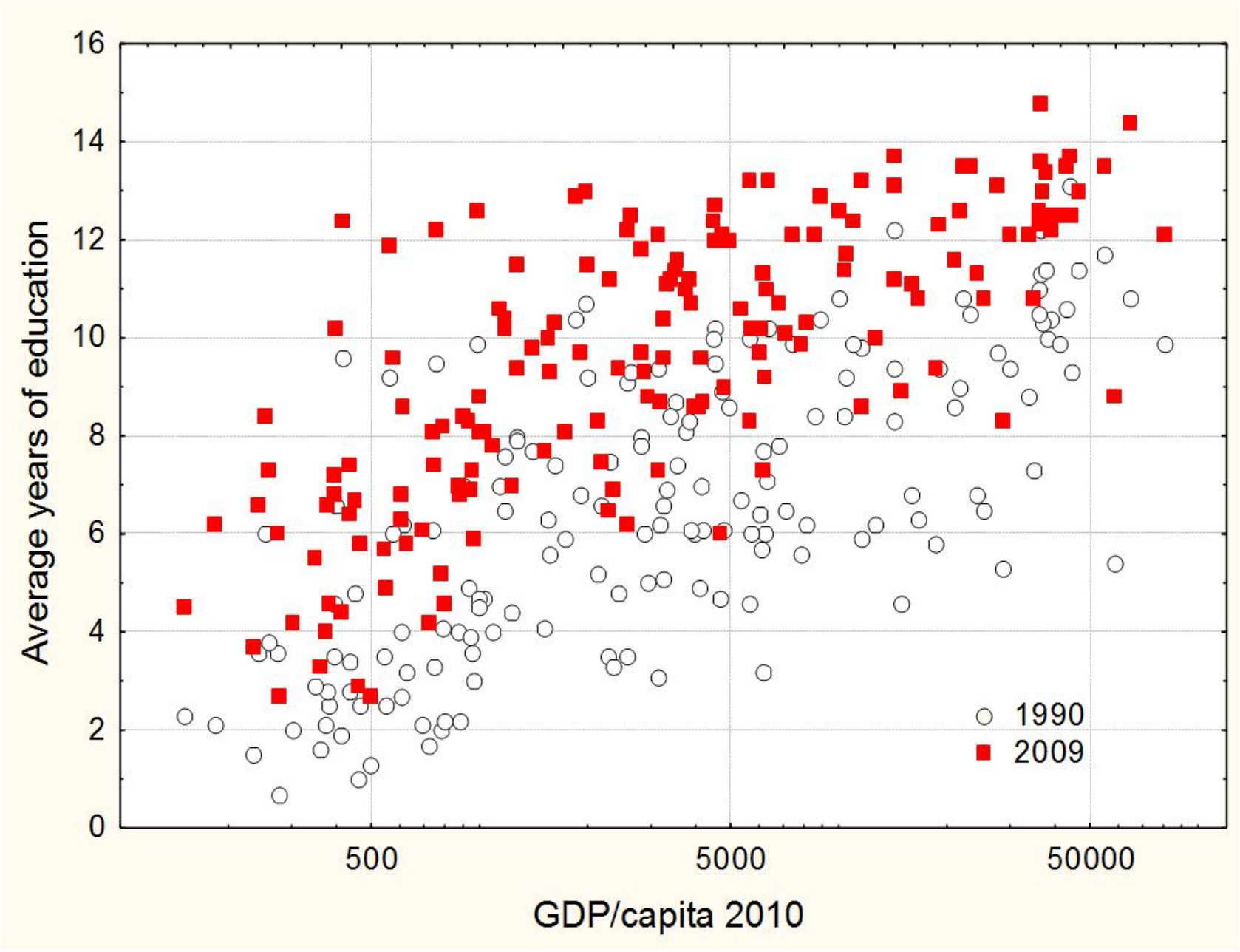

Undoubtedly, all counties, rich and poor, improved the level of education of their populations in the last decades, although richer countries improved more. The data presented in Figure 9.1 is for males. A similar analysis for females leads to the same overall conclusions. The scatter among countries, however, is wider, and the absolute numbers of female students finishing school in many counties is lower than for males.

Rich countries that are members of the OECD have most of their population schooled and continue to increase their level of formal education of their population significantly. Countries showing high economic growth, as in East Asia, have levels of schooling and of education lower than rich countries, but much higher than Latin American and Caribbean countries. This last group of countries has low economic growth, and has populations with low levels of education. Their educational levels increased during the last decades, but much slower than that in the other two groups of countries mentioned.

How can we achieve higher and better levels of schooling? At first glance, it appears that investment in education should be a perfect indicator of levels of development. However, this is not so. In Figure 9.2, we see the relationship in each of the countries of the world where data is available (data points on the graph) between the expenditure made by a country to educate its citizens (vertical axis) and its level of development (horizontal axis).

**Figure 9.2: Does spending improve education?**

Relationship between spending on education as % of GDP allocated to educational activities (vertical axis) and the wealth each country produces (Real GDP per capita in US $ on the horizontal axis). Data from the World Bank.

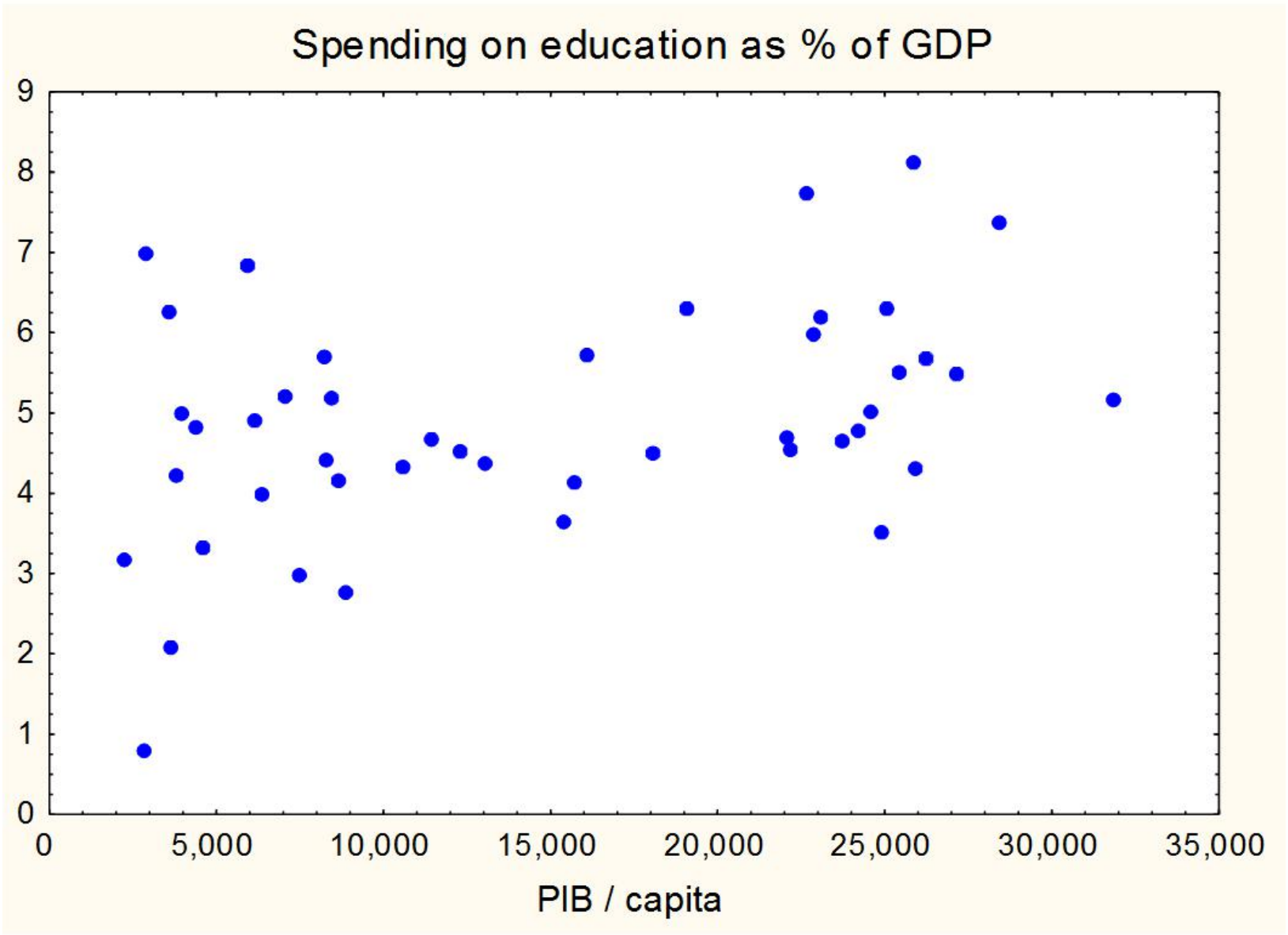

This figure shows no convincing correlation between spending in education and the level of development. Other ways of assessing spending in education reveal similar trends. This result suggest that perhaps resources assigned to education are often misspent or that different countries count spending on education in different ways.

A better correlation between the degree of development of a country and an indicator related to education is obtained when we compare the coverage in education with the GDP per capita (Figure 9.3). Clearly, in countries with a high GDP/capita, 90% of the population has over 10 years of schooling. In contrast, populations in countries with a low GDP/capita have on average less than 8 years of schooling.

**Figure 9.3: The length of schooling affects the wealth of the nation**

Relationship between the years of formal education that the 90% more educated population has on average (vertical axis) and the wealth the country produces (Real GDP in US $ per capita on the horizontal axis). Data from the World Bank.

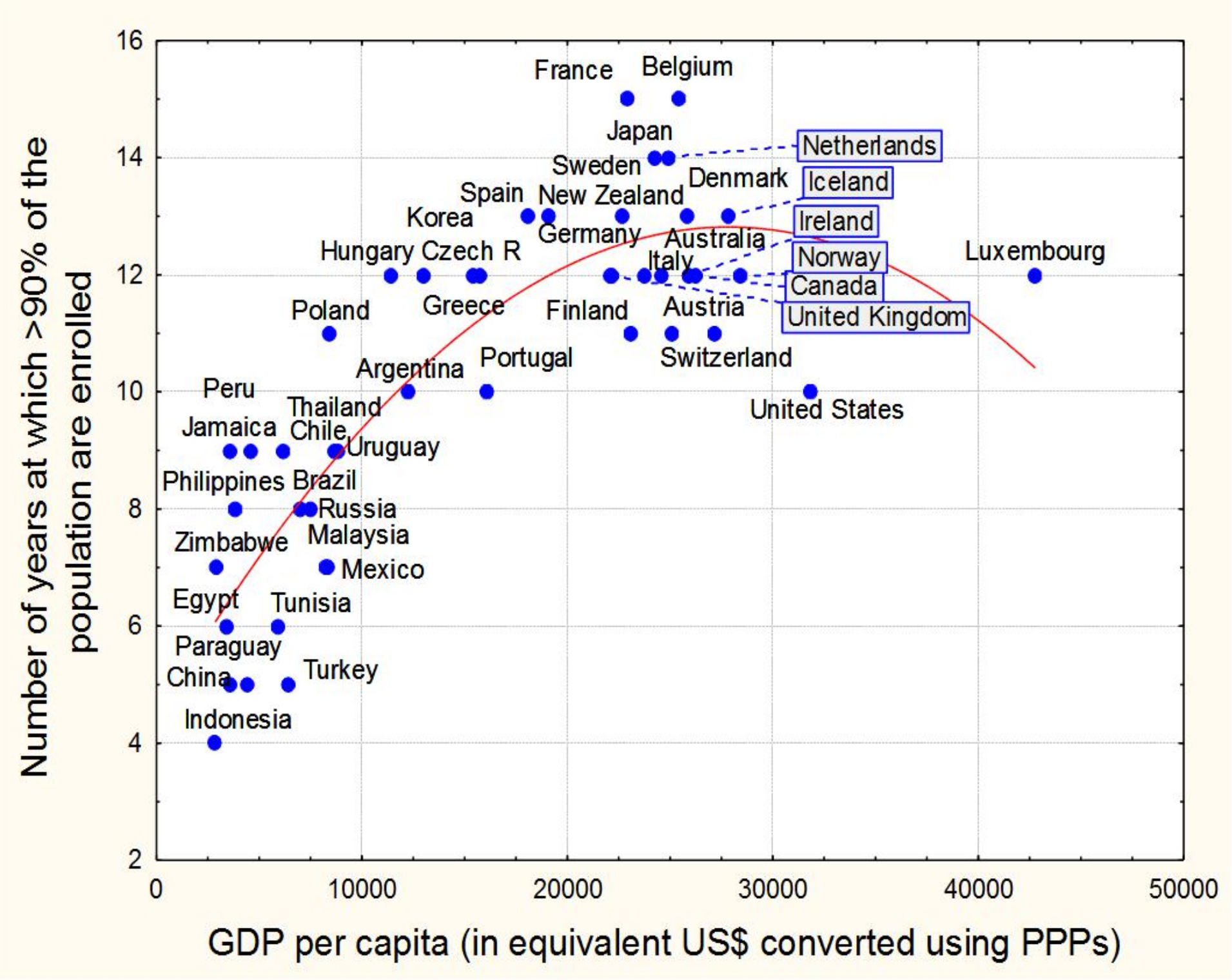

Not all education has the same effect on the wealth of a nation. The quality, often affected by attempts to use education to influence politically the citizenry, is important in establishing the synergistic relationship education – wealth generation. There are nations like Cuba and North Korea, for example, which provide the vast majority of its population with formal education, and yet fail to assemble an economic system that generates wealth for the nation. High quality education is required to power an economy that produces wealth.

One of the most efficient investments that can be made is in primary education. The OECD tests children at different levels of education in various countries using the same methodology (PISA tests), so as to render the results comparable between countries. In Figure 8.4 the results of the PISA test for mathematics from 15 year old students are compared with regards to the investments made in primary education. The 15 year old students are supposed to have mastered primary education a few years ago. The results are very convincing. The data converge on a non-linear regression line showing a positive correlation between investment in primary education and fewer failures in 15 year olds in tests for skills in mathematics.

**Figure 9.4: Verbal and mathematical abilities of students from Latin America and the wealth of the nation**

Relationship between PISA test scores on mathematical abilities (vertical axis) of 15 years olds in OECD countries, compared to the investment made in primary education as % of GDP the year before (horizontal axis). Data from OECD and the World Bank.

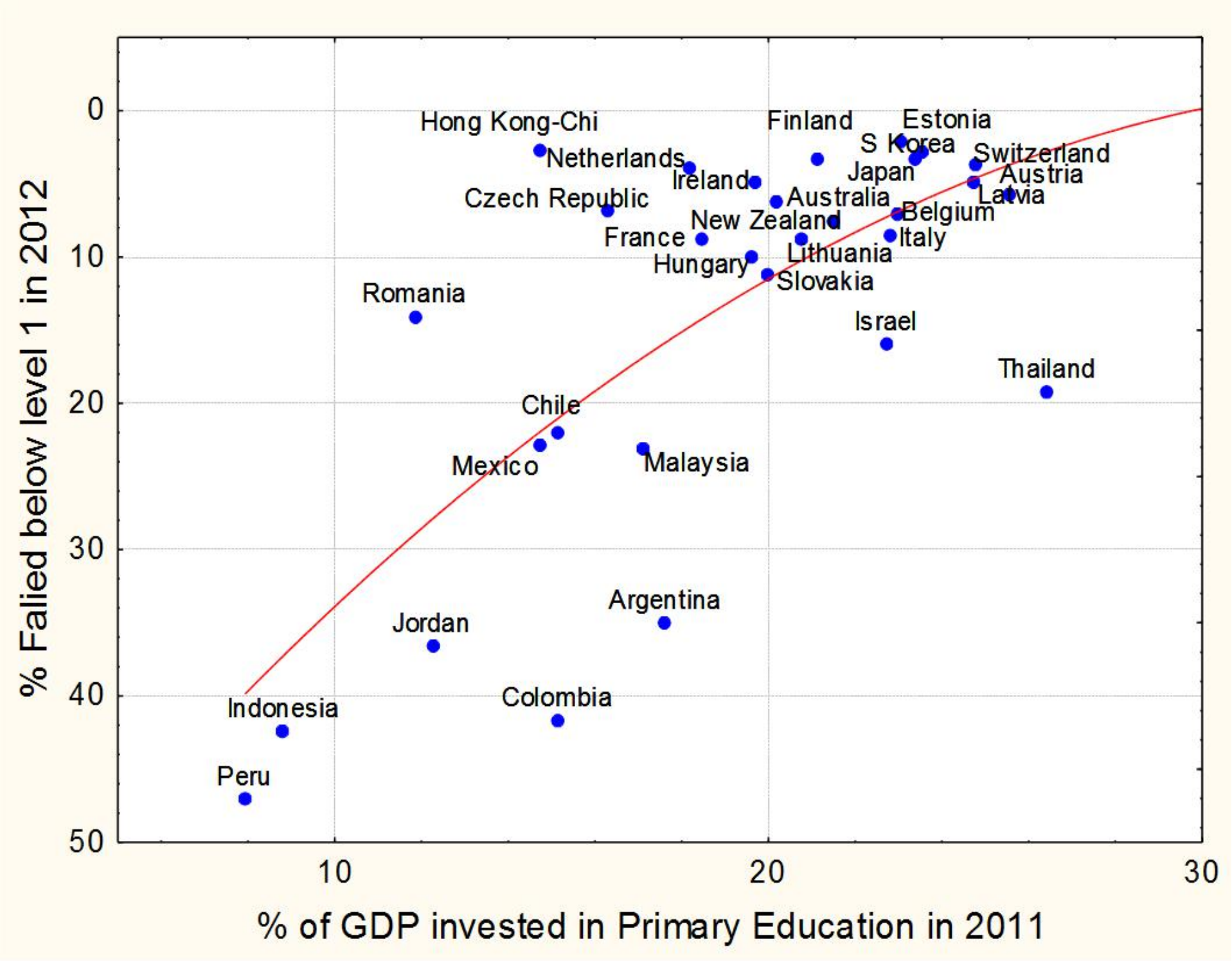

When we applied a meta-analysis with several hundreds of indicators and indexes that measure different aspects of education, we found that the indexes that are more correlated to the welfare of the inhabitants of a country are related to science and technology (see Figure 7.4). This result suggests a strong correlation between the ability of a country to produce science, i. e., its scientific development, and its level of wealth. This correlation also exists with other variables, but to a lesser degree. Using a Spearman correlation analysis, we measured the strength of the correlation between the scientific production of a country and the average wealth of its inhabitants. For the number of publications in the natural sciences, it was 0.93 (1 is the maximum and 0 the minimum), while the correlation between the productivity of the social sciences and the average wealth of the country was only 0.61. Another measure of creativity, the number of feature films produced, achieved lower levels of correlation than science when analyzing its co-dependence on the wealth of the country (correlation coefficient of 0.73). Although statistical correlations cannot be used as evidence of any cause or effect, they show that the conditions that favor the wealth of a country would appear to be more related to the natural rather than the social sciences. It would be extremely interesting to find out the real causes of these statistical correlations.

## *Education and Science*

Science and technology are closely related. For example, Ricardo Hausmann argued that the knowledge embedded in technological skills is a prime driver of economic growth. We have shown that basic science is an even better predictor of economic growth (see article in Plos One mentioned in Chapter 7). Certainly, both affect the potential and actual economic growth of a society. Both, technological skills and scientific development can be quantified. Estimates of the degree of scientific development of country can be made by calculating the number of scientific paper per capita in that country as reported by the Scopus scientific database (SPS). The technical skills can be quantified by estimating the economic complexity (EC) of a country by counting how many different and difficult production processes are hosted in that country. In Figure 9.5, the relationship between the gross national product per capita (GDP/capita) of different countries, the number of scientific papers per capita published by that nation in international journals (SPSc), and the economic complexity index (ECI) of the country are presented.

**Figure 9.5: Science and wealth**

Relationship between the Economic Complexity Index (ECI on the horizontal axis), Scientific Productivity per capita (SPSc) in 1998 in logarithmic scale (vertical axis), and economic wealth, expressed as GDP per capita for 2008 reflected in the size of the data points. Data from SCImago-Scopus and the World Bank.

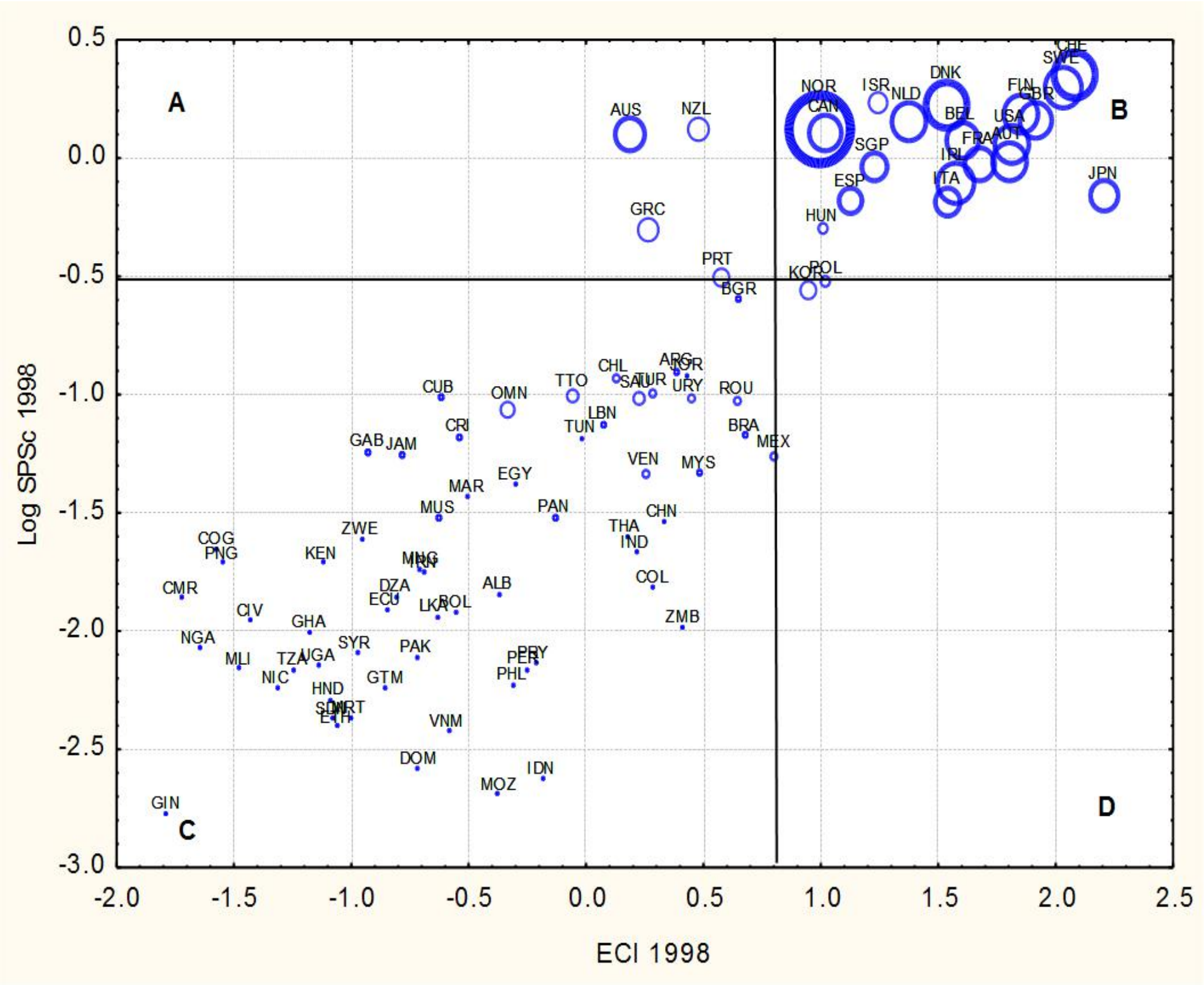

Conspicuously, Figure 9.5 shows that rich countries produce much more scientific papers and have a more complex economy than poor ones. Rich countries fall in the upper-right quadrant of the graph (quadrant B), with some exceptions, such as Australia and New Zealand, which have high scientific productivity, are rich, but have economies with intermediate economic complexity (quadrant A).

If we focus however on economic growth, rather than on the acquired wealth, a different picture emerges. In Figure 9.6, the relationship between economic growth during the decade from 1998-2008 of different countries is compared with the number of scientific papers per capita published by that nation in international journals (SPSc), and the economic complexity index (ECI) of the country.

**Figure 9.6: Science and economic growth**

Relationship between ECI (horizontal axis), SPCc (vertical axis) and economic growth expressed as the % change in GDP/capita for 1998-2008. Data points with negative values are revalued to 0. The size of the data point is proportional to % difference GDP/capita for 1998-2008. Data from the World Bank and SCImago-Scopus.

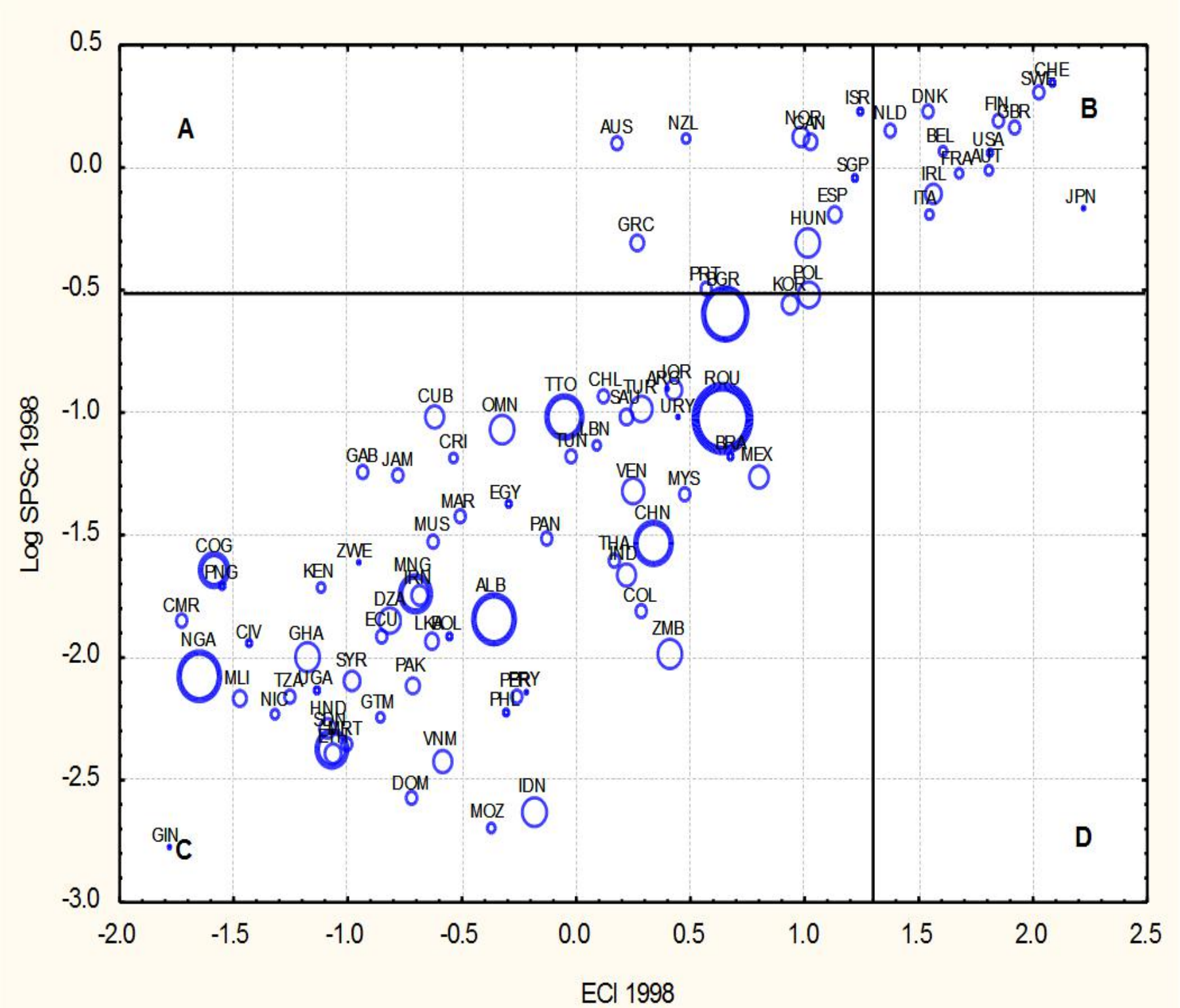

Now, the countries with the most sophisticated economies and the highest scientific development were not those with the highest growth. High growth was reserved for developing countries with intermediate complex economies and about average scientific productivity located in quadrant C in the graph.

Additional analyses revealed that natural sciences are the academic activity that most correlated with economic growth. This correlation does not show causation. It might suggest that the conditions required to produce science are similar to those that produce economic growth, or that the ability to produce science determines the wealth of a country, or alternatively, that countries that have accumulated wealth are able to invest heavily in science. We compare the relative development of the different sciences in a country with their wealth at the time, and with their economic growth during the next 10 years (see Table 9-1). We found that middle income countries that had a relatively higher scientific productivity in basic sciences also had a much higher economic growth during the next 10 years, compared to countries which invested relatively more efforts in applied sciences or in the social sciences and humanities.

**Table 9-1: Basic natural science is better predictors of higher future economic growth**

Comparisson between economic wealth and economic growth and the relative publishing activity of various academic disciplines, estimated using the number of papers reported by Scopus-SCImago for the year 2000 in different academic subject areas. One hundred countries with populations of more than 5 million inhabitants were included in the study. Numbers in bold represent statistically significant Spearman correlation coefficients (rejection level for the null hypothesis of $p < .01$). Publication effort was estimated as the percentage of all publications of the country in a given academic area. Economic wealth was estimated using GDP/capita for the year 2000. Economic growth was estimated as the percentage difference between GPD/capita in 2010 with that in 2000.

| Area | GDPc 2000 vs SPSc 2000 | % GDPc 2010-2000 vs SPSc 2000 |
|---|---|---|
| **Chemistry** | **0.35** | **0.47** |
| **Earth-Planet Sci** | -0.11 | **0.40** |
| **Mathematics** | **0.38** | **0.39** |
| **Physics-Astronomy** | **0.50** | **0.33** |
| **Material Sci** | **0.44** | **0.33** |
| **Engineering** | **0.48** | 0.22 |
| **Chemical Engine** | **0.54** | 0.21 |
| **Dentist** | **0.27** | 0.07 |
| **Art & Humanities** | -0.01 | -0.01 |
| **Veterinary** | ***-0.29*** | -0.05 |
| **Business-Manag-Accounting** | **0.27** | -0.05 |
| **Immunology & Microbiology** | ***-0.52*** | -0.10 |
| **Multidisciplinary** | 0.02 | -0.12 |
| **Comp Sci** | **0.69** | -0.12 |
| **Health** | **0.35** | -0.13 |
| **Decision Sci** | **0.55** | -0.14 |
| **Agriculture & Biology** | **-0.67** | -0.15 |
| **Nursing** | -0.08 | -0.16 |
| **Environmental Sci** | ***-0.27*** | -0.17 |
| **Social Sci** | ***-0.33*** | -0.20 |
| **Economy -Finance** | 0.13 | -0.22 |
| **Psychology** | **0.28** | -0.22 |
| **Pharmacy** | **0.23** | -0.24 |
| **Biochem-Genet-Molecular Biology** | **0.59** | -0.25 |
| **Neuroscience** | **0.72** | ***-0.33*** |
| **Medicine** | -0.22 | ***-0.40*** |

The results are unexpected in the light of the prevailing mainstream economic dogma. Jeffry Sachs, for example, recommended health, energy, agriculture, climate and ecology as the areas of science where investments were most likely to promote economic growth. None of them came out as positively correlated as the example shown in the figure. On the contrary, countries that knowingly or unknowingly complied with Sachs's recommendations achieved very poor economic growth. It is investment in hard sciences and basic sciences, such as physics and chemistry that correlate strongest with economic growth. This result strongly suggests a role of rational thinking as cultivated in natural science in promoting economic development.

Cultural secular values and tolerant moral attitudes favor the expansion of sciences and sustained economic development. The 2002 Pew Global survey reveals the countries whose citizens show greater religious and moral tolerance. These same countries are the ones that show greater scientific and economic productivity. Economic growth and scientific and creative productivity are generally favored by liberal and tolerant attitudes. This is possible in an open society where human creativity flourishes unhindered and where society can implement novel mechanisms of economic productivity that will support even more science and creativity.

# 10. SOCIAL DYNAMICS

Human beings with the same genetic background, equivalent history, inhabiting similar geography and having a comparable culture, placed in different social and educational networks, can produce different economies. The issues discussed so far, although do explain part of the process of the creation of wealth of nations, are not sufficient to complete our view of the phenomenon. The way individuals interact and the mechanisms available to carry out these relationships are factors that undoubtedly affect the economic behavior of any society. The functioning of the social dynamics and the elements that determine the efficiency of the network that links the different agents of a society at local, national or global level are important aspects to understand the problem to which we are committed here. How do the type of social structures and the way individuals in a society relate to one another affect the ability to produce wealth? We know little of this dynamic and of the underlying mechanisms, but the little we know reveals its enormous importance in the matter that concerns us.

## *Social Capital*

One of the mysteries that have astounded economists is the fact that due to similar historical, geographical, economic and political conditions, societies often differ markedly in their ability to produce and accumulate wealth. On the other hand, economic systems that have been successful in one country, after being implemented in another, do not achieve the expected goals. This indicates that there are factors that we often associate with culture, beyond the historical, geographical, economic and political, that determine the economic behavior of a society. Some of these factors are identified as causes that influence the ability to acquire and accumulate wealth in our societies. One of them is what has come to be called Social Capital, parts of which we have discussed previously.

We refer here to Social Capital as the collective value of networks of personal relationships that reflects the collection of attitudes, traditions and customs that facilitate commercial transactions, labor relations and investment of economic capital. Trust, as mentioned before, is an element that enhances the economic capital because it lowers transaction costs and makes economic investments more efficient. Societies with little interpersonal trust limit their potential to do business with relatives, while in societies with high levels of trust in the community, generally supported by laws respected by all, individuals have a greater range of possibilities to start and run a business.

Within this category of entities that favor Social Capital, we can include all those cultural traits that promote the creation and accumulation of wealth. However, the creation and accumulation of wealth can arise, at times, in several different ways, and the behaviors, attitudes and traditions that favor it in an environment not necessarily do so in another, and may even prevent it in a different setting. This is the case of the habit of accumulating goods or resources. In a society that lives in a temperate climate with distinct seasons, it is generally advantageous to accumulate resources to survive the winter. This same accumulation of goods in a tropical society can be very harmful. The goods, if they are perishable, rot, and attract pests and diseases. Therefore, their storage wastes effort better employed in another activity. This climatic constraint on the behavior of capital accumulation is known to affect not only humans as explained in Chapter 3.

The attitudes, values and social traditions may also represent a kind of negative social capital. That is, many values of a society prevent the accumulation of wealth and progress. This applies to several traditional societies, with a strong influence of the Catholic Church or the Muslim tradition, which consider financial gains as a kind of usury. Considering the interest charged is usury has an implicit assessment that capital investment must not produce more wealth than the one invested. It is, in other words, to deny the possibility that synergistic forces combine to produce wealth. It is no wonder that societies that consider profits as usury and, as is the case of some of them, even prohibit charging interest on capital, have difficulty accumulating wealth and producing sustained economic growth. Witty financiers have found ways around these taboos in recent years developing what is called Islamic Banking.

In the case of usury, there is no doubt that high interest rates deter investment and development. But exorbitant interest rates only emerge when there are monopolies or arbitrary regulations of the States. The flow of money, in a free and rationally regulated society, follows the laws of diffusion. That is to say, it flows to the places where better interests are paid, there where capital can generate more wealth. An analogy is the regulation and distribution of nutrients and energy-carrying molecules (ATP) in living organism that follow the laws of diffusion. A higher consumption of energy in a given tissue or organ creates a reduction of the concentration of the molecules transporting energy and therefore, thanks to the diffusion-driven forces, a greater flow of energy to the tissue or organ that requires it. This is the way in which a free capital market should operate, where the gradient of flow of resources is modulated by differential rates of interest.

The role of a regulator (government) in a market of goods and services, according to this organic view, is to prevent the consolidation of commercial monopolies and to keep costs of transactions and the costs of entry to the various businesses low. Social attitudes that promote these goals and the operation of these regulatory mechanisms, such as transparency, trust, simplicity, the rejection of corruption, and promotion of entrepreneurship, can then be considered as a capital, which acts analogously to investment of resources.

## *Sustained Economic Growth*

*"Any man can make mistakes, but only an idiot persists in his error."*

Marcus Tullius Cicero (106-43 BC)

How a society and its individuals react to problems and how it reacts to catastrophes is an important element in economic growth. We can guess some of the details of the underlying psychosocial mechanisms to these reactions, but we know its effects on specific economic phenomena much better. Figure 10.1 illustrate the point.

**Figure 10.1: Sustained growth versus prolonged crises**

Changes in GDP per capita in 1990 International Geary-Khamis $ per capita (vertical axis) during 70 years (horizontal axis) in Bolivia and in the USA. Arrows indicate the onset of a recession. Data from Maddison Historical Database.

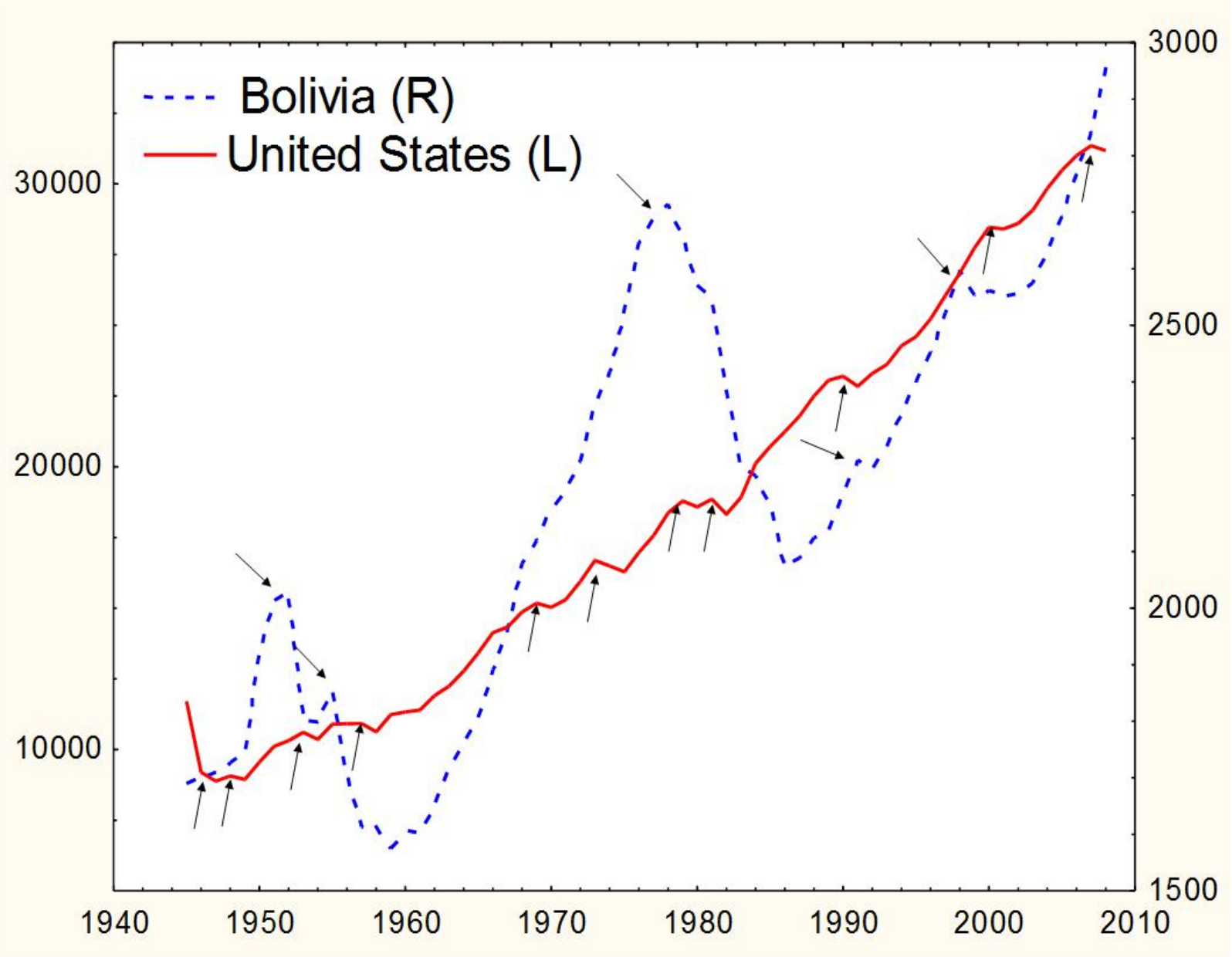

The figure shows two curves, which if analyzed in detail show that they have a given number onset of recessions or years were GDP started to decreased. Bolivia suffered 5 of these onsets of recessions or inflection points where economic growth became negative from one year to another during this period. In contrast the USA achieved sustained economic growth despite suffering 11 onsets of recessions over time. Bolivia has made significant economic growth only for short periods of time and endured long periods of economic recession. The USA achieved growth shortly after the onset of a recession, whereas Bolivia had difficulties in managing economic recession so as to re-initiate growth.

This ability to achieve sustained economic growth by handling economic recession efficiently so as to re-initiate growth, is not given by a greater ability to avoid economic crises or avoid errors in the handling of the economy. The number of years in which the economy of the USA started recessions was nearly twice that of Bolivia during this period. The difference is that the economy of the USA, once a recession has started, recovers in a very short time and restarts its growth. In contrast, the economy of Bolivia suffered prolonged recessions before they could restart growth. That is, US society reacted more quickly and coherently to crises and corrected its economic policies much faster and more radically than that of Bolivia. These examples suggest that the wealth of a nation is influenced by its learning capacity. It is not easy to avoid errors, but it is possible to correct them. This flexibility and dynamism in managing the economy ensures sustained economic growth over time, allowing the accumulation of wealth and hence a high standard of living of its inhabitants.

**Figure 10.2: Effect of long-term economic growth**

Economic growth of 10 countries over the last 100 years. Real GDP in 1990 International Geary-Khamis $ per capita (vertical axis) vs. time (horizontal axis). Data from Maddison Historical Database.

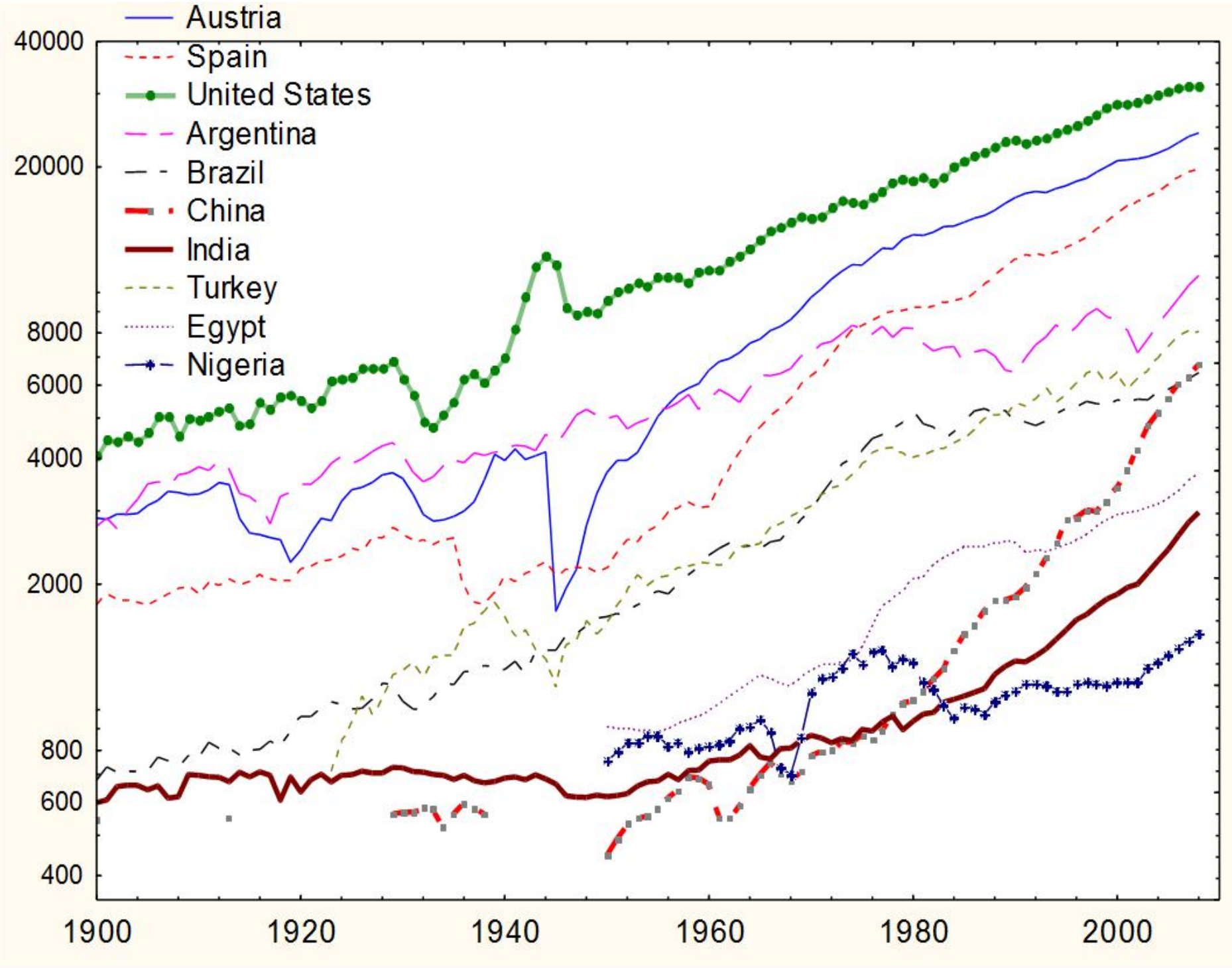

Many so-called "economic miracles" are nothing more than a steady growth over long periods of time. This is the case of Spain and Austria after 1960, or China after 1980. These countries achieved a prolonged sustained growth for several decades, distancing themselves from the poorest countries and moving closer to the rich nations of the world at the end of last century or beginning of this one. In contrast, politically unstable countries, such as Nigeria, Argentina and Egypt show an unstable economic development. This is reflected in Figure 10.2, which shows how even small changes in the rates of sustained growth, maintained over long periods of time produce large differences between the wealth of the nations, as in the case of the USA.

## *Wars*

Another version of the argument of the importance of continuity in economic growth was developed by economists at the World Bank, the United Nations and the Copenhagen Consensus 2004, among others. Of all the factors analyzed that cause poverty in the world today, the most important proved to be armed conflicts, civil wars and international war. War and epidemics had the greatest negative economic impact among the cases analyzed. War disintegrates social institutions or limits their functioning and creates the environment for the production of epidemics, and both promote misery and poverty. It is not easy to avoid wars, both civil and international. Civil wars aiming for the independence of a part of a country can be avoided with a good dose of democracy, as in Scotland, although a success of this recipe is not guaranteed, as shown by the cases of Northern Ireland and the Basque Country. Especially Africa and the Middle East are suffering the ravages of war nowadays. The war not only destroys already accumulated wealth, but also ends lives, eradicates factories and aborts the chances of future economic production. Peace and prosperity are associated very strongly. This has been expressed in different forms by mythologies worldwide and recognized by almost all modern human religions.

Achieving lasting peace is not easy. Many have been the failures but also many the successes of attempts to prevent war and build peace in recent decades. The example of the European Union deserves special attention. Europe has been the protagonist of wars and violence since humanity has memory. European wars have been particularly vicious and deadly in the last century. European countries acted assertively, even at the cost of some of their sovereignty, when they joined the European Union, ensuring peace and prosperity in the long term. Many sub-regional integration efforts try to emulating the European Union. We wish them success, as they might assure long-term prosperity.

## *Imperialism and International Conspiracies*

We often hear in Third World auditoriums the comment that the cause of poverty in developing countries is the imperialism of the rich capitalist countries that ruthlessly exploit poor countries. Too often, a leader has been heard foisting the failure of their economic policies on the intransigence of the global economic power of the time, or a plot of international capitalism and transnational oligarchies.

Undoubtedly, in terms of nations and societies, the strongest takes advantage of the weaknesses of others. European colonialism impoverished many nations and eliminated entire societies. The Indian Prime Minister in 2005, Manmohan Singh, estimated that in 1700, before the arrival of the British, India perceived approximately 22.6% of the revenues of the world; at the time of its independence, the British managed to place India in a position of perceiving only 3.8% of these global revenues. Imperialism neither is nor was the monopoly of Europeans, as attested by archaeological remains and the history of cultures such as the Aztecs, Ethiopians, Mayans, Incas and Mongols, among many others.

Today's wealthiest nations, with their protectionist and subsidy policies, significantly slow down the export potential of agricultural products and thus the potential growth of developing countries. Combating injustices in international relations is necessary and requires constant and coordinated efforts of the countries. However, no rational policy bases the success of the economy of a country, or conditions its future, on the goodwill of others. History teaches us that freedom and wealth are not achieved by the generosity of the powerful, but by the decisive and sustained effort of each country. Blaming the cause of its own failure to external forces is a pathological reaction rather than an economic rationality. Poverty produces feelings of inferiority that cause resentment and insecurity and trigger innate behaviors in search of an aggressor, the agent causing the evil. In psychology, the projection of the problems on an *external locus* is used. That is, we seek to blame others for our problems to unburden our conscience. It is used to get rid of feelings of guilt, and assigns the causal reason for the ailment to an inaccessible external entity. Unfortunately, pathological reactions have not been good in designing successful economic policies.

We know of examples that demystify imperialism as the sole cause of underdevelopment and poverty. The losing countries in the Second World War were conquered and subjugated by the winning powers, especially the USA, the greatest military and economic power nowadays. A few decades after the war ended, two of the richest nations in the world economic ranking were Japan and Germany, the two nations subjugated by the USA. Victorious powers of the Second World War such as Russia and Britain occupied places in the ranking of wealth far below the losers Japan and Germany. Certainly, in this case, American imperialism did not prevent these nations from becoming wealthy.

Pseudo-scientific explanations of the causes of poverty abound. Great Capital's evil plans, conspiracies of intelligence institutions of large nations and/or strategic military plots are sometimes named as factors interested in keeping poor many nations and causing poverty in developing countries. These hypotheses lack economic rationality in a modern society. When collecting resources and exploiting land for agriculture are the main sources of wealth, wars and territorial conquest may be beneficial to a conquering civilization. But in modern economies based on technology and trade this calculus does not hold (see next chapter). In any business interaction, the richer the client is, the richer the merchant may become. It is not possible now to create sustainable wealth based on the poverty of other nations. Therefore, any ultra-rational plot to enrich a group of individuals or nations will only be successful if it encourages the wealth of the allegedly exploited.

It is the inability of politicians in charge of designing and implementing plans and economic measures that are responsible for the poverty of developing nations. The search for scapegoats across national borders only seeks to hide this fact. The economic irrationality of many politicians and a large part of the population of developing countries shows the lack of education on economic issues and an absence of basic knowledge of economics. Ignoring basic concepts like the relationship between wealth, work and savings, or accepting irrational populist promises, such as expanding public spending by reducing taxes, will only lead to delaying industrial and economic growth. The lack of economic education in both the population and the elites responsible for managing the economy of a country constitutes another cause of poverty.

When things go wrong, it is very comfortable to assign blame to evil men or evil forces beyond our control. By doing this, we believe that the world is a good and peaceful place, but for some nasty and selfish individuals. Populism, the easy way, ignorance, underdevelopment and resentment are elements that favor an economic vision in which the poor are victims of the unbridled greed of the rich. This attitude, although it may have elements of truth, is not constructive. Freedom and well-being is the achievement of the individual or the nation, not a gift from the powerful. Only with our action, will we succeed in setting ourselves on the road to progress. Let us remember the wise words of and Asians intellectual giant: “The superior man seeks in himself all he wants; the inferior man seeks it in others” Confucius (551-479 AD)

## *Human Behavior and Economics*

The sciences that try to bridge the gap between microeconomics and the macroeconomics are several. We can mention behavioral economics, economic psychology, game theory, ecological economics, bio-economics, evolutionary economy, the study of artificial societies, and computer simulation of societies. All these approaches are interdisciplinary and draw their paradigms and foundations from more than one discipline. Developments such as evolutionary stable states or Nash equilibria, chaotic dynamics describing attractors and phase changes, game theory and adaptive evolutionary dynamics, were originally conceived by researchers of living systems and theoretical biologists (examples include John Maynard Smith 1920-2004, William Hamilton 1936-2000, J.B.S. Haldane 1892-1964, R.A. Fisher 1890-1962 and John Louis von Neumann 1903-1957). These theories have been used to explain important animal behaviors related to competition for territory, food or reproduction, being able to predict behaviors observed in nature with amazing accuracy. They have helped us understand the asymmetries in parental investment; conflicts among genders; the adaptive value of sex and optimal strategies for mate selection; and the emergence and functioning of animal societies. The application of this knowledge to the study of economics has opened the door to evolutionary economics, behavioral economics, experimental economics and bio-economics.

Economists have reaped the fruits of this current. Many have won the Nobel Prizes working in these new sciences. The laureates Herbert Simon (1916-2001), Thomas C. Schelling (1921-), Vernon L. Smith (1927-), John F. Nash (1928-), Gary S. Becker (1930-), Robert Aumann (1930-) and Daniel Kahneman (1934-) deserve special mention as pioneers of this heterodox activity in economics. Becker resorted to psychology by studying individual preferences and their impact on the economy. Simon studied how humans make decisions and how it affects the economy. The general conclusion from these studies is that humans, in economic terms, do not always act rationally. This is because our economic rationality is limited. Our perception of risks is distorted and our mental algorithms to establish priorities have not been optimized through biological evolution for economic problems. Such recognition of the limitations of our innate economic rationality is fundamental for the construction of better economic theories. Classical economic theory assumes a rational behavior of the average human to simplify their analysis. Surprisingly, these simple assumptions produced quite acceptable results. However, by studying progressively more complex problems and with increasingly more detail, these simplifications are no longer justifiable. We require new economic theories that can incorporate the human individual with the way of reasoning he actually expresses. An illustrative example of this upgrading of the economic theory was suggested by Herbert Simon. He proposed that individuals, rather than optimizing and maximizing economic variables, find satisfactory solutions with minimal effort. He called it “sufficing”, a concept that describes the behavior of choosing the first acceptable solution at hand, optimizing the use of time and increasing the speed of decision-making. Including this new vision of human economic rationality significantly affects the analysis of economic processes.

The 2002 Nobel Prize winners in economics were Vernon L. Smith and the sociologist Daniel Kahneman. The contribution of Vernon Smith is having established laboratory experiments as tools in the empirical economic analysis, especially for the study of alternative market mechanisms. The psychologist-economist Daniel Kahneman, on his part, succeeded in integrating psychological research into economic science, especially regarding the reasoning that guides decision-making under uncertainty. Kahneman's work shows how human judgment takes shortcuts that systematically depart from the principles on which the calculations of probabilities are based. These contributions, and many others, have expanded decisively our potential to study the effects of the activity of the individual on the micro-economic variables, which in turn influences the macro-economic dynamics.

# 11. COMPLEX SYSTEMS ECONOMICS

## *From Micro to Macro*

Complex phenomena are the results of events that occur at lower levels of organization. For example, the spatial structure of a colony of flamingos, or a herd of cattle, or of a human crowd, emerges from the interaction of its members, which in turn is limited by the properties of their anatomy and physiology. In this sense, macroeconomic indicators are aggregates, adding the contributions of each of the individuals or agents that make or participate in an economy. The individual contribution constitutes the basis of the economic phenomena studied. But social interactions by themselves affect complex systems. It is time to pay attention to the underlying phenomena in macro-economic based on micro-economic.

Scientific research can focus on humans for example. Nevertheless, each research differs in the spatial scale at which the subject is observed. Research might highlight genetics, anatomy, physiology, psychology, behavior, sociology, ecology, history, or other aspects. Each scale of spatial and temporal observation opens up a new world of relationships, laws and phenomena, and yet, all possible perspectives present us the same entity. Through the different scales, we can detect the emergence of particular phenomena that are not present in the previous levels. These are called emergent phenomena. When we detect an emergent phenomenon, it is worth stopping the journey and deepening the analysis of the mechanisms that produce such phenomena. This cognitive activity has been called the study of emergence, of self-organization or of the dynamics of complex systems.

In economic sciences, the scales of analysis converge on two different sub-disciplines: Microeconomics and Macroeconomics. These two worlds have endured their own developments for a long time and have created their analytical tools independently. At the interface of these two sciences, we will continue the search for the causes of poverty and for the mechanisms of collective generation of wealth.

## *Game Theory and Econophysics*

Another important approach for analyzing the dynamics of wealth gets its tools from physics and mathematics. A pioneer of this approach was the German Gottfried Achenwall (1719-1772). He introduced the concept of statistics as the mathematical treatment of issues concerning the State. The Italian sociologist Vilfredo Pareto (1848-1923) developed this modern form of analysis by studying the distribution of income among the citizens of a nation, finding that the distribution is not Gaussian (i.e., governed by completely random phenomena) but follows a law of power. This difference may seem a simple mathematical triviality, if it were not because it indicates the presence of an extremely relevant phenomenon for our subject of study (see Chapter 6, for example). If the components of a system are acting independently or interacting randomly, they produce distributions in their properties that we recognize as normal or Gaussian. This distribution disappears if there are strong interactions between the parties. Pareto, however, did not possess this knowledge. The investigative work of the sociologist George Kingsley Zipf (1902-1950) and of many other researchers from various disciplines was necessary to enlighten us about how small variations in the characteristics of the interaction between individuals affect the aggregate variables of society.

The properties of the particles and their effect on their interactions have been studied by physicists for centuries and they have therefore been able to develop appropriate tools for their study. For example, the properties of matter and of its phases of gas, liquid and solid are studied by statistical mechanics with amazing success with these tools, which tempts us to try to apply them to problems of sociology and economics. This is how econophysics was born.

Mathematics has also had important contributions for economics, especially in statistics, probability theory and game theory, and this has helped to advance the capacity of quantitative analysis of sociology and economics. An example of these is the game, developed in 1950 by Merrill Flood and Melvin Dresher to explore strategic alternatives in the so-called *Cold War*. The game, called *The Prisoner's Dilemma*, consists of two players (prisoner buddies), which have two alternatives of action each. Either they denounce the other (they do not cooperate) or cooperate with the other prisoner and do not talk. If both players cooperate, none will be punished and both benefit from this action. If one of them does not cooperate and denounces his buddy, he will be harmed and the non-cooperator will benefit. If the two do not cooperate, no one benefits but the punishment may be less than in the case of a cooperator is denounced. This asymmetry of benefits resulting from the action of each player (see Table 11.1), is analogous to many real-life situations. The solution to the problem without using better communication between the actors is not simple. The rational recommendation, from the point of view of game theory, is that none of the players cooperates, since then they minimize their risks and their losses, although the gains are not maximized. To maximize the profits of the duo, both are required to cooperate, but they risk being denounced. An equivalent situation is called the *Tragedy of the commons*, proposed by Garrett Hardin in 1968, where several shepherds compete for the grass of a common savanna. The most beneficial action for the individual in the short term is not the one that benefits the whole community.

**Table 11.1: Benefits for Agent A in the Prisoner's Dilemma game**

(The table for Agent B is symmetrical to the A)

| Agent A is > | **Cooperative** | **Uncooperative** |
|---|---|---|
| Agent B is ∨ | | |
| **Cooperative** | ++ | +++ |
| **Uncooperative** | - | + |

In late 1970, Robert Axelrod at the University of Michigan promoted an international effort to find the best mathematical solution to the prisoner's dilemma and selected a strategy called *Tit For Tat* as the most successful for playing the game repeatedly. This strategy involves cooperating when the other cooperates and not cooperating when the other does not. Years of research in games and strategies to play them have led Axelrod to propose four characteristics that promote not only cooperation among players, but also the maximum accumulation of wealth among them. These characteristics of successful long-term strategies are as follows: 1: Be kind (start any new interaction with cooperation), 2: Reciprocate (play *Tit For Tat*), 3: Be restrained (not trying to be smarter than the other) and 4: Do not be envious (no matter if the other earns more than I do, as long as I win). It appears that these recommendations to win computer games of the *Prisoner's Dilemma* also seem to be applicable to situations of social and economic cooperation in real situations. Several examples of modern experimental economics suggest this. The crux of this game is the way rewards and costs are distributed. Cooperation in games with reward distributions were all players benefit is very stable. Biological evolution, of course, discovered this trick and many symbioses, animal societies and social interactions are based on mutualism and/or synergistic exchanges.

Other studies by econo-physicists reveal several underlying ordered structures in the data and economic and financial phenomena, such as attractors, repulsors and areas of catastrophic discontinuity, which help us understand these complex phenomena, as they are known from statistical mechanics and the physics of linear phenomena. These characteristics detectable by numerical, graphical and statistical methods reveal inherent properties of the systems components. These features are best studied using computer simulations.

## *Sociodynamics*

Adam Smith in his book The Wealth of Nations described the operation of the market as follows: “Every individual necessarily labours to render the annual revenue of the society as great as he can”. He, however, does not have the slightest intention of promoting the public interest or is aware that he is promoting it. He intends only his own gain and is led, as in many other cases, by an invisible hand that makes him promote a cause that does not form part of his intentions. This is not a disadvantage for society. By pursuing his own interest, he frequently promotes that of the society more efficiently than if his interest were the latter. I do not know of much good dispensed by those who strive to represent the common good. It is not from the benevolence of the butcher, the brewer, or the baker, that we can aspire to our dinner, but from their attention to their own interests". This is a beautiful description of phenomena where the interactions at the individual level bring as a consequence dynamics significant only at the social level, without individual activity being conscious of it. It is a fascinating phenomenon but difficult to study using traditional experimental techniques.

The discovery of *the invisible hand of the market* is a major achievement of humankind. It recognizes the absence of centralized social cohesive forces and discovers forces of the market that explain our social dynamics. Often, fundamental advances in sciences are based on discovering the absence of certain entities. For example, by accepting the absence of phlogiston, the mysterious element that scholars during the early XIX century believed was released from burning objects, allowed the discovery of oxygen. Likewise, it was the demonstration of the nonexistence of ether on which electromagnetic waves are dispersed, what allowed the development of quantum mechanics and the relativity theory. The absence of central coordination in smoothly working markets is a fundamental discovery in economics.

The effect of the behavior of the individual on the performance of the social aggregate can be studied using "artificial societies" or computer simulations of social dynamics. Modern computer simulations, among which agent based simulations stand out, allow us to integrate the various aspects discussed in this book in a single model. This model allows us to explore the effects on the economy of changes of the different variables. In these simulations, virtual societies are created constituted by hundreds of thousands of individuals living in the computer memory. These models are used to explore the importance of concepts such as social capital, social investment, public policies, environmental conditions and other variables on the dynamics of the accumulation of wealth of a society or nation.

The power of computer simulation can be illustrated in Figure 11.1 Here, we represent one of the ways we can integrate three different elements in one body, in this case, a sphere. It is only when we have built the sphere that we can study the impact of small variations in one of the elements on the bearing speed, the flotation properties or any of the emergent properties of the object under study. These properties of the object only appear once the object is constructed and are not apparent from the properties of the component parts individually. Similarly, the properties of a society or an economy emerge when the individuals that constitute it are added. The macro-economic variables, for example, are not apparent from the isolated study of the individuals composing the society, but appear only after individuals interact to form a society.

**Figure 11.1: Reconstructing complexity**

An example of the emergence of novel properties. The properties of the sphere cannot be deducted from the properties of its components analyzed in isolation, although are dependent on these. Drawing inspired by sketches from Alida Ribbi for the Author.

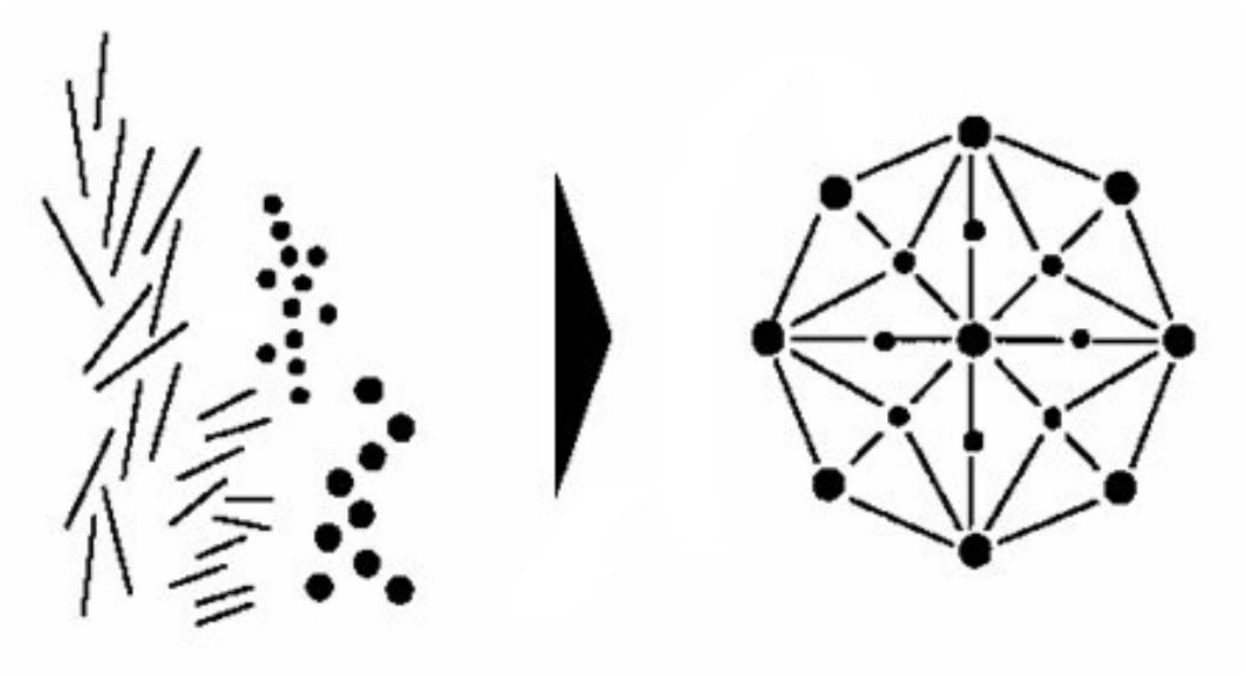

## *Social Simulations and Emergent Properties*

The computer simulation allows us to reconstruct complex systems to understand them better, to determine if we have a full understanding of it. Simulations allow us to determine if with the elements we know to be part of the system, we can reconstruct the phenomena we are interested in studying. Once the reconstruction of an aspect of reality is achieved, the simulation will allow us to explore its characteristics and eventually will serve to make quantitative predictions.

Although this field of knowledge is vast, I will just present a harvest from my own vineyard: the effect of interpersonal relationships in the generation of aggregate wealth of a society.

I have developed a computer model called *Sociodynamica*, which creates virtual societies within the machine to our taste and liking. These imaginary worlds are constituted by individuals or agents, which can be defined or characterized by the experimenter at will by assigning specific properties. For example, we can create societies of identical agents, where everyone is altruistic and generous with others. We can also create societies of exploiters, where each agent will try to take the wealth from others. *Sociodynamic*a also allows creating societies composed of mixtures of different agents or more complex societies, with a structure of division of labor, where some agents are "farmers" and exploit renewable natural resources, others are "miners" and collect non-renewable natural resources, and a third group of agents act as traders, exchanging renewable natural resources for non-renewable resources between farmers and miners, increasing the value of the non-renewable resource in the process.

This artificial society allows exploring the effect certain behaviors and the consequences of ways of relating of individuals have on the aggregate wealth of the society. For example, the simplest interaction that we can simulate is the transfer of resources among individuals. This transference defines a donor of the resources that makes a transfer or “donation” valued with a utility K. This utility will produce in the recipient a benefit A. We can also conceive transactions where the donor agent invests K in the donation, and for business reasons recovers, thanks to this investment a benefit B in the future. Playing or working with these three variables with *Sociodynamica* and exploring the parameter space of the computational model, we obtain a range of types of interactions between agents. These interactions can be classified based on their effect on aggregate wealth that the artificial society achieves. These types of possible interactions are presented in Table 11-2.

| Type of interaction | Balance donor - receptor | Auto-balance donor | Effect on the donor | Effect on the receptor | Effect on the aggregate |
|---|---|---|---|---|---|
| | | | | | |
| dissipative altruism | K > A, A>0 | K > B | - | + | - |
| **synergistic altruism** | **K < A, A>0** | **K > B** | **-** | **+** | **+** |
| dissipative business | K > A, A>0 | K < B | + | + | - |
| **synergistic business** | **K < A, A>0** | **K < B** | **+** | **+** | **+** |
| altruistic punishment | K>0, A<0 | K > B | - | - | - |
| prophylactic punishment | K>0, A<0 | K < B | + | - | - |
| exploitation | K<0, A<0 | K < B | + | - | - |

K = cost to the donor, A = benefit to the recipient, B = benefit to the donor. Adapted from the article by Jaffe 2002 in the Journal of Artificial Societies and Social Simulations (JASSS).

Table 11.2 summarizes the range of possibilities of economic interactions that can occur in a society and its long-term effect on the aggregate wealth of the virtual society. The results of the simulations reveals that, in most cases, the effect of the interaction on the aggregate wealth of the society is negative, even in simulations where all agents are altruistic. Only in two situations could the aggregate, that is, the virtual society as a whole, benefit from the action of the donor. They are the rows shown in bold print. In both cases, the simulations allowed the appearance of synergies. In summary, the results of these simulations show that the conditions for the interaction between agents to produce positive effects on the aggregate wealth of the society are very specific and can be analytically summarized as follows:

1. For there to be an increase in the aggregate wealth of the society, the cost K of the utility donated must be less than the sum of all benefits B obtained thanks to this donation in the future: $K < \Sigma_t (A_t + B_t)$

2. Interactions, in order to be called altruistic, must comply with the condition: $\Sigma_t A_t > \Sigma_t B_t - K$ That is, the costs incurred by the donor must be greater than the utility recovered by him.

3. Society increases its aggregate wealth only if the actors create or add value.

These results can be illustrated graphically in Figure 11.2.

**Figure 11.2: Types of interpersonal relationships in terms of its social utility**

Schematic representation of the different forms of social interaction and their effect on the individual and the society, based on the results of the simulations with Sociodynamica.

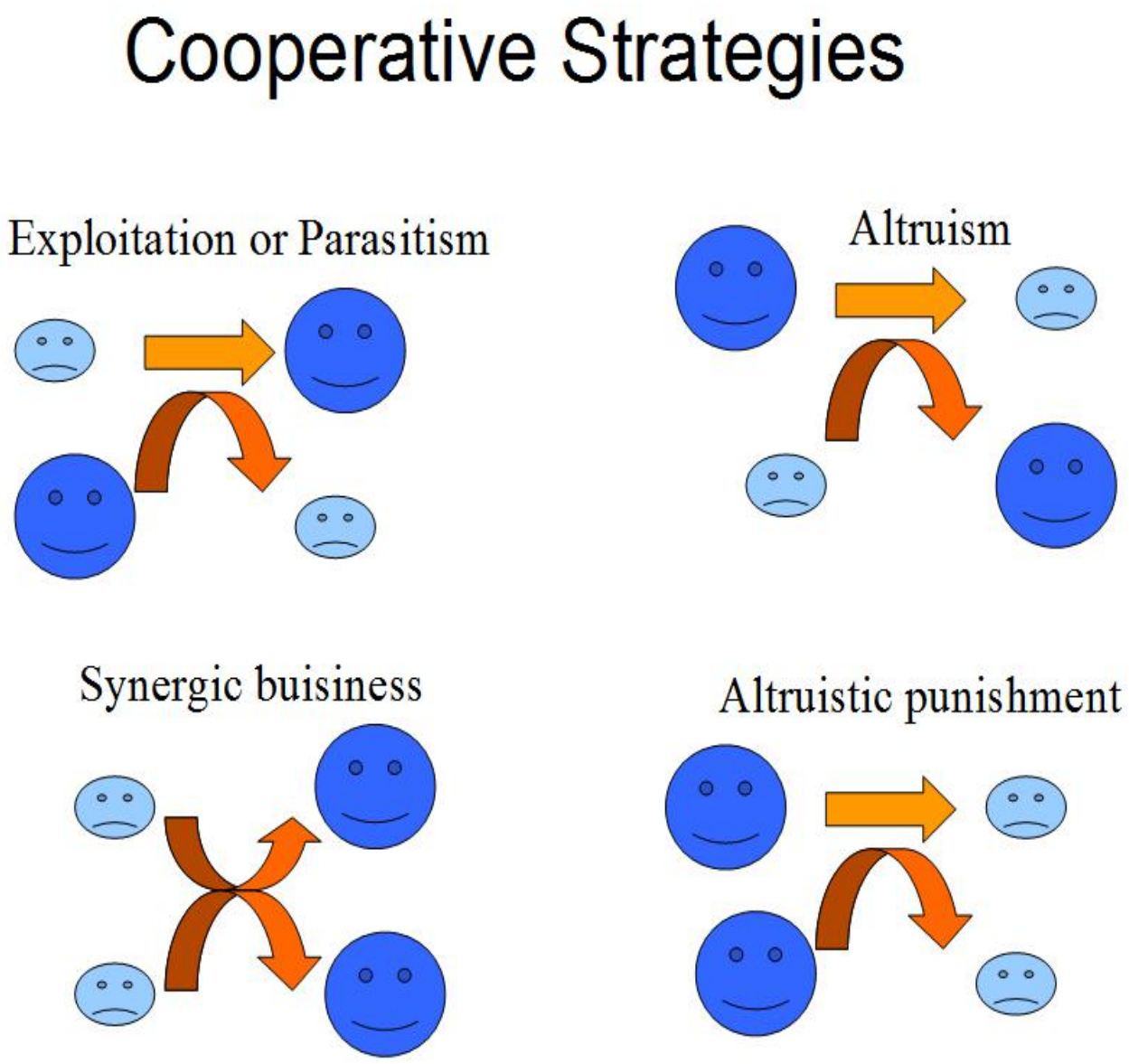

The figure illustrates that in their overall effect on society, parasitic interactions and the pure altruistic actions are detrimental. They both represent interactions of “zero sum”. That is, what one loses, the other wins and vice versa. The qualifier altruistic or parasitic only depends on the point of view of the agent who suffers the interaction, and in both cases, the result is negative on the aggregate. It is only with the presence of some synergistic effect, which creates value, that interaction can increase the aggregate utility of the system. In this case, the interaction is not zero sum and both agents win. It is the so- called “win-win” interaction. These are the interactions that produce wealth. In real life, most useful interactions or exchanges are rather dissipative, since the transfer of utility or wealth, by fulfilling the second law of thermodynamics in a closed system, produces losses. That is, interactions that do not produce any benefit to the recipient and cost energy or time to the actor. These interactions lower the aggregate wealth of the system since they waste or dissipate it.

In short, the diversity of actions and their effects on society can be classified into four broad categories as listed in Table 11.3.

**Table 11.3: Four basic types of social interactions**

| Action | Effect | Behavior |
|---|---|---|
| **Wise** | Both the individual and society win | Behavior of social investment |
| **Selfish** | Individual wins at the expense of society | Destructive selfishness |
| **Altruistic** | Society wins at the expense of the individual | True altruism |
| **Stupid** | The individual and the society lose | Destructive Behaviors |

Now then, the interactions we humans carry out have been selected and decanted by biological evolution and our cultural history. Consequently, they must have an adaptive reason. Purely dissipative interactions can be considered non-adaptive and clearly all societies reject them. They are considered as social aberrations, antisocial conducts or negative behaviors.

Interactions of "zero sum", especially the selfish ones, are the type that has dominated the social interactions of *H. sapiens* for most of its history. It is the interaction of the hunter-gatherer with non-genetically related peers. The prey or fruit that an individual donates or takes is not available to others. The fish he extracts from the river cannot be caught by another fisherman. These limitations of the environment favor mutualistic interactions in which every act of generosity is expected to be rewarded in the future. It is the basis of economic ideologies that prioritize the distribution of wealth. This way of seeing the world can be summarized as "God creates the resources, we human hand them out."

On the other hand, synergistic interactions have very powerful properties and the actions that promote them can be considered as an investment in the short and/or long term, as they are creators of wealth. Its omnipotent presence is relatively recent in human societies. Its operation is based on technology and knowledge. The creation of technological empires is not possible by the simple addition of wills. It also requires the complementation of knowledge and skills, the enhancement of cooperative interaction with a broad understanding of sciences and appropriate technologies. It is a product of the scientific and industrial revolution. For example, the optimized cooperation between workers, technicians, managers and financiers in a technology company, creates an added value that is greater in orders of magnitude (10 to 1000 times greater), calculated per capita, than the value created by the interaction among the same number of subsistence farmers.

## *Adam Smith's Invisible Hand and F. Hayek's Economic Calculus*

Another revelation of the simulations of Sociodynamica is that the optimal behaviors of individuals in a society without division of labor are different from the optimal behaviors in a more sophisticated society, where specialists perform different tasks. The creation of synergies is much more likely in societies with division of labor. The diversity of the behaviors of individuals in those societies favors the creation of wealth and the strength of the society. That is, the economic and social development of a society is conditioned by the development of behaviors and values of its members.

To visualize this effect Sociodynamica was run to simulate money. While doing so, fundamental questions had to be addressed. What is money? Classical economic theory was of little help. An orthodox economic description would read as follows: Assuming the economic environment as a multidimensional hyperspace in which economic consequences of human action are represented, and the exchange as a way to promote social processes; then money does not appear suddenly in the economy, but as the result of pre-existing values. This is clearly the case with certain objects that assume a monetary function. In this sense, we define money as the purest expression of the concept of the economic value. Money is then more than a simple economic concept: it is an abstraction of a social relationship. Money is a token, a kind of abstraction that acts like a medium of exchange and which assumes some knowledge and trust granted on it from the ones who use it. Obviously, the use of money eases transactions because it is a well-known abstraction, which summarizes -at the moment of its acceptance- the whole group of elements that supports it. As token, money is an appropriate tool in systems of growing complexity. Theoretically speaking, in economic orthodoxy, the central function of money is to be a means of payment; therefore it is no more than ‘a veil ' in the economic world, an element that only facilitates the exchange. On the contrary, post-Keynesians and Circuitists, emphasize money as being generated through the credit process and its performance as token to be decisive in the economy.

These definitions of money are not very useful for building simulation models. Money is popularly used to value things. This valuation however is mostly flawed. A diamond of one gram might cost fortunes and is of little practical use, whereas a pill of one gram might save the life of a person and costs cents. It is difficult to specify what money is or which asset can be considered as money, because money is usually defined by its functions. It is used to facilitate commerce and credit. Those who emphasize the credit process point out that what we call money is in permanent change, in continuous flow through the economy, and no predictable relationship among the quantity of money and the behavior of economic agents exists. Thus, the velocity of circulation of money is not constant, but volatile and essentially unpredictable. Continuous financial innovations induce a growing credit readiness and an increase of the velocity of monetary circulation. In the sophistication process, the financial innovation generates cost reductions as well as a diversification of risks, increasing the level of liquidity. In practice, however, this process has proved to be crisis prone.

Simulations allow testing different levels of sophistication or financial depth as to their economic characteristics. We simulated economic systems in heterogeneous environments, based on barter, going through those of monetized economies (money as species), self-financing, simple intermediation, until arriving to systems of complex intermediation with external sources characterized by an active management of assets and equities. Here, each agent adjusted prices according to the supply and demand of the resources he had access to and to its needs. These experiments allowed us to track the path of money and/or of any resource, including its velocity, as experienced for each individual actor, while simultaneously monitoring macroeconomic variables. An advantage of simulations is that they do not have to assume pure competitive markets and general equilibrium models. They can focus on the effects of imperfections in the market and on asymmetric transaction costs. Simulations comparing barter economies that have no financial instrument, with economies that use money to ease trades, and with financial economies with the possibility of money creation thorough credit processes, are very revealing. For example, very simple economies with agents dividing skills for farming, mining and trading can be contrasted with economies constituted by multi-skill or omnipotent agents, using each of the three types of economic scenarios just mentioned. The virtual world simulated included fields covered with “Food” and fields containing “Minerals”. Each agent was visualized so that its size was proportional to its wealth, where the width was determined by the food and minerals accumulated and the height by the amount of money the agent possesses. The thickness of the border of the data point was proportional to the perceived cost of living calculated as prices of food and minerals; and the color of the border reflected the ratio food price/mineral price, which allowed identifying agents that paid more for minerals and those willing to pay more for food. The color of the body of spheres described the type of agent. In this form, the simulation allowed following micro-economic events and at the same time have a macroeconomic picture. Color pictures and actual simulations are presented in:

http://atta.labb.usb.ve/Klaus/EC/ECVideos.html , and simulations can be run using http://atta.labb.usb.ve/Klaus/Programas.htm

**Figure 11.3. Example of a snapshot of a simulation using Sociodynamica**

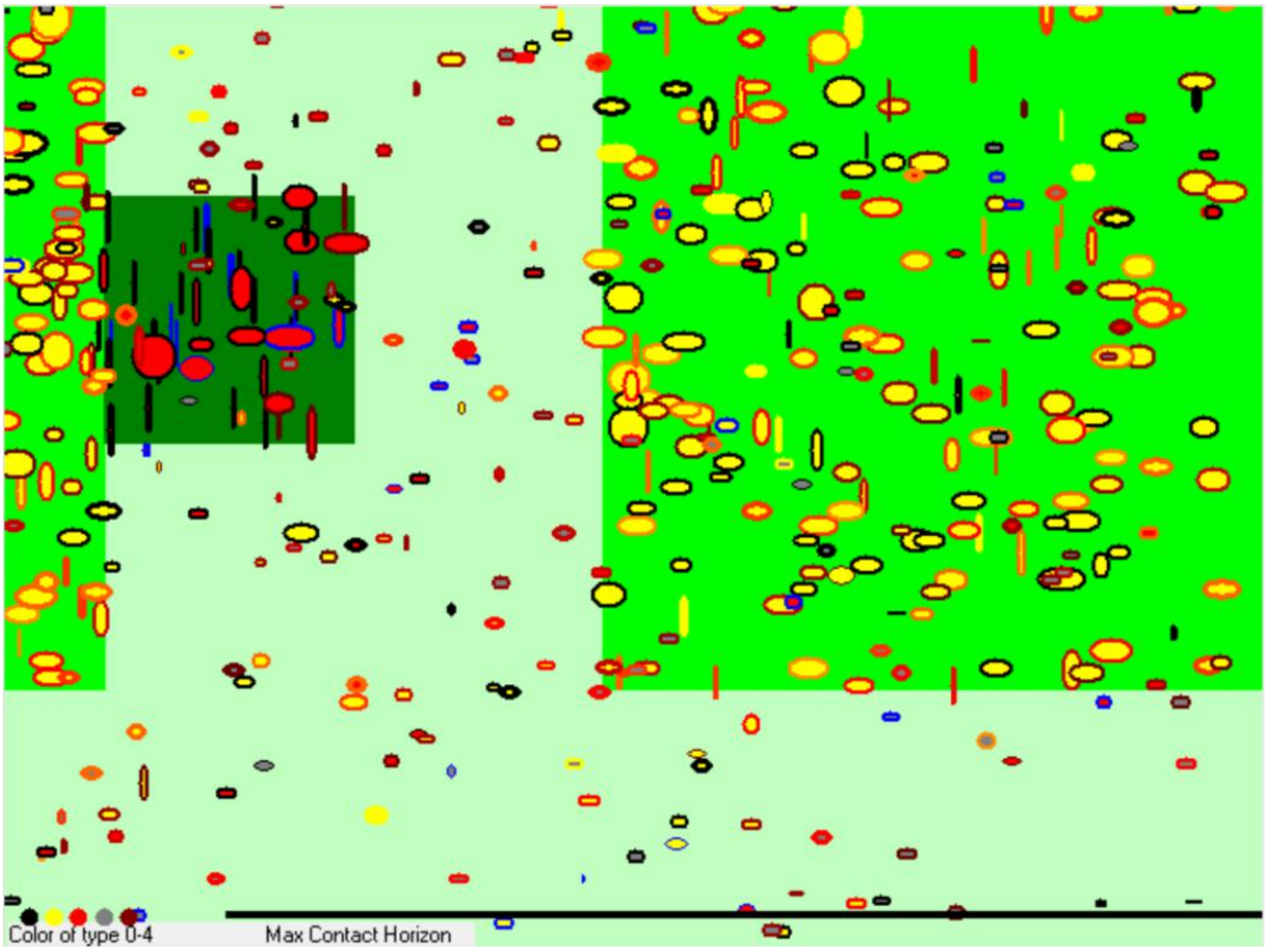

- Virtual resource landscape populated by different type of agents agentsBright green field is covered with “Food”; darker green field is covered with “Minerals”; the lightest green is devoid of resources.

- The size of the bubble is proportional to the wealth of the agent and the height by the amount of money the agent possesses.
- The thickness of the border is proportional to the perceived cost of living calculated as Food Price + Mineral Price
- The color of the border is more reddish or even yellow the higher the ratio Food/Mineral Price. Agents with red and yellow pay more for minerals, whereas those with blue or black pay less for minerals compared to what they are willing to pay for food. The color of the body of sphere described the type of agent: yellow were Farmers, red were Miners and gray were Traders.

The main results obtained were:

1. *Wealth distribution follows Zipf's law and economic dynamics is heterogeneous and non-linear:* Evidently, classical economic analysis is too simple to capture many relevant aspects of real economies, such as price asymmetries, skewed wealth distribution functions, and heterogeneities in the economic landscape.

2. *The most important invention in economic history is credit, not money:* GDP levels achieved were similar in barter and

monetary economies, the second a little higher. The big difference arose in simulations of financial economies where credit was allowed. Here, production increases almost ten times. Traders' contribution to social production increases while farmers and miners decrease their share.

3. *Division of Labor and Adam Smith's Invisible Hand:* Among the hundreds of different combinations of economic features explored, the single feature that increased the total production of wealth of the virtual society most was the existence of division of labor founded on a diversity of skills. Simulation with omnipotent agents performing all tasks, produced much less aggregate wealth than simulations where different agents performed different tasks, such as farming, mining or trading. This counter-intuitive result was due to the fact that optimal prices and conditions for trade are different for each agent, depending on its spatial position in the virtual world. Omnipotent agents had to assume average solutions to balance their different tasks. Therefore, they never traded at optimal prices and optimal quantities according to their spatial position.

In such a simple system, the working of the invisible hand, first postulated by Adam Smith, becomes visible. In this virtual world with only two resources and three type of agents, division of labor promotes cooperation over competition, favoring win-win interactions over zero-sum games. The win-win cooperation allows synergies to emerge that endow the market with its extraordinary economic force. Simplifying the virtual world one bit more, by reducing the number of resources or of type of agents, make the synergistic network to collapse. We have found the simplest system where Adam Smith's invisible hand becomes visible and where it can be quantified. Evidently, division of labor produces synergistic increases in economic terms in more complex settings. It is time to study them quantitatively in more detail. Specific simulations of real scenarios making visible the actions of each agent, might even allow predicting the price dynamics, and by doing so solving Friedrich Hayek's Economic Calculus.

These are just a few selected examples of the power of agent based simulations. Simulations in experimental economic research are growing are as they are a fantastic tool to make complex phenomena visible to human understanding. They have an enormous potential in teaching economics. With the proper adaptations, didactic games simulating real economies may become an indispensable tool for the teaching of economics at all levels of educational and academic specialization. Science learned through simulations might overcome complexity and beat self-serving cognitive biases.

# 12. A NEW ECONOMY

## *What have we learned?*

Simulations allowed us to analyze the economic dynamics in human societies that have already disappeared. They showed us the characteristics of societies where barter prevails and allows us to compare them with societies that use money for its trade. We can model modern or postmodern societies, where all human activity can be securitized and valued monetarily or otherwise. From simulations, we can deduce the characteristics of agents that are more likely to thrive in each of these economies. This exercise already revealed that behaviors that optimize the wealth of hunters and collectors of natural resources differs from the behavior of successful farmers, and these two differ from the optimal behaviors agents need to display in a technological advanced society.

The adaptations that evolution imposed on humans to optimize their agricultural societies included means to enforce tradition, respect for the family, saving, private property, investment and work, all values that favor success in an agrarian society. These qualities are enhanced by cyclic climate changes to which agricultural practices and other human activities must adapt. These values differ from those evolved by hunters and collectors. Many of the attitudes and economic behaviors we call zero-sum are adaptations that served humanity for a long period of its history when it practiced hunting and gathering fruits and tubers. These adaptations value individual strength, alertness, opportunism, gang organizing, chieftaincy and the pursuit of personal gain regardless of the effect on others. There are still large human populations in the world engaged in these economic practices, or in economic activities of rentier economies. Even humans living in developed industrial societies may find niches were paternalistic government handouts or particular economic incentives optimize these attitudes. These values hinder the establishment of modern, technologically advanced and globalized economies.

The society dominated by science and technology is dependent on the conduct of their agents. Here, the agents with capacity for innovation, critical scientific analysis, statistical perspective, and ability to cooperate to produce synergies and long-term investment are the most efficient at producing wealth. Technology allows us access to sources of wealth hidden in nature. The application of these technologies requires diverse information and complex social interactions between individuals. The interactions between science, technology, information and social organizations may produce wealth far greater than that achieved by the sum of individual efforts. If so, we speak of synergistic interactions. Technology and the modern enterprise or companies are the catalysts that allow these synergies to occur. Therefore, they are considered by most economists as the basis of a modern economy. Evolutionary economics studies how the technologies and the companies that exploit them transform themselves and the societies from which they emerged. It is this dynamic of mutual interaction directed by the evolution of modern economies as described by Carlota Perez in her book *Technological Revolutions and Financial Capital*), that we need to master in order to advance our economies in the future.

Analyzing the industrial and technological revolutions that various countries have experienced at different times is illuminating. We saw that the phenomenon of massive increase of wealth accessible to the average human inhabitant of the country occurs only after a technological revolution (Figure 2.3). It is with the creation of the *Homo industrialis* that human societies began to accumulate wealth steadily. This revelation allows us to attack the problem of the wealth of nations with a new perspective: What differentiates the pre-technological human (i.e. pre-agricultural, pre-industrial, pre-info-technical) from the post-technological human is decisively associated with the conditions that allow citizens of a given country to form wealth-creating societies, or fail in that attempt.

In the case of the industrial revolution, this phenomenon is better understood. Figure 12.1 shows the first industrial power, the United Kingdom, increasing slowly its wealth in the middle of the IX Century. The USA started its increase in wealth due to industrialization at the end of the IX Century and overtook the United Kingdom very soon afterwards. All countries starting their industrialization later managed their industrialization much faster. South Korea, the last one to have completed it, achieved high wealth of its citizens in much shorter time spans than the United Kingdom or the USA. China, the last curve in the plot, has just started recently its industrialization.

**Figure 12.1: Four examples of industrial revolutions.**

GDP per capita in 1990 International Geary-Khamis $ per capita (vertical axis) during 200 years (horizontal axis) in 7 countries. Data from Maddison Historical Database.

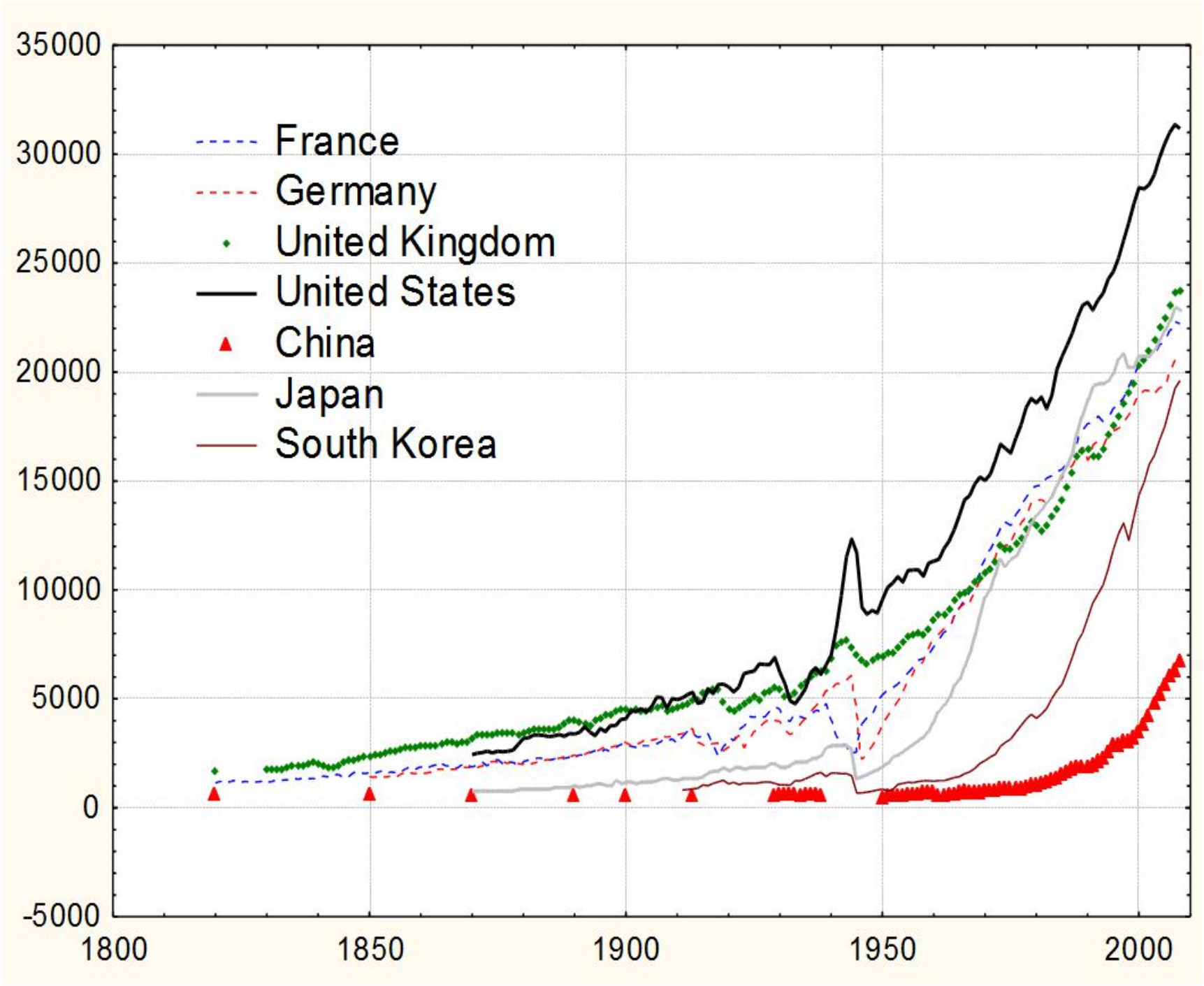

We know now that the behavior and values that optimize the economic efficiency of an agent in a society of hunters and gatherers, with little or no division of labor, differs markedly from those that optimize the action of the agent in agricultural societies or in a hunter and gatherer group. This phenomenon seems to have deep roots in the anthropological reality and even in our genes. Studies in pre-literate societies around the world reveal that the communities that keep economies based on hunting and fishing, emphasize independence as a core value in the education of their children, while communities that get their livelihood from agricultural activities, maintain educational systems with great emphasis on obedience. Thus, we cannot rule out the existence of a genetic component in some behaviors that promote or inhibit the formation of a modern technological society. The *agricultural Homo* cannot be created overnight. Nor can the *agricultural* Homo, the *hunter Homo* or the *Neolithic Homo* be transformed into *Homo industrialis* without profound changes in the bio-social relationships on which society is based. All humans do not necessarily have to follow the same path to progress. Some groups of people will want to avoid the economic progress and maintain lifestyles compatible with the *Neolithic Homo*. Others will want to create post-technological societies or societies governed by information technologies in an accelerated way.

We must repeat, at the risk of being annoying, that a genetic regulation of a behavior does not involve an inability to modulate or change behavior by learning or by other cultural methods. Genetic regulation of a behavior implies that its ramifications and effects on other behaviors are complex and not discernible by simple observation or by computer simulations. Methods developed by population genetics and ethology are required to understand and grasp the scope and significance of the genes that regulate behavior. These techniques, although long ignored by economist, will be more frequently applied to the evolutionary study of our societies and economies.

The switch from a pre-industrial society to one engaged in an industrialization process may not be the same to that of other technological revolutions. Future technological revolutions, such as health services, communication, entertainment and others, might have a different dynamic as their motor of economic activities differs from the industrialization processes we have experienced so far. We might expect, however, that the economic success or failure of a nation will continue to depend on prevailing individual attitudes. There are many ways to metamorphose a collector of resources, such as a gold digger (called Garimpeiro in Brazil) working with artisanal methods, to a technological entrepreneur like the industrial "Robber Barons" in the USA or high tech gurus such as Bill Gates. We all have to learn how to better synergize individual action by building better company that will generate wealth steadily in the future. Our societies, in different degrees, are composed of “Garimpeiros” and “Gates”. Understanding the relationship among attitudes, habits, values and economic productivity better will eventually allow us to deepen our comprehension of how nations generate their wealth and how they might do so in the future. Only interdisciplinary research will help us move in this direction, overcoming the limitations of our minds and unraveling complex relationships of variables. We need more and better empirical evidence, experimental validation, and clever novel observations to convert economics in a real science.

Sustained investment of efforts, resources and time spent on researches, both theoretical and practical, will enable to clarify doubts and will help us understand the dynamics of the transformations in our society. A greater knowledge of social phenomena will increase our individual freedom and the successful planning and selection of our future. We all have a right to know why, how and where we are limited and when, where and how we can expand our potentials. It is the development of the science of the social and of the dynamics of its transformations that will ensure us a future with freedom.

## *The Edge of Chaos*

Theoretical scholars of complex systems have coined the term "*at the edge of chaos*". This concept stems from the search for the optimal conditions for stimulating creativity in a complex system. It was found that those conditions are the same that lead the system into chaos, only that the optimum in terms of creativity of the system is achieved just before falling into chaos. Simulations with artificial societies also show this effect. Creative societies meet the optimum conditions for generating wealth. They must be isolated, but not too much; they must undergo changes and receive new stimuli, but not too much; they must maintain traditions, but not always; they must have a centralized structure but must also foment decentralization; they must innovate and retain; they need to concentrate efforts and must expand possibilities.

Wars might spur innovation. In the preparation and development of an armed conflict, both nationally and internationally, there are individuals, companies and societies that can benefit both by the stimuli to certain sectors of production favored by the conflict, and by the expectations or the results of the war that will affect future trade. It is even conceivable that in particular cases, the incentives outweigh the losses, especially if the conflict fails to develop fully. Others argue that it is only with the intense emotional pressure that wars cause that humans develop new cutting-edge technologies. Past world wars induced the development of modern aviation and improved our communications systems, for example. Even Galileo financed part of his studies on the forces of gravity with ballistic technological developments with warfare purpose. On the other hand, the intensely competing modern pharmaceutical industries are a clear example of how there can be investment in research and development in peacetime without the spur of a war.

The vision that opens with the systematic study of complex systems, together with the recent and not so recent history of the nations of the world, shows that no extreme or simple solution will be viable nor produce development in any country. Magical policies to overcome poverty do not exist. A country determined to take the path of economic development requires social development, democracy, strengthening of its institutions, citizen participation, education, rule of law and a lot of knowledge, but also discipline and motivation. It is easy to determine when the limit is passed, it is difficult to predict how long it will take before the optimal balance is achieved.

This intrinsic characteristic of any complex system and therefore, of human and non-human societies, requires countries that want to achieve their goals of economic and social development to keep policies open to criticism and rectification. It is timely learning and suitable knowledge, continuous monitoring and constant adjustment, which must guide the development policies of a modern nation. The yardstick by which the success or failure of a given economic policy will be measured will not be its elegance, theoretical bases or ideological conceptualization but the maximization of the creation of wealth.

## *Evolution of Creativity*

Understanding the process of human creativity associated to wealth creation is essential to build societies capable of sustaining economic growth for long periods, and thus reduce poverty levels to a minimum. Let us use our time windows of analysis to learn from the past 4000 years of history of the Hominids on earth represented schematically in Figure 12.2. We see that during this period, progress in knowledge and in the power of domain humanity has over nature has not been uniform. We have had very productive periods in terms of the production of knowledge in antiquity. These periods were replaced by others in which the spread of religions dominated, accompanied by attitudes that value less the acquisition of new knowledge, but emphasizes tradition and the continuity of values. These were times with conservative social dynamics seeking uniform distribution of knowledge at the expense of creating new knowledge. In the graph, we can identify at least three different periods for these characteristics. The period includes the golden age of classical Greece, the period that saw the birth of Moses, Christ and Muhammad and the dissemination of their ideas, and the period that saw the birth of science and the European cultural renaissance.

**Figure 12.2: The Story of Mankind**

Representation of time windows with some of the most important intellectual advances of the last four thousand years. Designed by the author.

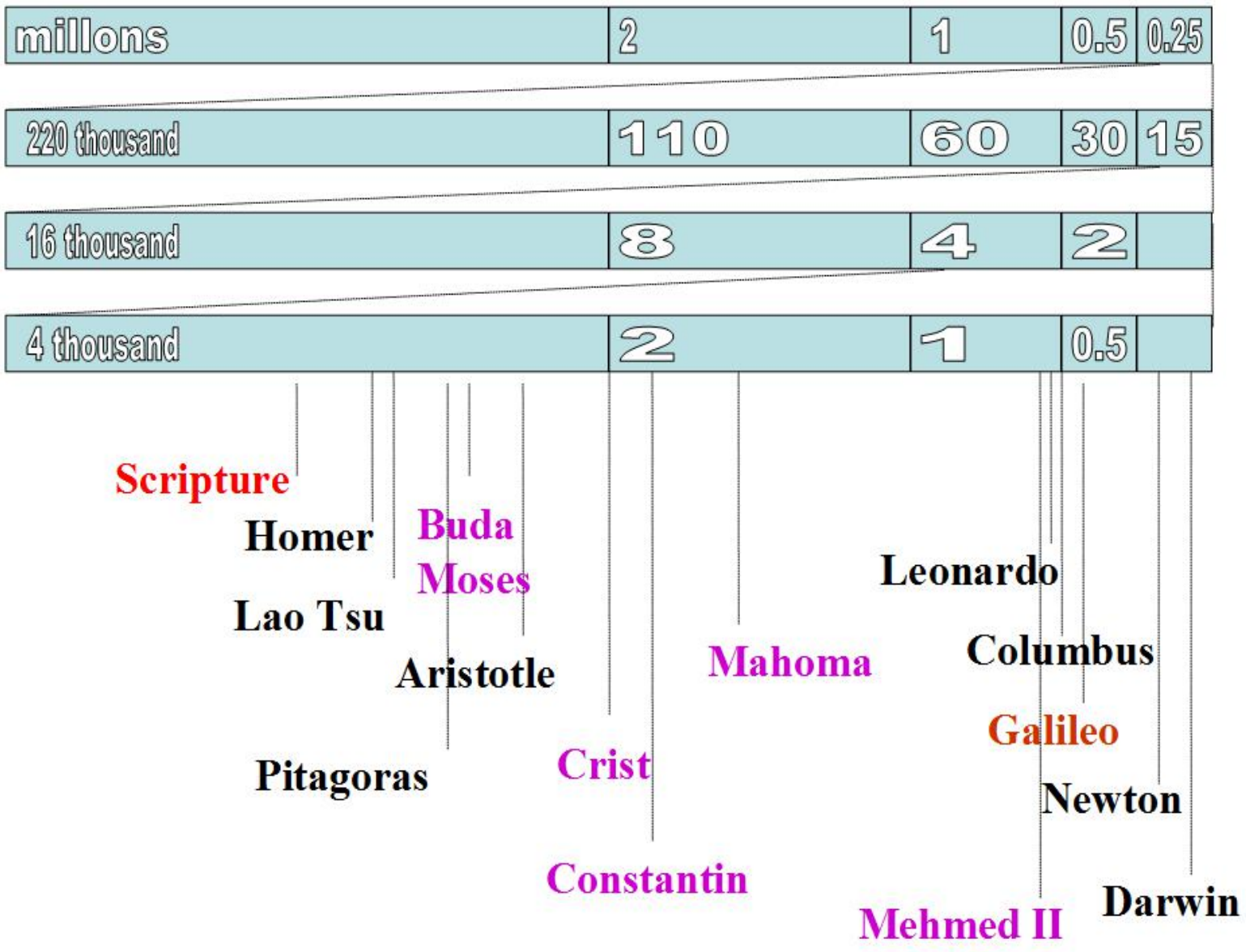

From a historical perspective of the economy, humankind has endured major stages of economic development, each tied to a particular optimization of values, skills and motivations of the individuals. The first of these stages, concomitant to the end of the hunter-gatherer period, was described by Homer and is associated with an ideal individual of great physical ability and good sense of observation. The second stage, concomitant with a prevalence of farmers, shepherds and trader, represented by the classical Greek and Chinese period, is associated with a significant sophistication of socialization and the emergence of institutions that foment conservation of traditional knowledge that promotes respect for authority and age. This era gave rise to another characterized by the emergence of the dominant contemporary religions. At this time scale, it is only in relatively recent times that we experienced the European Renaissance, which allowed the emergence of experimental sciences, which gave birth to the technology that triggered a new economic era through an industrial revolution. This revolution produced exponential economic and population growth with important implications for humanity.

**Figure 12.3: History of Science**

Representation of time windows with some of the most important technical advances of the last four hundred years. Designed by the author.

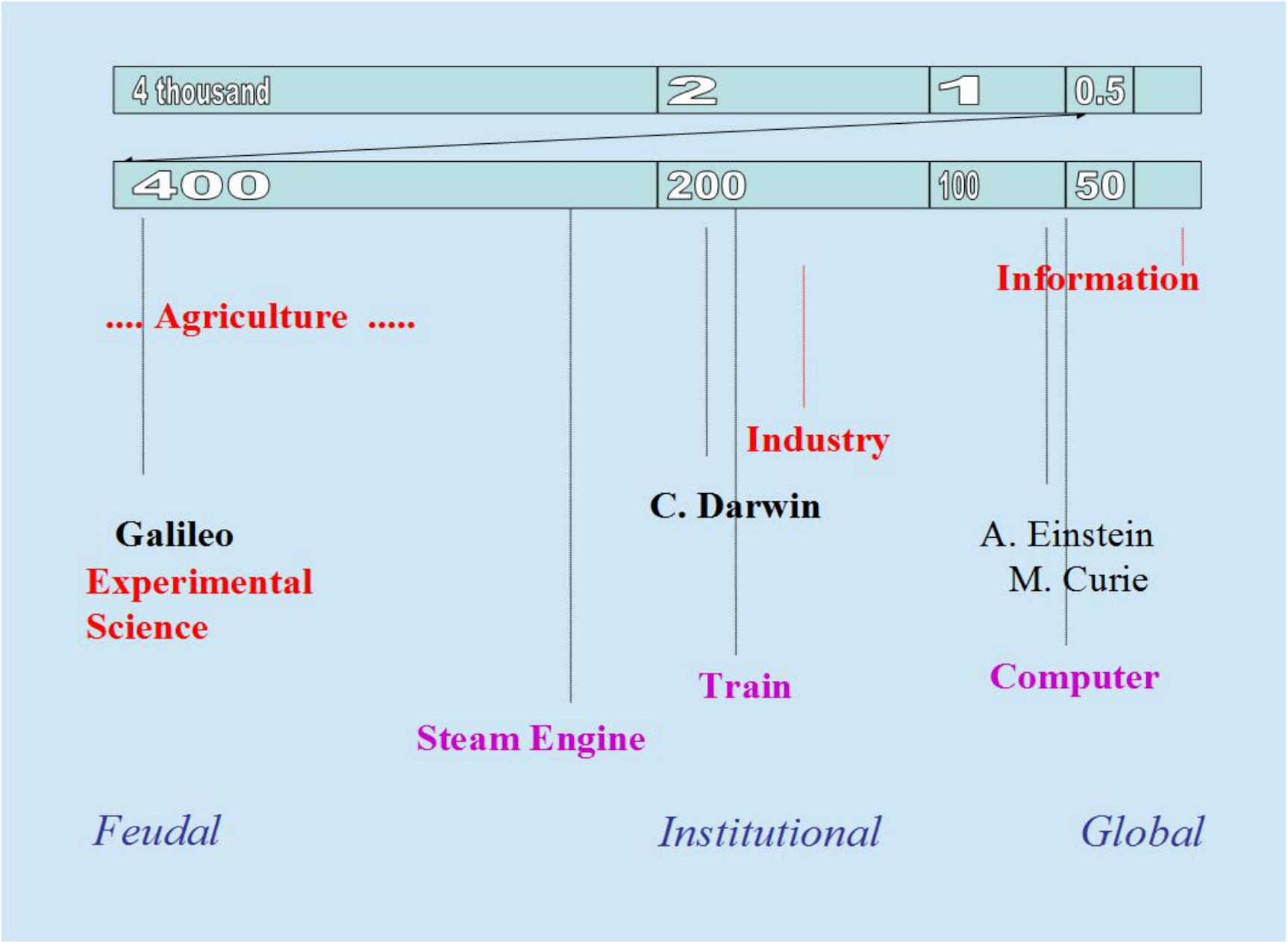

A historical analysis of events in more recent times (Figure 12.3) reveals in more details this Renaissance revolution that allowed the rapid emergence of technologies that transformed the economies of the societies that adopted them. As we saw before, societies that produce and maintain each of these stages of economic development are markedly different. Hunters and gatherers develop the institution of the family and sometimes the clan or extended family, with authority figures that are an extrapolation of the paternal or maternal authority. Farmers, due to their sedentary habits, established by the farming activity, develop cities and kingdoms that in turn are based on behaviors and values that enable to maintain the network of relationships that sustains society. More recently, it was the scientific and technological development, triggering an industrial revolution, which provided the conditions that allowed the emergence and development of modern democracies in some countries. From this last analysis, the importance of the emergence of industry driven by technology in the history of human civilization is evident. The phenomenon that has higher correlation with the onset of that rapid technological and economic growth in rich nations is the emergence of science, which in turn enables the development of technologies that affect the values, skills and motivations of the individual and modulates society.

A well described experiment of the role of science in the modernization and industrialization of a nation is reported by Japanese history. It is the scientific value of medical treatises from the Dutch, experimentally proven by Japanese scholars, including Sugita Gempaku (died in 1817), that prompt Japanese society to discard the Chinese classic texts in 1774 and initiate a development of knowledge based on real facts rather than on empty theories. This process leads Japan to eventually become a modern technological power capable of maintaining the most long-living population in the planet at levels of wealth among the highest in the world.

## *Skepticism and Humility*

The view of human history sketched above has and ever improving science as the motor of recent human economic and social progress. Economics itself must become a more scientific endeavor if it wants to increase its explanatory and predictive powers. Specifically is has to firmly ground its heuristic thinking on the following premises:

a- Our mind suffers from varied and serious limitations, which makes it impossible that, by itself, it can understand the world it perceives. That is, our mind is limited in its perceptual and cognitive abilities and is unable to see itself and the physical surroundings in all their details, relationships and properties.

b- All explanatory theory of the world around us should be rational and logical, so that is understandable and usable by any other human.

c- A theory, to be called scientific, must be testable and falsifiable. That is, any theory must be testable by methods and systems external to the mind, and every theory must be formulated in such a way that it can be declared false based on observations of nature or on experiments. The experiment, the empirical observation and the manifestations of nature prevail over any virtue that a theory developed by our mind may have. In other words, reality exceeds imagination.

A philosophical-poetic image by Plato, describes the relationship between man and his environment, and the reason of the effectiveness of science to increase our knowledge and our mastery over nature. According to a famous metaphor of Plato, the activity of searching for the truth about man is like that of beings locked in a cave and seeing only shadows of what happens outside the cave. These shadows, although partly reflect reality, are not. The tools, the technique and the scientific method rigorously applied allow us to advance in our description and understanding of nature, but there can always be another level of complexity, for which our observations are only some deformed shadows of this new reality. It is by creating hypotheses, making observations and experiments, discarding hypotheses and creating new ones, that science advances our knowledge. Empirical science is not the holder of the truth but is the only human heuristic construct that has allowed and will allow real progress in our understanding of ourselves and of the world around us.

We could paraphrase the British Prime Minister during World War II, Winston Churchill (1874-1965), when referring to the vicissitudes of the democratic system, calling it a system with many flaws, but the least bad of all systems of government. Science and its daughter technology is a company with many flaws, but is the most efficient of the ones we know to produce technological and economic progress, and is the only one that has so far achieved to produce benefits and improvements in the quality of life tangible to a vast majority of human populations.

To the extent that societies have a greater number of individuals who accept and use the scientific method as defined previously, to that extent their capacity for technological innovation and thus their capacity for economic growth increases. We are in the midst of a profound transformation of humanity. There are societies that have gone through a technological revolution and are possibly starting a new one, that of information or knowledge that will allow the articulation of a global economy. There are other societies dominated by myth and religion, which are formed by large sectors of the population living with a subsistence economy, characteristic of the hunters and gatherers or of subsistence farmers who first appeared on our planet about ten thousand years ago. It is, therefore, a function of the modern State to provide the necessary education to start a life in an economy dominated by science and technology to vast sectors of society. This transformation of the individual through formal and informal education may allow backward societies to provide opportunities for decent and productive lives for their citizens.

Easily exploitable resources are running out in the planet. The later the task of leading society to embrace a modern economy is undertaken, the more technology requirements this transformation will need. It is the challenge of our generations, and we have an advantage, we can learn from the experience of others, if it is our will.

## *Creating Synergies*

The existence of human societies is justified only if the action of the group produces greater benefits than the sum of individual efforts could produce. That is, society bases its very existence in the creation of synergies. This is the *raison d'être* of the family, of the company, trade association, corporation, of every institution. It is the *Value Creation*, the *Economic Benefit*, the *organizational Income*, the *Production of Profits*, the *Value Added*, the production of a greater amount of wealth than the one that comes as inputs of capital and labor, or the one that would be produced if each works on its own, which justifies the existence of every business. Synergy in economy equals the *Capital Gain* of Karl Marx. If a society fails to overcome the difficulties imposed by its environment more efficiently than its members separately would, it would not exist. The slogan *in unity there is strength* reflects the fact that today nearly all assume the relationship between society and synergy as obvious. However, it is not necessarily so. Not all forms of organizing ourselves and regulating our social coexistence produce additional benefits to all involved.

The emergence of synergistic phenomena as a result of the structuring of societies is not yet fully understood by scholars in any discipline. Science is only now beginning to develop tools, such as the analysis of complex systems, for example, that could enable us to dispel the fog that does not let us see this phenomenon clearly. Computer models have helped us measure and detect the presence of these synergies that emerge from the interactions between individuals, but in the background, we do not yet understand them enough to handle, use, exploit and optimize them.

The creation of synergies to produce wealth is the main issue that motivates every student of business and management. It is in very specific situations, in the construction of a factory, in establishing a business, in identifying a business partner, in the search for a partner, that elements that may or may not produce synergies become evident. This discovery of synergies seems to be a process in which the sum of individuals is more efficient than that of the isolated individual. Therefore, it is a process that requires the concerted effort of researchers from various disciplines acting in a consilient way that can wake up sources of synergies of the group.

At the level of economies and nations, history has allowed, more by chance than by scheduling and planning, synergies to emerge among creativity, science, technology, society and economy that produce rapid growth of wealth and fast increase of the welfare of a society. We expect that with sustained investment in scientific research oriented to analyzing this phenomenon, rational action can gradually displace chance. This will eventually allow the emergence of a truly modern social engineering that can create the conditions required for the emergence of societies that allow boosting the creative and cooperative capacity of its members, increasing the welfare of each and every one of them.

This phenomenon, in which synergistic effects transform the system that allowed the emergence of these effects, is not unique. It is known by physicists studying atomic sub-particles, by chemists studying the properties that emerge from the compounds formed of many different atoms, by biologists who try to understand the physiology of tissues composed of cells that interact in complex ways with each other and by evolutionary biologists studying the emergence of societies of living organisms. The so-called information revolution is also based on creating synergies between different realms of knowledge and the economy. It is the interaction of many sciences, in a creative, free, and plastic way, with economic, psychological and social sciences, which may eventually elucidate and understand these so fundamental and important problems for the future of humanity.

# 13. EPILOGUE

More than one reader will be disappointed because we did not get after all this intellectual journey, to a simple, unique and indisputable conclusion. This often happens with scientific progress in general and this is the fate of those who undertake the analysis of complex systems. The human mind seeks simple and clear explanations of the phenomena it observes. But forcing to reach these conclusions without the appropriate knowledge that would allow doing it in a consilient manner is a matter of dogmas, not science. The scientific mind must settle for "knowing that we do not know". It is the recognition of our ignorance on so many topics that will orient new intellectuals in their research efforts to advance our knowledge. I hope this analysis has aroused more doubts than the ignorance it might have overcome. The main objective of this work is to open minds rather than settle dogmas.

The history of humanity has not stopped or will stop in an imaginable future. *Homo sapiens* are in full transformation from gregarious animal to eusocial or sophisticatedly social animal. This transformation will still take a few millennia, during which we will experience setbacks and progresses in our route towards societies that meet our natural needs. Science, technology and the generation and distribution of wealth may be problems that only historians will remember. But until this happens, it is our duty to understand these processes and guide our actions for the benefit of societies more attuned to the needs and desires of humanity.

Before concluding I would like to note that I have omitted many key elements in this analysis. The origin of the wealth of nations can be analyzed from many other viewpoints not explored here. It is possible that even the most appropriate approach must await the appearance of a science that will emerge in the future. However, I think it is clear that a complex problem, as the one that has occupied us here, requires interdisciplinary approaches and attitudes, and both orthodox and heterodox methods of analysis. I also believe that the need to expand the application of the scientific method to analyze situations that have so far been monopolized by branches of the academy that apply more narrative methods was made evident. There is lot to do but I think the direction of the road ahead is traced.

It is my conviction that the phenomenon of the production of wealth and the processes that produce poverty in the world are understandable only with a broad interdisciplinary scientific vision. We know of practical experiences that achieved to reduce poverty in relatively short times. We also know the economic policies that allowed rich countries to generate sustained economic growth. We know that we have failed many times in predicting or handling complex economic processes. But successes seem to be more frequent and failures less so. The health and wealth of humans have increased on average in the past and is still increasing. That is why we can be confident to eventually produce ever better tools to design policies for a successful and inclusive economic development, where all or the great majority participate. Nevertheless, it is only with rationality, management of relevant information, capacity to process criticisms, ability to produce synergies, value merit and knowledge and intensive use of science, that we will be able to guide developing countries and the world towards a future that ensures a better standard of living and welfare of all its inhabitants. These efforts, by the way, will reduce war, terrorism and violent conflicts that breed on economic and political failures.

To similar conclusion have arrived countless thinkers. Let me quote one in particular, before finalizing: The thinker, cosmopolitan educator and polymath Simon Rodríguez (1771-1854) wrote: *"... Republics are Established but not Founded. ... safe means of reforming customs, to avoid revolutions - starting with SOCIAL ECONOMICS, with a POPULAR EDUCATION, deducting the DISCIPLINE proper of the economy to two principles: destination to its USEFUL exercises, and aspiration to PROTECTED property. ... ";* and two hundred years later we have little to add to this insight. Hopefully, a new science of economics will help us to be more assertive in guiding our political actions.

The final destination of the evolution of our societies is not known, nor is the success of modern technology guaranteed. Uruk, Tikal, Machu Picchu, Angkor, Persepolis, Troy, are only a few of countless examples of cities, societies, cultures and civilizations that have disappeared in the course of human history. It is in our hands to “load the dice” so as to improve our chance to reach a better destination. It is a responsibility that emerges from exercising freedom.

# Acknowledgments

With their comments, criticisms, ideas, motivations and corrections to the manuscript, in one way or another the following persons have given their contribution to this work, but they are not responsible for its content. They are in chronological order: Alfredo Rosas, Raúl Parra, Asdrúbal Baptista, José Luis Cordeiro, Werner Jaffé, Diana Ajami, Miguel Rodríguez, Angelina Jaffé, Arturo Gutiérrez, Evaldo Vilela, Flavio H. Caetano, Sary Levy, Violeta Rojo, who read and criticized the entire manuscript twice and made many corrections meticulously, Orlando Albornoz, Harold Vladimir Zavarce and Maxim Ross, Friedrich Welsh, Herbert Koeneker, Daniel Vernagy, Sandra Caula, and several anonymous reviewers. The support of Carlos Pacheco and Domingo Maza Zavala were essential for the edition of this book. The English version profited from the input of Silvya Baes, Sarah Lawrence, Leila Campoli and especially Jennifer Bernal. Many important contributions were probably made by people I forgot to mention, because their suggestions were so well formulated that I integrated them into my thoughts unconsciously.

*The writing of this book began on the banks of the Miranda River, Pantanal, Mato Grosso do Sul, Brazil in February 2002 and concluded in Saint Francoise, Guadeloupe, French West Indies, in February 2005. They were retaken at the banks of the Congaree River in South Carolina and finalized in the Valley of Caracas, at the feet of the Avila in October 2014.*